\documentclass[a4paper,fleqn,usenatbib]{mnras}

\usepackage[T1]{fontenc}
\usepackage{ae,aecompl}

\def \aap {A\&A}

\def \aj {AJ}
\def \apj {ApJ}
\def \apjs {ApJS}
\def \apjl {ApJL}

\def \mnras {MNRAS}
\def \pasp {PASP}
\def \nat {Nature}

\def \chisq {\ifmmode \chi^2 \else $\chi^2$ \fi}
\def \spose#1{\hbox to 0pt{#1\hss}}
\def \lta{\mathrel{\spose{\lower 3pt\hbox{$\sim$}}\raise 2.0pt\hbox{$<$}}}
\def \gta{\mathrel{\spose{\lower 3pt\hbox{$\sim$}}\raise 2.0pt\hbox{$>$}}}

\def \kms {\ifmmode \,\rm km\,s^{-1} \else $\,\rm km\,s^{-1} $ \fi }
\def \kpc {\ifmmode {\rm~kpc} \else ${\rm~kpc}$\fi}
\def \pc {\ifmmode {\rm~pc} \else ${\rm~pc}$ \fi }
\def \Gyr {\ifmmode {\rm~Gyr} \else ${\rm~Gyr}$\fi}
\def \Msun {\ifmmode M_{\odot} \else $M_{\odot}$ \fi}
\def \Lsun {\ifmmode L_{\odot} \else $L_{\odot}$ \fi}
\def \Rsun {\ifmmode R_{\odot} \else $R_{\odot}$ \fi}
\def \Msunpyr {\ifmmode M_{\odot}{\rm~yr}^{-1} \else $M_{\odot}{\rm~yr}^{-1}$ \fi}
\def \hMsun {\ifmmode h^{-1}\,\rm M_{\odot} \else $h^{-1}\,\rm M_{\odot}$ \fi}

\def \LCDM {\ifmmode \Lambda{\rm CDM} \else $\Lambda{\rm CDM}$ \fi}
\def \sig8 {\ifmmode \sigma_8 \else $\sigma_8$ \fi}
\def \OmegaM {\ifmmode \Omega_{\rm M} \else $\Omega_{\rm M}$ \fi}
\def \OmegaL {\ifmmode \Omega_{\rm \Lambda} \else $\Omega_{\rm \Lambda}$\fi}
\def \Deltavir {\ifmmode \Delta_{\rm vir} \else $\Delta_{\rm vir}$ \fi}
\def \rhocrit {\ifmmode \rho_{\rm crit} \else $\rho_{\rm crit}$ \fi}
\def \rhou {\ifmmode \rho_{\rm u} \else $\rho_{\rm u}$ \fi}
\def \zc {\ifmmode z_{\rm c} \else $z_{\rm c}$ \fi}

\def \rhos {\ifmmode \rho_{\rm s} \else $\rho_{\rm s}$ \fi}
\def \rs {\ifmmode r_{\rm s} \else $r_{\rm s}$ \fi}
\def \cvir {\ifmmode c_{\rm vir} \else $c_{\rm vir}$ \fi}
\def \Rvir {\ifmmode r_{\rm vir} \else $R_{\rm vir}$ \fi}
\def \Vvir {\ifmmode V_{\rm vir} \else $V_{\rm vir}$ \fi}
\def \Mvir {\ifmmode M_{\rm vir} \else $M_{\rm vir}$ \fi}
\def \Nvir {\ifmmode N_{\rm vir} \else $N_{\rm vir}$ \fi}
\def \Jvir {\ifmmode J_{\rm vir} \else $J_{\rm vir}$ \fi}
\def \Evir {\ifmmode E_{\rm vir} \else $E_{\rm vir}$ \fi}
\def \vvir {\ifmmode v_{\rm vir} \else $v_{\rm vir}$ \fi}
\def \lam {\ifmmode \lambda \else $\lambda$ \fi}
\def \lamp {\ifmmode \lambda^{\prime} \else $\lambda^{\prime}$ \fi}
\def \Vmax {\ifmmode V_{\rm max} \else $V_{\rm max}$ \fi}
\def \Mdm {\ifmmode M_{\rm dm} \else $M_{\rm dm}$\fi}

\def \Mgas {\ifmmode M_{\rm gas} \else $M_{\rm gas}$\fi}
\def \Mcg {\ifmmode M_{\rm cg} \else $M_{\rm cg}$\fi}
\def \Mhg {\ifmmode M_{\rm hg} \else $M_{\rm hg}$\fi}
\def \Mdisc {\ifmmode M_{\rm disc} \else $M_{\rm disc}$\fi}
\def \Md {\ifmmode M_{\rm d} \else $M_{\rm d}$ \fi}
\def \Mda {\ifmmode M_{\rm d,0\%} \else $M_{\rm d,0\%}$\fi}
\def \Mdb {\ifmmode M_{\rm d,20\%} \else $M_{\rm d,20\%}$\fi}
\def \Mdc {\ifmmode M_{\rm d,40\%} \else $M_{\rm d,40\%}$\fi}
\def \md {\ifmmode m_{\rm d} \else $m_{\rm d}$\fi}
\def \Mb {\ifmmode M_{\rm b} \else $M_{\rm b}$\fi}
\def \Mbh {\ifmmode M_{\rm b,pri} \else $M_{\rm b,pri}$\fi}
\def \Mbs {\ifmmode M_{\rm b,sat} \else $M_{\rm b,sat}$\fi}
\def \zo {\ifmmode z_{0} \else $z_{0}$ \fi}
\def \rd {\ifmmode r_{\rm d} \else $r_{\rm d}$\fi}
\def \rg {\ifmmode r_{\rm g} \else $r_{\rm g}$\fi}
\def \rb {\ifmmode r_{\rm b} \else $r_{\rm b}$\fi}
\def \rs {\ifmmode r_{\rm s} \else $r_{\rm s}$\fi}
\def \rc {\ifmmode r_{\rm c} \else $r_{\rm c}$\fi}
\def \rvir {\ifmmode r_{\rm vir} \else $r_{\rm vir}$\fi}
\def \rbh {\ifmmode r_{\rm b,pri} \else $r_{\rm b,pri}$ \fi}
\def \rbs {\ifmmode r_{\rm b,sat} \else $r_{\rm b,sat}$ \fi} 

\usepackage{graphicx}	
\usepackage{amsmath}	
\usepackage{amssymb}	
\usepackage{float}
\usepackage{tablefootnote}
\usepackage{lastpage}

\title[Bayesian inference of the IMF in Single Stellar Populations]{A hierarchical Bayesian approach for reconstructing the Initial Mass Function of Single Stellar Populations}

\author[M. Dries et al.]{
M. Dries,$^{1}$\thanks{E-mail: dries@astro.rug.nl}
S. C. Trager,$^{1}$
L.V.E. Koopmans$^{1}$
\\
$^{1}$Kapteyn Astronomical Institute, University of Groningen, PO BOX 800, 9700 AV Groningen, The Netherlands\\
}

\date{Accepted XXX. Received YYY; in original form ZZZ}

\pubyear{2016}

\begin{document}
\label{firstpage}
\pagerange{\pageref{firstpage}--\pageref{LastPage}}
\maketitle

\begin{abstract}
{Recent studies based on the integrated light of distant galaxies suggest that the initial mass function (IMF) might not be universal. Variations of the IMF with galaxy type and/or formation time may have important consequences for our understanding of galaxy evolution. We have developed a new stellar population synthesis (SPS) code specifically designed to reconstruct the IMF. We implement a novel approach combining regularization with hierarchical Bayesian inference. Within this approach we use a parametrized IMF prior to regulate a direct inference of the IMF. This direct inference gives more freedom to the IMF and allows the model to deviate from parametrized models when demanded by the data. We use Markov Chain Monte Carlo sampling techniques to reconstruct the best parameters for the IMF prior, the age, and the metallicity of a single stellar population. We present our code and apply our model to a number of mock single stellar populations with different ages, metallicities, and IMFs. When systematic uncertainties are not significant, we are able to reconstruct the input parameters that were used to create the mock populations. Our results show that if systematic uncertainties do play a role, this may introduce a bias on the results. Therefore, it is important to objectively compare different ingredients of SPS models. Through its Bayesian framework, our model is well-suited for this.}
\end{abstract}




\section{Introduction}
Since its introduction by \cite{Salpeter}, the initial mass function (IMF) has been a key parameter in the study of stars, stellar populations and galaxy evolution. Salpeter initially parametrized the IMF as a single power law. However, it was recognized later on that when the IMF was extended down to the lowest stellar masses that it did not follow a single power law. Instead, a lognormal distribution \citep{MillerScalo1979}, a multicomponent power law \citep{Kroupa1993} or a combination of a lognormal distribution for low masses and a power law for higher masses \citep{Chabrier} were proposed as alternatives. \cite{Dabringhausen}, among others, have shown that the latter two are in fact very similar. 

Measurements of the IMF have long been based on direct star counts and mass estimates of resolved stars. These kinds of measurements are not possible for stars beyond the Local Group. Therefore, for many astrophysical studies the IMF has been assumed to be universal and similar to the one of  the Milky Way. However, recent studies \citep{Dave, vanDokkum2008, Treu_et_al_2008, Grave_Faber_2010, Conroy_vanDokkum_2012a, Cappellari_et_al_2012, Spiniello_2012, Spiniello_2013, Ferreras, LaBarbera} suggest that the IMF might not be universal on a cosmological scale, indicating that the relative number of low-mass stars in the population changes as a function of galaxy mass or velocity dispersion. This may have important consequences for the many properties of galaxies that are derived on the basis of the IMF, such as their stellar content, chemical enrichment history and even their evolutionary history: see e.g. \cite{Tinsley1972}. 

Starting with \cite{Tinsley1968}, stellar population synthesis (SPS) models have been developed to transform the observable properties of a galaxy into a set of physical properties. Among the physical properties encrypted in the spectrum of a galaxy are its star formation history (SFH), the amount of gas and dust that it contains, its chemical composition and its IMF. However, deriving the low-mass end of the IMF on the basis of the spectrum of a galaxy is not straightforward. Dwarfs with masses $M < 0.4$ M$_{\odot}$ contribute only $\sim$1\% to the integrated light of an old stellar population \citep{Conroy_vanDokkum_2012a}. Nevertheless, they contribute 12 and 42\% of the total stellar mass for a standard Kroupa IMF \citep{Kroupa} and a Salpeter IMF with a low-mass boundary of 0.1 M$_{\odot}$ and a high-mass boundary of 100 M$_{\odot}$, respectively. For younger stellar populations, the relative contribution of low-mass stars to the spectrum is even less. In old stellar populations, the spectral similarity of low-mass stars and the most luminous stars (the K and M giants) further complicates the situation. However, a number of (gravity-sensitive) spectral features are known to be sensitive to either dwarfs or giants \citep{Faber_French, Schiavon_1997a, Wing_Ford, Schiavon_1997b, Gorgas_1993, Worthey_et_al, Schiavon_2007, Cenarro_2003, Spiniello_2012}. The challenge for a SPS model is to extract this information from a spectrum. 

Most SPS models are built upon three basic ingredients: a stellar evolution model in the form of isochrones as a function of age and metallicity, a stellar library, and an IMF. These ingredients form the basis of what is known as a single stellar population (SSP): a single, coeval population of stars with the same metallicity. The isochrone describes which stars are present in a stellar population, the stellar library provides a set of stellar spectra, and the IMF determines the distribution of stars along the isochrone. All of these ingredients have their own uncertainties. Models of stellar evolution are often one-dimensional codes and the results of these codes depend on the adopted prescriptions for uncertain factors, such as overshooting, rotation, interaction between binary stars, and mass loss. Stellar libraries may be theoretical, empirical or a combination of both. Both empirical and theoretical libraries have their own advantages and disadvantages. The assumption of a universal IMF is another source of uncertainty.

Real galaxies are not SSPs. Combining a set of SSPs with a SFH, a model for chemical evolution and possibly a dust model allows the construction of composite stellar populations (CSPs). To date, many different SPS models have been developed, e.g. \cite{BruzualCharlot}, \cite{PEGASEHR}, \cite{Maraston2005}, \cite{Conroy_vanDokkum_2012a}, and \cite{MIUSCAT}. Most of these models allow the user to change the IMF. Once an IMF or a set of different IMFs is defined, this allows the model to create synthesized spectra for a grid of different model parameters (including the IMF). The synthesized spectra are then compared with observed galaxy spectra to obtain values of, e.g., metallicity or IMF slope. Determining the best-fitting parameters is often done through a minimization technique, such as $\chi^2$ minimization in for example \cite{ulyss}. However, the ultimate goal of a SPS model would be a direct inference of the physical parameters from the spectrum.

Each SPS model uses its own set of ingredients, and the way in which these ingredients are combined also varies. This requires an objective manner to compare different SPS models with each other. A solution to this problem is provided by Bayesian inference. In this paper we develop a hierarchical Bayesian framework for SPS. Within our model, a parametrization of the IMF is used to construct a (flexible) IMF prior. Given this prior, our model allows for a direct inference of the piecewise IMF from the spectrum of an SSP. The outline of the paper is as follows. In Section 2 we discuss the Bayesian framework of our model. In Section 3 we describe how we construct a representative set of stellar templates as an input for our model. We then test our model by applying it to respectively mock SSPs and SSPs created by other SPS models in Sections 4 and 5.

\section{Hierarchical Bayesian inference}
\label{sec:modelDescription}
Within a hierarchical Bayesian model there are multiple levels of inference. In this paper, we have two levels. The first level of inference assumes that a certain model family $\mathcal{H}$ can provide a proper description of the truth and tries to obtain the best-fit for the free parameters within that model family (parameter estimation). The second level of inference allows us to compare a set of different model families $\{\mathcal{H}_i\}$ and tries to infer the most probable model family given the data (model comparison). In analogy to the analysis presented by \cite{MacKayBayesianAnalysis}, we derive a hierarchical Bayesian framework for modelling spectral energy distributions.

Neglecting the effect of extinction, which we will include in a future publication, the spectral energy distribution of a stellar population may be considered as the sum of the spectra of all the stars that it contains. This allows us to write the spectrum of the stellar population as a linear combination of a certain set of stellar templates. For an SSP, the stellar types that are present in the population are defined by an isochrone. The most important parameters that define an isochrone are its age $t$ and metallicity $[\mathrm{M/H}]$. An isochrone provides us with the stellar parameters (effective temperatures, surface gravities, luminosities, colors, initial masses, and current masses) of all the stellar types present in the corresponding SSP. These parameters are typically combined with a stellar library and an interpolator to create a spectrum $\mathbfit{s}$ for each of the isochrone stars. This procedure is discussed in more detail in Section \ref{sec:stellarTemplates}. 

Suppose that $\mathbfss{S}$ is a matrix with the spectra of all the isochrone stars in its columns, such that $\mathbfss{S}_{ij}$ corresponds to the i-th flux density bin $\mathbfit{s}_{i}$ of the spectrum of isochrone star $j$. Since the isochrone is defined by its age and metallicity, $\mathbfss{S} = \mathbfss{S}(t, [\mathrm{M/H}])$ is implicitly also a function of age and metallicity (i.e. the age and metallicity define the isochrone, the isochrone defines a set of stars and their parameters which in turn are used to create a corresponding set of stellar spectra that goes into $\mathbfss{S}$). Although here we consider SSPs, $\mathbfss{S}$ might equally well contain the spectra of the stellar templates of a CSP. In that case $\mathbfss{S}$ also becomes a function of the SFH of the stellar population.

If $\mathbfit{w}$ is a vector with the number of stars for each stellar template in $\mathbfss{S}$, the spectrum $\mathbfit{g}$ of the stellar population is given by:
\begin{equation}
\mathbfit{g} = \mathbfss{S}\, \mathbfit{w}.
\end{equation}
For each star, an isochrone provides in general both the initial mass and the current mass (taking into account a prescription for possible mass loss). Since $\mathbfit{w}$, hereafter called weights, represents the number of stars for each stellar template, the initial masses of the isochrone allow us to relate $\mathbfit{w}$ to the IMF of the stellar population and vice versa. The IMF,
\begin{equation}
\xi(M) \equiv \frac{\mathrm{d}N}{\mathrm{d}M},
\end{equation}
of an SSP is related to $\mathbfit{w}$ through
\begin{equation}
\label{eq:weights_IMF}
\xi (m_j) = \frac{\mathbfit{w}_j}{m_{\mathrm{high}} - m_{\mathrm{low}}},
\end{equation}
where $\mathbfit{w}_j$ is the number of stars of template $j$ in the stellar population and $m_j$ is the initial (rank-ordered) mass associated with template $j$ by the isochrone. The boundaries $m_{\mathrm{low}}$ and $m_{\mathrm{high}}$ of the mass bin are defined such that 
\begin{equation}
\label{eq:boundaries}
\begin{aligned}
\begin{split}
m_{\mathrm{low}} & = \frac{m_{j-1} + m_j}{2} \\
m_{\mathrm{high}} & = \frac{m_{j} + m_{j+1}}{2} .
\end{split}
\end{aligned}
\end{equation}
For the lowest mass template, $m_{\mathrm{low}} = m_{\mathrm{LMCO}}$ (low-mass-cut-off of the IMF) and for the highest mass template we take $m_{\mathrm{high}} = m_j$. In this way, $\mathbfit{w}_j$ corresponds to the number of stars in the mass bin ($m_{\mathrm{low}}$, $m_{\mathrm{high}}$). The way in which $m_{\mathrm{low}}$ and $m_{\mathrm{high}}$ are defined ensures that mass bins never overlap.

\begin{figure*}
	\centering
	\includegraphics[width=14cm]{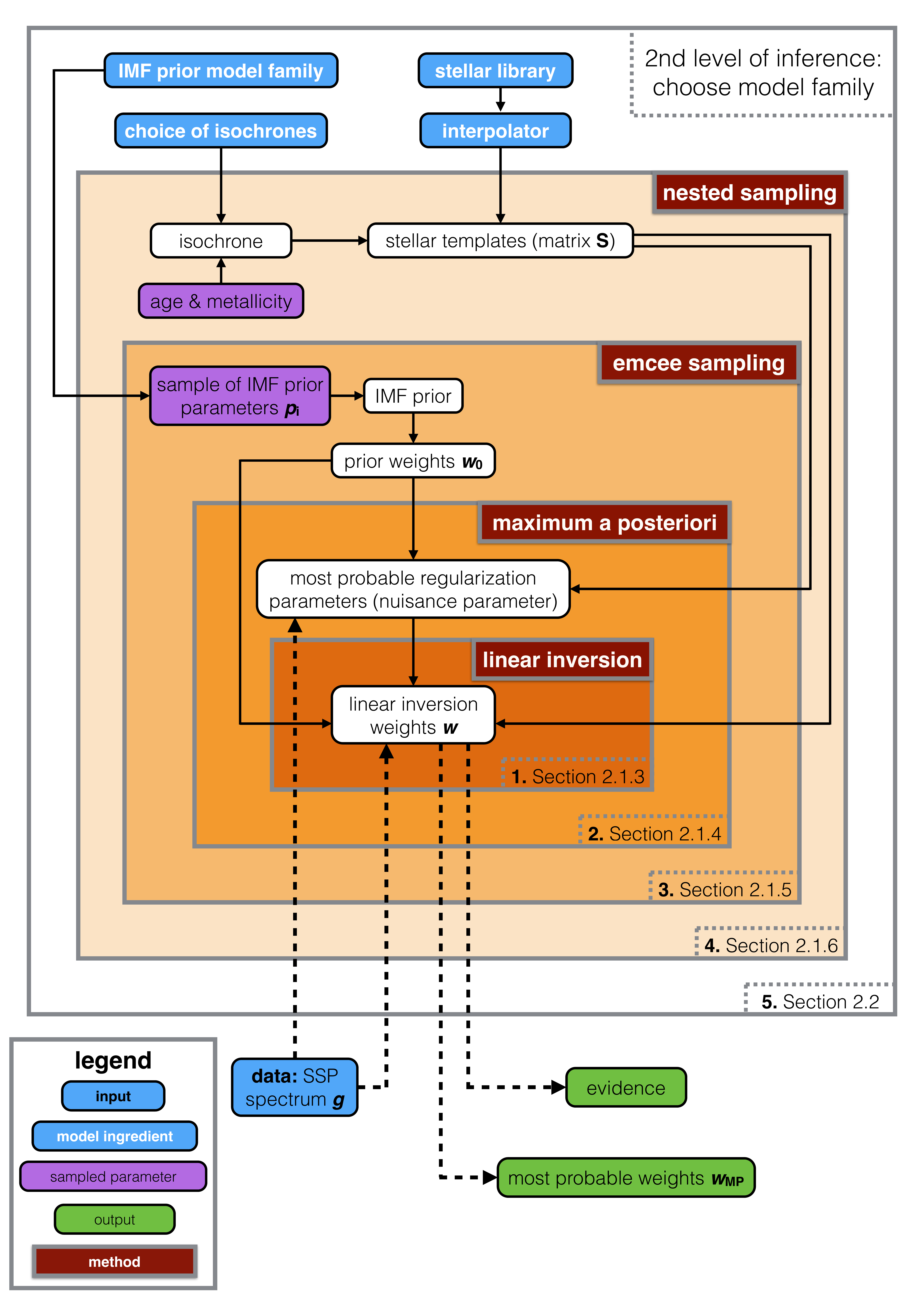}
	\caption{Flow diagram illustrating the hierarchical nature of our model for an SSP. We have a spectrum which forms the input data of the model. At the outermost level, we define a model family $\mathcal{H}$ by choosing a set of isochrones, a stellar library, an interpolator, a regularization scheme, and a parametrization for the IMF prior. One level below, an age and metallicity define an isochrone. This isochrone is combined with the stellar library and the interpolator to create a set of stellar templates $\mathbfss{S}$. The most probable age and metallicity are derived by calculating the evidence for every combination in a predefined age-metallicity grid. Going another level further down, a particular sample $p_{i,0}$ of the IMF prior model parameters $p_i$ is transformed into a prior on the weights $\mathbfit{w}_0$. The IMF prior model parameters are sampled using \texttt{emcee} \citep{EMCEE}. At the next level, the most probable value of the regularization parameter $\hat{\lambda}$ is determined given the data $\mathbfit{g}$, the stellar templates $\mathbfss{S}$ and the prior on the weights $\mathbfit{w}_0$. Finally, at the innermost level the data $\mathbfit{g}$, the stellar templates $\mathbfss{S}$, the prior on the weights $\mathbfit{w}_0$ and the most probable regularization parameter $\hat{\lambda}$ are combined to reconstruct the most probable weights $w_{\mathrm{\!_{MP}}}$ and calculate the evidence for that particular set of parameters. At the highest level, we also calculate the evidence for a model family by marginalizing over all the free parameters in that model family. This allows us to compare different model families with each other and is referred to as the second level of inference.}
	\label{flowDiagram}
\end{figure*}

In this section we first discuss how to find the most probable distribution of weights $\mathbfit{w}_{\mathrm{\!_{MP}}}$ by using the combination of regularization and hierarchical Bayesian inference. Subsequently we discuss how Markov Chain Monte Carlo techniques may be used to reconstruct the parameters of a certain IMF prior parametrization and to find the age and metallicity of the SSP. As a last step, we show how different model families may be compared on the basis of their Bayesian evidence. The hierarchical nature of the different steps in the model is illustrated in Fig. \ref{flowDiagram}.

\subsection{The first level of inference}
At the first level of inference, we assume that model family $\mathcal{H}$ is the correct model family and we try to infer the model parameters given the data $\mathbfit{g}$. A model family $\mathcal{H}$ is defined on the one hand by the set of stellar templates $\mathbfss{S}$ (e.g. a set of SPS  templates as a function of age and metallicity) and on the other hand by the parametrization of the IMF prior, which defines the space of possible priors on the weights $\mathbfit{w}$. In Section \ref{sec:stellarTemplates} we discuss how to construct a representative set of stellar templates whereas the parametrization of the IMF prior, and hence the prior on the weights, is discussed in more detail in Section \ref{sec:thePrior}. 

Given a certain model family $\mathcal{H}$, we infer the number of stars $\mathbfit{w}$ for each of the templates in our model given the spectrum of a stellar population $\mathbfit{g}$. Note that in this paper we consider SSPs but in principle this can also be done for CSPs if we include the SFH and chemical evolution of the stellar population in our model as well, as we plan to do in the future.

\subsubsection{The most likely solution}
In reality the spectrum of a stellar population contains noise, such that the observed spectrum of the stellar population becomes
\begin{equation}
\mathbfit{g} = \mathbfss{S}\, \mathbfit{w} + \mathbfit{n},
\end{equation}
in which $\mathbfit{n}$ represents the noise in the data. Assuming that the noise is Gaussian distributed, the likelihood of the data ${\cal{L}}(\mathbfit{g}|\mathbfit{w}, \mathbfss{S})$ given the weights $\mathbfit{w}$ and the stellar templates $\mathbfss{S}$ is 
\begin{equation}
\label{eq:likelihood}
{\cal{L}}(\mathbfit{g}|\mathbfit{w}, \mathbfss{S}) = \frac{\mathrm{exp}[-E_{\!_D}(\mathbfit{g}|\mathbfit{w}, \mathbfss{S})]}{Z_{\!_D}},
\end{equation}
in which 
\begin{equation}
\begin{aligned}
\label{eq:eqED}
\begin{split}
E_{\!_D}(\mathbfit{g}|\mathbfit{w}, \mathbfss{S}) & = \frac{1}{2}(\mathbfss{S}\, \mathbfit{w} - \mathbfit{g})^{\mathrm{T}} \mathbfss{C}_{\mathrm{D}}^{-1} (\mathbfss{S}\, \mathbfit{w}-\mathbfit{g}) \\
& \equiv \frac{1}{2}\chi^2,
\end{split}
\end{aligned}
\end{equation}
where $\mathbfss{C}_{\mathrm{D}}$ is the covariance matrix. The likelihood is normalized by 
\begin{equation}
\label{eq:normLikelihood}
Z_{\!_D} = (2\pi)^{N_{\!_D} / 2} (\mathrm{det\:} \mathbfss{C}_{\mathrm{D}})^{1/2},
\end{equation}
in which $N_{\!_D}$ is the number of data points in the spectrum. 

The most likely solution $\mathbfit{w}_{\mathrm{\!_{ML}}}$ may be found by maximizing the likelihood function ${\cal{L}}(\mathbfit{g}|\mathbfit{w}, \mathcal{H}_i)$ defined in equation \ref{eq:likelihood}. Maximizing the likelihood implies minimizing $E_{\!_D}$, so that we obtain
\begin{equation}
\nabla E_{\!_D}(\mathbfit{w}_{\mathrm{\!_{ML}}}) \equiv \frac{\partial E_{\!_D}(\mathbfit{w}_{\mathrm{\!_{ML}}})}{\partial \mathbfit{w}} = 0.
\end{equation}
The solution to this equation is given by
\begin{equation}
\label{eq:wML}
\mathbfit{w}_{\mathrm{\!_{ML}}} = (\mathbfss{S}^{\mathrm{T}} \mathbfss{C}_{\mathrm{D}}^{-1} \mathbfss{S})^{-1}  \mathbfss{S}^{\mathrm{T}}  \mathbfss{C}_{\mathrm{D}}^{-1}  \mathbfit{g}.
\end{equation}

Finding $\mathbfit{w}_{\mathrm{\!_{ML}}}$ is in general an ill-posed problem. Therefore we use a prior on the weights $\mathbfit{w}$ to regularize the solution that we obtain and to find the most probable distribution of weights $\mathbfit{w}_{\mathrm{\!_{MP}}}$.

\subsubsection{The prior}
\label{sec:thePrior}
Suppose that within a model family $\mathcal{H}$, the IMF prior is parametrized by a set of (non-linear) parameters which we call $p_i$. For example the IMF prior may be parametrized as a power law which is defined by its slope $\alpha$ and the normalization $C_{\mathrm{norm}}$: in that case $p_i = \{\alpha, C_{\mathrm{norm}}\}$. If we take one particular combination $p_{i,0}$ of the parameters $p_i$, this completely defines a prior $\xi_0(p_{i,0}, M)$ on the IMF. By using the initial masses associated to the templates through the isochrone, the prior $\xi_0$ on the IMF translates into a prior on the weights which we refer to as $\mathbfit{w}_0$. Since $\displaystyle \xi(M) \equiv \frac{\mathrm{d}N}{\mathrm{d}M}$, the number of stars that we have for template $j$ is given by
\begin{equation}
\label{eq:IMF_weights}
w_{0,j} = \int\limits_{m_{\mathrm{low}}}^{m_{\mathrm{high}}} \xi_0(p_{i,0},M) \mathrm{d}M,
\end{equation}
where $m_{\mathrm{low}}$ and $m_{\mathrm{high}}$ are defined in equation \ref{eq:boundaries}. So within one model family $\mathcal{H}$, there is a range of different models $\mathcal{H}_0$ that are defined by different priors $\mathbfit{w}_0$. The allowed range of priors $\mathbfit{w}_0$ within the model family is defined by the functional form of the IMF prior and its parameters $p_i$. Note that that the latter may have their own priors as well.

Once we have transformed the prior on the IMF $\xi_0$ into a prior on the weights $\mathbfit{w}_0$, we define the regularization function $E_{\!_S}(\mathbfit{w}| \mathbfit{w}_0, \mathbfss{C}^{-1}_{\mathrm{pr}})$ as
\begin{equation}
\label{eq:regFunction}
E_{\!_S} (\mathbfit{w}|\mathbfit{w}_0, \mathbfss{C}^{-1}_{\mathrm{pr}}) = \frac{1}{2} (\mathbfit{w}- \mathbfit{w}_0)^{\mathrm{T}} \mathbfss{C}^{-1}_{\rm{pr}} (\mathbfit{w} - \mathbfit{w}_0),
\end{equation}
where $\mathbfss{C}^{-1}_{\rm{pr}} = \nabla \nabla E_{\!_S}(\mathbfit{w})$ is the (constant) Hessian of $E_{\!_S}$. Hence the regularization function puts a penalty on $\mathbfit{w}$ for deviating from the prior distribution of weights $\mathbfit{w}_0$. Deviations are only possible if the data require it (i.e. if a deviation increases the likelihood more than it decreases the prior). Note that $\mathbfss{C}^{-1}_{\rm{pr}}$ is part of $\mathcal{H}$ as it is a model-dependent choice that relates to the form of regularization being used: we might for example use the identity matrix or enforce smoothness. Given the regularization function, the prior probability function may be expressed as
\begin{equation}
\label{eq:prior}
\mathrm{Pr}(\mathbfit{w}|\lambda, \mathbfit{w}_0, \mathbfss{C}^{-1}_{\mathrm{pr}}) = \frac{\mathrm{exp}[-\lambda E_{\!_S}(\mathbfit{w}|\mathbfit{w}_0, \mathbfss{C}^{-1}_{\mathrm{pr}})]}{Z_{\!_S}(\lambda)},
\end{equation}
where $\lambda$ is the regularization parameter and the prior probability function is normalized by $Z_{\!_S}$. A larger regularization parameter implies that there is more emphasis on the prior and less on the likelihood. In Section \ref{sec:regParameter} we show how the value of the regularization parameter may be derived in a Bayesian manner. The regularization parameter is therefore a nuisance parameter that must be marginalized over in the results.

\subsubsection{The posterior}
We have seen that a model family $\mathcal{H}$ is defined by the stellar templates $\mathbfss{S}$, the parametrization of the IMF $p_i$ and by the choice of the Hessian $\mathbfss{C}^{-1}_{\mathrm{pr}}$, such that $\mathcal{H} = \{\mathbfss{S}, p_i, \mathbfss{C}^{-1}_{\mathrm{pr}} \}$. Within such a model family there exists a range of models $\mathcal{H}_0 = \{\mathbfss{S}, \mathbfit{w}_0, \mathbfss{C}^{-1}_{\mathrm{pr}}\}$, in which $\mathbfit{w}_0$ is related to one particular choice $p_{i,0}$ of the IMF prior parametrization. In this way, each model $\mathcal{H}_0$ is defined by a different prior $\mathbfit{w}_0$. Hence the prior $\mathbfit{w}_0$ is not fixed  but should be considered as a flexible entity that is allowed to change within the boundaries of the IMF parametrization $p_i$. For every model $\mathcal{H}_0$, the likelihood and the prior $\mathbfit{w}_0$ are combined to find the most probable distribution of weights $\mathbfit{w}_{\mathrm{MP}}$. Defining $M(\mathbfit{w}|\mathcal{H}_0)$ as
\begin{equation}
\label{eq:eqM}
M(\mathbfit{w} | \mathcal{H}_0) = E_{\!_D}(\mathbfit{w} | \mathbfss{S}) + \lambda E_{\!_S}(\mathbfit{w} | \mathbfit{w}_0, \mathbfss{C}^{-1}_{\mathrm{pr}}),
\end{equation}
we apply Bayes' theorem to combine the likelihood function and the prior probability function into the posterior probability function
\begin{equation}
\begin{aligned}
\label{eq:posterior}
\begin{split}
P(\mathbfit{w}|\mathbfit{g}, \lambda, \mathcal{H}_0) & = \frac{\mathcal{L}(\mathbfit{g}|\mathbfit{w}, \mathbfss{S})\cdot \mathrm{Pr}(\mathbfit{w}|\lambda, \mathbfit{w}_0, \mathbfss{C}^{-1}_{\mathrm{pr}})}{P(\mathbfit{g}| \lambda, \mathcal{H}_0)} \\
& = \frac{\mathrm{exp}[-M(\mathbfit{w})]}{Z_{\!_M}(\lambda)},
\end{split}
\end{aligned}
\end{equation}
where the posterior is normalized by $Z_{\!_M}(\lambda)$. The last equation shows that the posterior probability distribution for the weights $\mathbfit{w}$ of the stellar templates is controlled by two functions. On the one hand there is the `goodness of fit' represented by $E_{\!_D}(\mathbfit{w}|\mathbfss{S})$ and on the other hand there is the deviation of the weights from the prior represented by the regularization function $E_{\!_S}(\mathbfit{w}|\mathbfit{w}_0, \mathbfss{C}^{-1}_{\mathrm{pr}})$. The balance between these two functions is set by the regularization parameter\footnote{Naively one might think $\lambda=0$ will give the highest posterior, but since the prior is normalized, lowering $\lambda$ makes the width of the prior very large, hence lowering the probability density at the position where the likelihood peaks. This lowers the posterior probability. Making $\lambda$ larger will increase the latter, but might make the fit to the data more difficult, lowering the likelihood. Balancing these is the Bayesian equivalent to Occam's razor, finding the simplest model that fits the data.}.

To find the most probable solution $\mathbfit{w}_{\mathrm{\!_{MP}}}$, we have to maximize the posterior probability density function (equation \ref{eq:posterior}). Maximizing $P(\mathbfit{w}|\mathbfit{g}, \lambda, \mathcal{H}_0)$ implies minimizing $M(\mathbfit{w}|\mathcal{H}_0)$, so that we have $\nabla M(\mathbfit{w}_{\mathrm{\!_{MP}}}) = 0$. Defining $\mathbfss{B} \equiv \nabla\nabla E_{\!_D}(\mathbfit{w}) = \mathbfss{S}^{\mathrm{T}} \mathbfss{C}_{\mathrm{D}}^{-1}\mathbfss{S}$ as the Hessian of $E_{\!_D}$ and using the definition of $E_{\!_S}$ from equation \ref{eq:regFunction}, we have for the most probable solution
\begin{equation}
\label{eq:wMP}
\mathbfit{w}_{\mathrm{\!_{MP}}} = \mathbfss{A}^{-1}(\mathbfss{S}^{\mathrm{T}} \mathbfss{C}_{\mathrm{\!_D}}^{-1} \mathbfit{g} + \lambda \mathbfss{C}^{-1}_{\mathrm{pr}} \mathbfit{w}_0),
\end{equation}
where $\mathbfss{A} \equiv \nabla \nabla M(\mathbfit{w}) = \mathbfss{B} + \lambda \mathbfss{C}^{-1}_{\mathrm{pr}}$ is the Hessian of $M(\mathbfit{w})$. In practice we solve equation \ref{eq:wMP} by using non-negative least squares\footnote{\cite{NNLS}} (NNLS) to ensure a physically meaningful solution (i.e. the number of stars cannot be negative: $\mathbfit{w} \nless 0$).

The most probable solution depends on the model $\mathcal{H}_0 = \{\mathbfss{S}, \mathbfit{w}_0, \mathbfss{C}^{-1}_{\mathrm{pr}}\}$ as well as on the regularization parameter $\lambda$ that regulates the balance between the `goodness of fit' and the penalty term resulting from the regularization function. The inversion of the most probable weights is represented by the inner block in Fig. \ref{flowDiagram}. To find $\mathbfit{w}_{\mathrm{\!_{MP}}}$, the inner block needs information from the outer levels: a set of stellar templates, a prior on the weights and a regularization parameter. In Section \ref{sec:regParameter} we show how to find the most probable value of the regularization parameter given the model and the data.

\subsubsection{Uncertainties of the most probable weights}
\label{sec:weightsError}
Using a second order Taylor expansion for $M(\mathbfit{w})$ around $\mathbfit{w}_{\mathrm{\!_{MP}}}$, we may approximate $M(\mathbfit{w})$ as
\begin{equation}
\label{eq:T3}
M(\mathbfit{w}|\mathcal{H}_0)  = M(\mathbfit{w}_{\mathrm{\!_{MP}}}) + \frac{1}{2} \Delta \mathbfit{w}^{\mathrm{T}} \mathbfss{A} \Delta \mathbfit{w},
\end{equation}
with $\Delta \mathbfit{w} = \mathbfit{w} - \mathbfit{w}_{\mathrm{\!_{MP}}}$. This allows us to approximate the posterior as
\begin{equation}
P(\mathbfit{w}|\mathbfit{g}, \lambda, \mathcal{H}_0) \approx P(\mathbfit{w}_{\mathrm{\!_{MP}}}) \cdot \mathrm{exp}\left[ -\frac{1}{2}\Delta \mathbfit{w}^{\mathrm{T}} \mathbfss{A} \Delta \mathbfit{w} \right].
\end{equation}
From this equation we see that the posterior may be approximated locally as a multivariate Gaussian distribution with covariance matrix $\mathbfss{A}^{-1}$. The marginalized errors on the individual weights $\mathbfit{w}_{\mathrm{\!_{MP}}}$ resulting from the linear inversion may be obtained by taking the square root of the diagonal elements in $\mathbfss{A}^{-1}$.

\subsubsection{The regularization parameter}
\label{sec:regParameter}
To find the optimal regularization parameter $\hat{\lambda}$, we have to find the maximum value for the probability density function $P(\lambda|\mathbfit{g}, \mathcal{H}_0)$. According to Bayes' theorem $P(\lambda|\mathbfit{g}, \mathcal{H}_0)$ is written as
\begin{equation}
P(\lambda | \mathbfit{g}, \mathcal{H}_0) = \frac{P(\mathbfit{g} | \lambda, \mathcal{H}_0) \cdot P(\lambda)}{P(\mathbfit{g} | \mathcal{H}_0)} \propto P(\mathbfit{g}|\lambda, \mathcal{H}_0) \cdot P(\lambda).
\end{equation}

Neglecting the normalization constant $P(\mathbfit{g}|\mathcal{H}_0)$, the function to consider for optimizing $\lambda$ is the product of the likelihood $P(\mathbfit{g} | \lambda, \mathcal{H}_0)$ and the prior $P(\lambda)$. Note that the likelihood term $P(\mathbfit{g} | \lambda, \mathcal{H}_0)$ appears as the normalizing constant of equation \ref{eq:posterior}: this term is often referred to as the evidence. Using equations \ref{eq:likelihood}, \ref{eq:normLikelihood}, and \ref{eq:prior}-\ref{eq:posterior} we  have
\begin{equation}
\label{eq:lam1}
P(\mathbfit{g} | \lambda, \mathcal{H}_0) = \frac{Z_{\!_M} (\lambda)}{Z_{\!_D} \cdot Z_{\!_S}(\lambda)}.
\end{equation}

Using the definition of $E_{\!_S}$ from equation \ref{eq:regFunction}, the normalization of the prior becomes
\begin{equation}
\begin{aligned}
\label{eq:lam3}
\begin{split}
Z_{\!_S} & =\int \mathrm{d}^{N_\mathrm{w}} \mathbfit{w} \cdot \mathrm{exp}(-\lambda E_{\!_S}) \\
& = \left( \frac{2\pi}{\lambda} \right)^{N_{\mathrm{w}} / 2} (\mathrm{det \: \mathbfss{C}^{-1}_{\mathrm{pr}}})^{-1/2},
\end{split}
\end{aligned}
\end{equation}
where $N_{\mathrm{w}}$ is the number of stellar templates in model $\mathcal{H}_0$. Using the Taylor expansion from equation \ref{eq:T3} we have for $Z_{\!_M}$
\begin{equation}
\begin{aligned}
\label{eq:lam4}
\begin{split}
Z_{\!_M} (\lambda) & = \int \mathrm{d}^{N_\mathrm{w}} \mathbfit{w} \cdot \mathrm{exp}(-M(\mathbfit{w})) \\
& = e^{-M(\mathbfit{w}_{\mathrm{\!_{MP}}})} (2\pi)^{N_{\mathrm{w}}/2} (\mathrm{det \: \mathbfss{A}})^{-\frac{1}{2}}.
\end{split}
\end{aligned}
\end{equation}
Combining equations \ref{eq:normLikelihood} and \ref{eq:lam1}-\ref{eq:lam4} allows us to write the logarithm of $P(\mathbfit{g} | \lambda, \mathcal{H}_0)$ as
\begin{equation}
\begin{aligned}
\label{eq:evidence}
\begin{split}
\log P(\mathbfit{g} | \lambda, \mathcal{H}_0) & =  - M(\mathbfit{w}_{\mathrm{\!_{MP}}}) - \frac{1}{2}\log(\mathrm{det\: \mathbfss{A}}) \\
& + \frac{N_{\mathrm{w}}}{2} \log \lambda\: + \frac{1}{2}\log(\mathrm{det \: \mathbfss{C}^{-1}_{\mathrm{pr}}}) \\
& - \frac{N_{\mathrm{d}}}{2} \log 2\pi + \frac{1}{2}\log(\mathrm{det \:	} \mathbfss{C}_{\mathrm{D}}^{-1}).
\end{split}
\end{aligned}
\end{equation}

Since we do not know a priori the value of $\lambda$ nor its order of magnitude, we choose a flat prior in $\log \lambda$ such that $P(\lambda) \propto 1/\lambda$. The optimal regularization parameter is then found by solving $\displaystyle \frac{d}{d\log \lambda} \log \left( P(\mathbfit{g} | \lambda, \mathcal{H}_0)\cdot P(\lambda) \right) = 0$, which results in the following non-linear expression for the most probable value of the regularization parameter $\displaystyle \hat \lambda$
\begin{equation}
\label{eq:regParameter}
\hat{\lambda} E_{S}(\mathbfit{w}_{\mathrm{\!_{MP}}}) = \frac{N_{\mathrm{w}}}{2} - \frac{1}{2}\hat{\lambda} \mathrm{Tr}(\mathbfss{A}^{-1} \mathbfss{C}^{-1}_{\mathrm{pr}}) - 1,
\end{equation}
where the last term in this equation originates from the prior on $\lambda$. This equation may be solved by using a non-linear solver. The process of finding the most probable regularization parameter is represented by block 2 in Fig. \ref{flowDiagram}. Note that for every step in finding the solution to equation \ref{eq:regParameter}, the model has to go to the inner block to find the most probable weights $\mathbfit{w}_{\mathrm{\!_{MP}}}$. Instead of solving for the most probable regularization parameter, $\lambda$ can in principle also be sampled as a nuisance parameter together with the other non-linear parameters of the model.

\subsubsection{Reconstructing the IMF model parameters}
\label{sec:reconstruction_IMFmodelparameters}
To reconstruct the parameters $p_i$ of the IMF parametrization in a model family $\mathcal{H}$, we compare the different models $\mathcal{H}_0$ in that model family with each other. The posterior probability of a certain model $\mathcal{H}_0$ is given by
\begin{equation}
P(\mathcal{H}_0 | \mathbfit{g}) \propto P(\mathbfit{g} | \mathcal{H}_0) \cdot P(\mathcal{H}_0).
\end{equation}
In the case of a flat prior $P(\mathcal{H}_0)$, models may be compared on the basis of the likelihood term $P(\mathbfit{g}| \mathcal{H}_0)$. Taking into account that $P(\mathbfit{g}| \mathcal{H}_0)$ is actually a marginalization over $\lambda$, we may write it as
\begin{equation}
P(\mathbfit{g}| \mathcal{H}_0) = \int P(\mathbfit{g} | \lambda, \mathcal{H}_0) \cdot P(\lambda | \mathbfit{g}, \mathcal{H}_0) \mathrm{d}\lambda,
\end{equation}
where $P(\mathbfit{g} | \lambda, \mathcal{H}_0)$ is the evidence derived in equation \ref{eq:evidence}.

If we make the assumption that $P(\lambda|\mathbfit{g}, \mathcal{H}_0)$ is a strongly peaked function at the most probable value $\hat{\lambda}$ \citep{MacKayBayesianAnalysis}, we may approximate it by a delta function centred on $\hat{\lambda}$ so that we obtain:
\begin{equation}
\begin{aligned}
\begin{split}
P(\mathbfit{g}| \mathcal{H}_0) &= \int P(\mathbfit{g} | \lambda, \mathcal{H}_0) \cdot P(\lambda| \mathbfit{g}, \mathcal{H}_0) \mathrm{d}\lambda \\
& = \int P(\mathbfit{g} | \lambda, \mathcal{H}_0) \cdot \delta(\hat{\lambda}) \mathrm{d}\lambda \\
&= P(\mathbfit{g} | \hat{\lambda}, \mathcal{H}_0),
\end{split}
\end{aligned}
\end{equation}
and we may rank the different models on the basis of $P(\mathbfit{g} | \hat{\lambda}, \mathcal{H}_0)$, i.e. the evidence from equation \ref{eq:evidence} evaluated for the most probable regularization constant $\hat{\lambda}$. Note that if we compare two models $\mathcal{H}_{0,1}$ and $\mathcal{H}_{0,2}$ on the basis of their evidence, we are actually interested in the ratio of the evidence for model $\mathcal{H}_{0,1}$ by the evidence for model $\mathcal{H}_{0,2}$. This ratio is called the Bayes factor $K$. Values of $K > 10^{1/2}$ and $K > 10^1$ may be considered as, respectively, substantial and strong evidence in favour of $\mathcal{H}_{0,1}$ whereas $K > 10^2$ is in general considered as decisive evidence in favour of $\mathcal{H}_{0,1}$ \citep{Jeffreys}.

The ability that we now have to quantify the posterior of a model $\mathcal{H}_0(p_i)$ on the basis of the evidence allows us to use Monte Carlo sampling techniques to reconstruct the posterior probability distribution of the non-linear IMF prior model parameters. In this paper we use \texttt{emcee} \citep{EMCEE} to sample the IMF prior parameters $p_i$. The reconstruction of the IMF prior model parameters is visualized by block 3 in Fig. \ref{flowDiagram}. For every sample of the IMF prior model parameters $p_{i,0}$, the model constructs a corresponding prior $\mathbfit{w}_0$. Then the model finds the most probable regularization parameter at the level below. Finally the model determines the most probable weights and calculates the evidence which may in turn be used to compare different samples $p_{i,0}$ of the IMF prior model parameters.

\subsubsection{The age and metallicity of the SSP}
\label{sec:ageMetallicitySSP}
Before we can actually sample the IMF prior model parameters $p_i$, we have to define the set of stellar templates that we are going to use. For SSPs, the stellar templates are defined by an isochrone of a certain age and metallicity. If we want to compare a combination of two different ages and metallicities (i.e. $\mathbfss{S}(t_1, [\mathrm{M/H}]_1)$ vs. $\mathbfss{S}(t_2, [\mathrm{M/H}]_2)$) we may once again use the Bayes factor
\begin{equation}
K_{12} = \frac{P(\mathbfit{g}|\mathbfss{S}(t_1, [\mathrm{M/H}]_1))}{P(\mathbfit{g}|\mathbfss{S}(t_2, [\mathrm{M/H}]_2))}.
\end{equation}
In this equation, $P(\mathbfit{g}|\mathbfss{S}(t_j, [\mathrm{M/H}]_j))$ is defined as
\begin{equation}
\label{eq:ageMetallicityEvidence}
P(\mathbfit{g}|\mathbfss{S}(t_j, [\mathrm{M/H}]_j)) = \int P(\mathbfit{g}|p_i, \mathbfss{S}(t_j, [\mathrm{M/H}]_j)) \cdot \mathrm{Pr}(p_i) \mathrm{d}p_i,
\end{equation}
which represents the evidence (or marginal likelihood) for the templates defined by $\{t_j, [\mathrm{M/H}]_j\}$ (i.e. in determining the evidence for the age and metallicity combination $\{t_j, [\mathrm{M/H}]_j\}$ we marginalize over all parameters in the inner layers, among which the IMF prior model parameters $p_i$). To find the most probable age and metallicity, we determine the evidence for each entry in a predefined age-metallicity grid. The age-metallicity combination that results in the highest evidence is then used to refine the sampling of the IMF prior model parameters $p_i$ with \texttt{emcee}. Note that we are only allowed to do this if there is a clear peak for the evidence in our age-metallicity grid. Otherwise, the resulting distributions for the IMF model parameters should be marginalized over all ages and metallicities. 

An efficient method for determining the integral in equation \ref{eq:ageMetallicityEvidence} is provided by nested sampling \citep{NestedSampling}. We calculate evidences by using \texttt{Multinest} \citep{Multinest2008, Multinest2009, Multinest2013}. To implement the Monte Carlo sampling techniques that we use in this paper, we have developed our code as a pipeline of \texttt{cosmoSIS} \citep{CosmoSIS}. \texttt{CosmoSIS} is a cosmological parameter estimation code that brings together different inference tools, including \texttt{Multinest} and \texttt{emcee}.

The reconstruction of the (most probable) age and metallicity is represented by block 4 in Fig. \ref{flowDiagram}. For every age and metallicity, the model selects an isochrone which is combined with the stellar library and the interpolator to create a set of stellar templates. At the level below, \texttt{Multinest} requires a complete sample of the IMF model parameters $p_i$ which allows it to marginalize over these parameters and calculate the evidence for the corresponding age and metallicity. This step still belongs to the first level of inference as we are trying to determine a set of parameters (i.e. age and metallicity).

\subsection{The second level of inference}
For the first level of inference we assume a certain model family $\mathcal{H}$. This model family is defined by the choice of isochrones, the stellar library, the interpolator, the regularization method and the parametrization of the IMF. The choice we make in this work to restrict ourselves to SSPs is also a model-dependent choice. We define $p_{\mathcal{H}}$ as the set of parameters that defines a model family. The second level of inference allows us to compare different model families with each other. As an example, for a given dataset we might want to compare two different IMF prior parametrizations: e.g. a double power law parametrization versus a lognormal parametrization.

According to Bayes' theorem, the posterior of a model family $\mathcal{H}$ given the data $\mathbfit{g}$ is
\begin{equation}
P(\mathcal{H}|\mathbfit{g}) \propto P(\mathbfit{g}| \mathcal{H}) \cdot P(\mathcal{H}).
\end{equation}
Assuming a flat prior $P(\mathcal{H})$ over the model families, the posterior of a model family is proportional to the likelihood $P(\mathbfit{g}|\mathcal{H})$. This likelihood term is a marginalization over the free parameters of the model family
\begin{equation}
P(\mathbfit{g}|\mathcal{H}) = \int P(\mathbfit{g}|p_{\mathcal{H}}, \mathcal{H}) \cdot P(p_{\mathcal{H}}|\mathcal{H}) \mathrm{d}p_{\mathcal{H}}.
\end{equation}
This integral may be determined by using e.g. \texttt{Multinest} and gives us the evidence for a certain model family. 

If we return to the example where we want to compare a double power law parametrization of the IMF with a lognormal parametrization, the relevant model parameters $P_{\mathcal{H}}$ to marginalize over are the parameters of the IMF prior parametrization $p_i$. This is however only true if we find a sufficiently strong peak in the evidence of the age-metallicity grid that justifies the use of an SSP. If this is not the case, we should in principle also marginalize over all ages and metallicities to obtain the evidence for a model family. In the current paper, since we test only SSP models, there is no strong need to do this, but we plan to further expand the code to sample directly over the space of age and metallicity and then expand the code to enable modelling CSPs as well. The second level of inference in our model is represented by the outer shell in Fig. \ref{flowDiagram}. 

Now that we have discussed the general setup of our model, in the next section we discuss the particular set of ingredients that we use to apply our model in this paper.


\section{Stellar templates}
\label{sec:stellarTemplates}
In this section we describe the basic ingredients that we use to create the stellar templates as an input for our model and that form the columns of the matrix $\mathbfss{S}$. Here we consider the example of an SSP as a function of age and metallicity. Therefore the age and metallicity are two free parameters of the model, although they are here solved on a regular grid of values and therefore are not part of the `continuous' set of parameters $\mathbfit{w}$, $\lambda$ and the IMF prior model parameters $p_i$. Those continuous parameters can be inferred and the evidence obtained for each chosen age and metallicity via nested sampling as discussed in Section \ref{sec:ageMetallicitySSP}. Since that evidence is the probability density for a chosen age and metallicity, it can be used for model comparison. 

Note that although we describe one particular set of ingredients in this section, in principle these ingredients may be substituted by any other set of ingredients. The approach described in Section \ref{sec:modelDescription} will still be valid, as long as we are able to construct a representative set of stellar templates.

\subsection{Isochrones}
The stars that are present in an SSP are defined by an isochrone. For a given age and metallicity, an isochrone provides us with the effective temperatures, surface gravities, masses and luminosities of the stars in an SSP corresponding to that particular age and metallicity. 

Within our model we use the Padova isochrones described in \cite{Padova3}. These isochrones may in principle be replaced with other models and different isochrone models may be assessed based on the evidence. The age and metallicity that define an isochrone are in principle continuous parameters. However, for every combination of age and metallicity we need to create an isochrone and for every isochrone star we need to interpolate a corresponding spectrum. Since isochrone determination and spectrum interpolation is time consuming, we create the stellar templates before running the model. Therefore we model age and metallicity as discrete parameters. We define a grid of ages and metallicities such that $\log \mathrm{age}= \{8.0, 10.11\}$ and $[\mathrm{M/H}] = \log \left( Z/Z_{\odot} \right) = \{-1.0, 0.4\}$ with $Z_{\odot} = 0.019$. The spacings of the grid are $\Delta \log t \mathrm{\:[Gyr]} \approx 0.062$ and $\Delta \mathrm{[M/H]} = 0.05$, respectively.

\subsection{Stellar library}
To construct a spectrum for a given set of stellar parameters, the starting point is a stellar library. Currently we use the (empirical) MILES stellar library (\cite{MILES}), consisting of approximately 1000 stars. Once again, note that this is only one particular choice, and in future work we plan to extend our model to include the X-Shooter Spectral Library \citep{XSL_DR1}. Although the MILES library covers a broad range in atmospheric parameters, empirical libraries have the disadvantage that they provide a limited coverage of the Hertzsprung-Russel (HR) diagram. 

\begin{figure}
	\resizebox{\hsize}{!}{\includegraphics[width=\columnwidth]{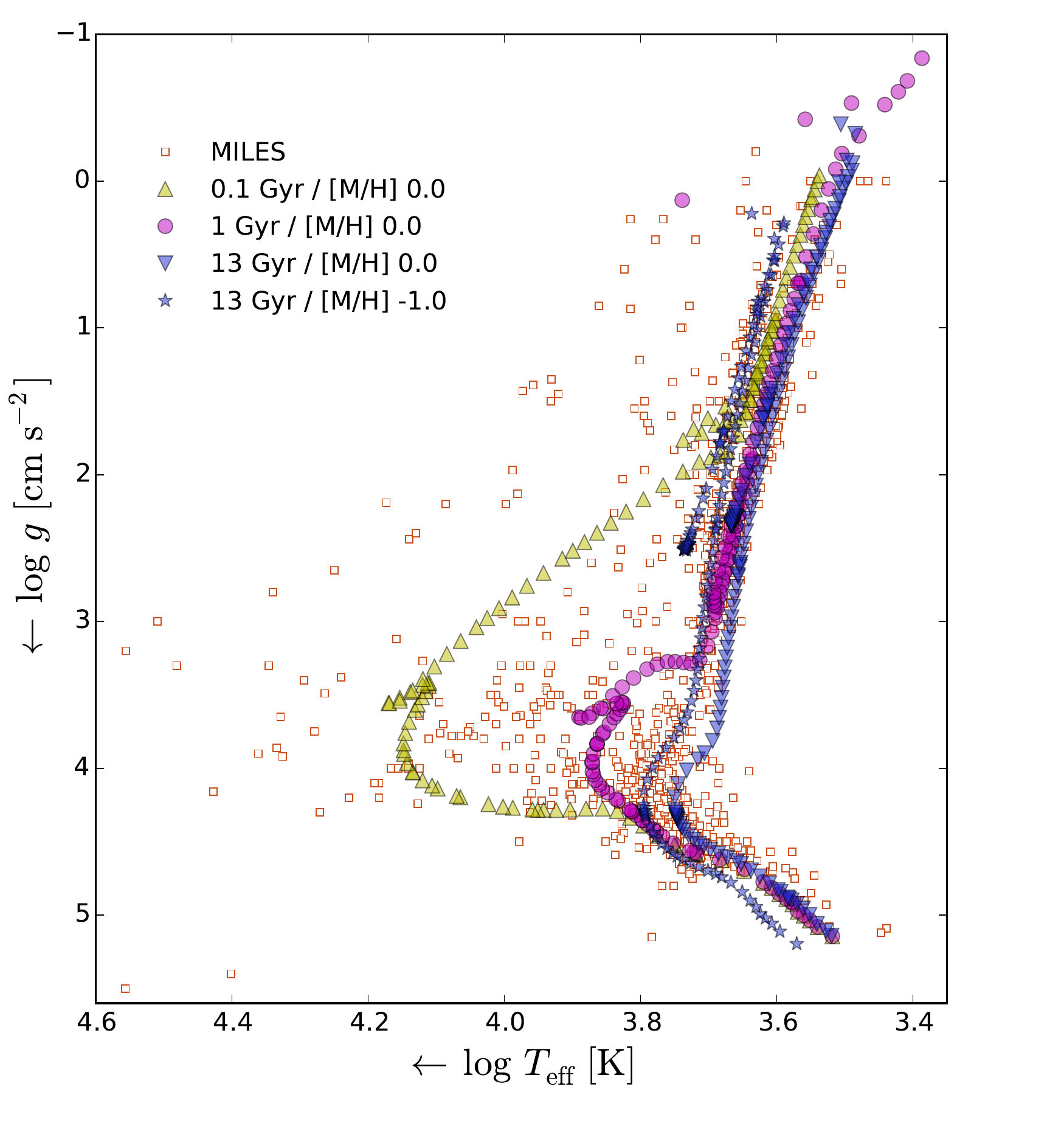}}
	\caption{Effective temperatures and surface gravities of stars in the MILES library (\textit{red squares}). Also shown are three isochrones for solar metallicity with ages 0.1 Gyr (\textit{yellow upper triangles}), 1 Gyr (\textit{magenta circles}), and 13 Gyr (\textit{blue lower triangles}). For the 13 Gyr isochrone, a low metallicity variant with $\mathrm{[M/H]} = -1.0$ is also shown (\textit{blue stars}).}
	\label{figureMILESparameters}
\end{figure}

Fig. \ref{figureMILESparameters} shows the HR diagram of the MILES stars. The figure also shows four isochrones used in our model. One can see that, although in general the coverage of the isochrones is quite good, there are some regions of parameter space where there is clearly a lack of stars. Especially for the low-mass end and the upper giant and asymptotic giant branches, it is apparent that the stars in the library do not fully cover the parameter space defined by the isochrone stars. As a consequence, there can potentially be significant uncertainties in the stellar spectra that are constructed in these regions.

\subsection{Interpolator}
The limited coverage of the HR diagram by empirical libraries requires a method to attach the stars in the library to the isochrones. We use an interpolator to do this, which for a given set of stellar parameters tries to interpolate between the surrounding spectra to create a representative stellar spectrum.

The idea behind such an interpolator is to create a function that interpolates between the spectra in the stellar library, allowing us to construct stellar spectra at all relevant locations in the HR diagram. Before creating an interpolator, one has to define the parameters that are required to model the spectrum of a star. In addition to the effective temperature and surface gravity, this would in principle require detailed knowledge of all chemical abundances in the star. However, the isochrones only define the overall metallicity $[\mathrm{M/H}]$, whereas the stars in the MILES library have measured values of $[\mathrm{Fe/H}]$ available. Therefore we choose to use an interpolator $S_{\lambda}(T_{\mathrm{eff}}, \log g, [\mathrm{Fe/H}])$ that interpolates the spectra of the stars in the three dimensional space of effective temperature, surface gravity and $[\mathrm{Fe/H}]$. In addition, we assume that $[\mathrm{M/H}] = [\mathrm{Fe/H}]$ to make the conversion from the isochrones to the stellar library straightforward. For the parameters of the stars in the MILES library, we use the values derived by \cite{Cenarro}.

Interpolating between stellar spectra may be done by using either a local approach or a global approach. Within the local approach described in \cite{Vazdekis_interpolator}, the spectra in the library that surround the point for which we want to create a spectrum are weighted and combined to create a representative spectrum for that particular point. The global approach described in \cite{Prugniel_interpolator} fits a polynomial to each of the spectral bins individually. This polynomial may then be used to determine the flux in each of the bins for the required set of atmospheric parameters. 

In this work, we use a local approach very similar to that described in \cite{Vazdekis_interpolator}. Before we build the interpolator, we normalize the stars in the MILES library such that they have the same magnitude in the (Johnson) V-band. Suppose that we want to create a spectrum for the point $\{\theta_0, \log g_0, \mathrm{[Fe/H]_0}\}$ (where $\theta = 5040 / T_{\mathrm{eff}}$). Within the three dimensional space of $\theta$, $\log g$ and $\mathrm{[Fe/H]}$, this point is surrounded by eight cubes. The first step of the interpolator consists of finding the nearby stars that are present in each of these eight boxes. The initial size of each of these cubes is $1.5\sigma_{\theta}\times 1.5\sigma_{\log g} \times 1.5\sigma_{\mathrm{[Fe/H]}}$, in which $\sigma_p$ corresponds to the typical uncertainty in the parameters $p = \{\theta, \log g, \mathrm{[Fe/H]}\}$. These typical uncertainties are defined on the basis of the local density of stars $\rho$, such that
\begin{equation}
\sigma_{p} = \sigma_{p,\mathrm{m}} \cdot \exp\left({\left( \frac{\rho - \rho_{\mathrm{M}}}{\rho_{\mathrm{M}}} \right)^2 \ln\frac{\sigma_{p,\mathrm{M}}}{\sigma_{p,\mathrm{m}}}}\right).
\end{equation}
In this equation $\rho_{\mathrm{M}}$ is the maximum density of stars in the grid, which is taken as the 99.7 percentile of all densities in the grid. For the minimum uncertainty $\sigma_{p,\mathrm{m}}$ and maximum uncertainty $\sigma_{p,\mathrm{M}}$ we use the same values as \cite{Vazdekis_interpolator}: $\sigma_{\theta,\mathrm{m}} = 0.009$, $\sigma_{\theta,\mathrm{M}} = 0.17$, $\sigma_{\log g,\mathrm{m}} = 0.18$, $\sigma_{\log g,\mathrm{M}} = 0.51$, $\sigma_{\mathrm{[Fe/H]}, \mathrm{m}} = 0.09$ and $\sigma_{\mathrm{[Fe/H]},\mathrm{M}} = 0.41$. As an additional constraint for $\sigma_{\theta}$, $\sigma_{T_{\mathrm{eff}}}$ should lie within $60$ $\mathrm{K} \leq \sigma_{T_{\mathrm{eff}}} \leq 3350$ $\mathrm{K}$. If no stars are found in one of the boxes, the size of the box is enlarged in steps of $0.5\sigma_p$ along each of its axes until at least one star is found or the axes reach a size of $10\sigma_p$. Note that the metallicity parameter is only taken into account for stars with 4000 K $\leq T_{\mathrm{eff}} \leq$ 9000 K. Outside of this range the uncertainty in the metallicity is relatively large and in addition there is a significant number of stars with unknown metallicity. 

As a next step, we create a representative spectrum for each of the boxes that contain stars. To create the spectrum of a box, each of its stars is assigned a weight $W_{\mathrm{s}}$ such that
\begin{equation}
\begin{aligned}
\begin{split}
W_{\mathrm{s}}  = & \frac{\mathrm{SN}_{\mathrm{s}}^2}{\mathrm{SN}_{\mathrm{max}}^2} \cdot \exp\left({-\left( {\frac{\theta_{\mathrm{s}} - \theta_0}{\sigma_{\theta}}}\right)^2}\right) \\
&\cdot \exp\left({-\left({\frac{\log g_{\mathrm{s}} - \log g_0}{\sigma_{\log g}}}\right)^2}\right)\\
& \cdot  \exp\left({-\left({\frac{\mathrm{[Fe/H]}_{\mathrm{s}} - \mathrm{[Fe/H]}_0}{\sigma_{\mathrm{[Fe/H]}}}}\right)^2}\right),
\end{split}
\end{aligned}
\end{equation}
where $\{\theta_{\mathrm{s}}, \log g_{\mathrm{s}}, \mathrm{[Fe/H]_s}\}$ are the parameters of the star and $\mathrm{SN}_{\mathrm{s}}$ is the signal-to-noise ratio (SNR) of the star. The maximum SNR, $\mathrm{SN}_{\mathrm{max}}$, is defined to be the 99.7 percentile of the the SN ratios of all the stars. If $\mathrm{SN}_{\mathrm{s}} > \mathrm{SN}_{\mathrm{max}}$, $\mathrm{SN}_{\mathrm{s}}$ is set equal to $\mathrm{SN}_{\mathrm{max}}$. Once we have a weight for each of the stars in a box, the spectrum of that box $S_\mathrm{B}$ is calculated as
\begin{equation}
S_{\mathrm{B}} = \frac{\sum_{\mathrm{s}} W_{\mathrm{s}} S_{\mathrm{s}}}{\sum_{\mathrm{s}} W_{\mathrm{s}}},
\end{equation}
with $W_{\mathrm{s}}$ the weight of star ${\mathrm{s}}$ and $S_{\mathrm{s}}$ its spectrum. In the same way, we also determine a corresponding set of of parameters $p_{\mathrm{B}}$ for each of the boxes, such that
\begin{equation}
p_{\mathrm{B}} = \frac{\sum_{\mathrm{s}} W_{\mathrm{s}} p_{\mathrm{s}}}{\sum_{\mathrm{s}} W_{\mathrm{s}}},
\end{equation}
with $p = \{\theta, \log g, \mathrm{[Fe/H]}\}$.

Each box that contain stars is assigned a weight $W_j$ based on the distance of the box parameters $p_{\mathrm{B}}$ to the point $p_0$, so that
\begin{equation}
\begin{aligned}
\begin{split}
W_j = & \exp\left({-\left( {\frac{\theta_{\mathrm{B}} - \theta_0}{\sigma_{\theta}}}\right)^2}\right) \cdot \exp\left({-\left({\frac{\log g_{\mathrm{B}} - \log g_0}{\sigma_{\log g}}}\right)^2} \right)\\
& \cdot\exp\left({-\left({\frac{\mathrm{[Fe/H]}_{\mathrm{B}} - \mathrm{[Fe/H]}_0}{\sigma_{\mathrm{[Fe/H]}}}}\right)^2}\right),
\end{split}
\end{aligned}
\end{equation}
and as a final step the spectrum $S_0$ of the point $p_0$ that we want to interpolate is calculated as
\begin{equation}
S_0 = \frac{\sum_{j} W_j S_{\mathrm{B},j}}{\sum_{j} W_j},
\end{equation}
where the sum runs over all of the boxes that contain stars. For more details we refer to \cite{Vazdekis_interpolator}.

\subsubsection{Polynomial correction of the MILES stars}
\label{sec:polCorrections}
\begin{figure}
	\centering
		{\includegraphics[width=\columnwidth]{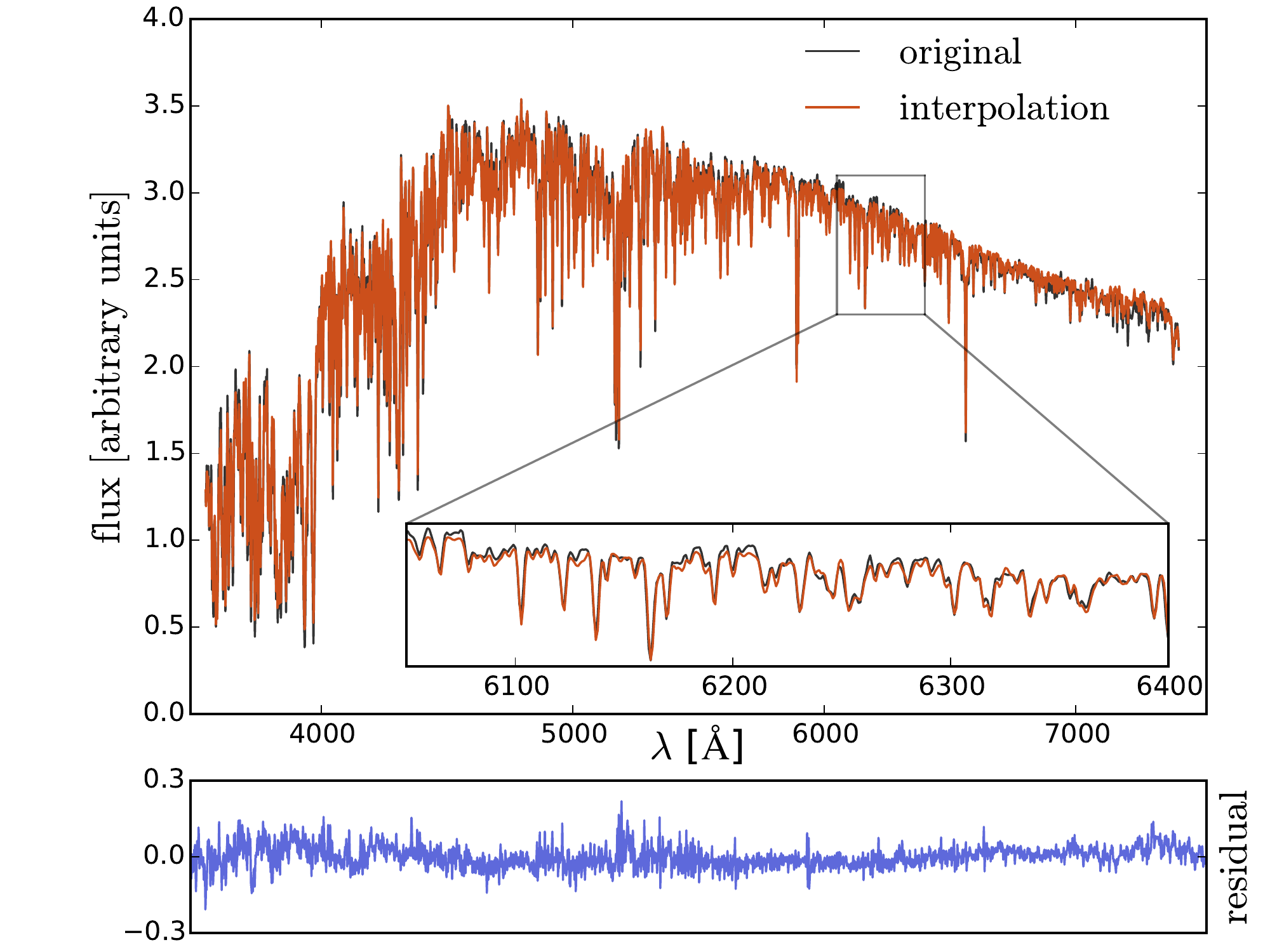}}
		{\includegraphics[width=\columnwidth]{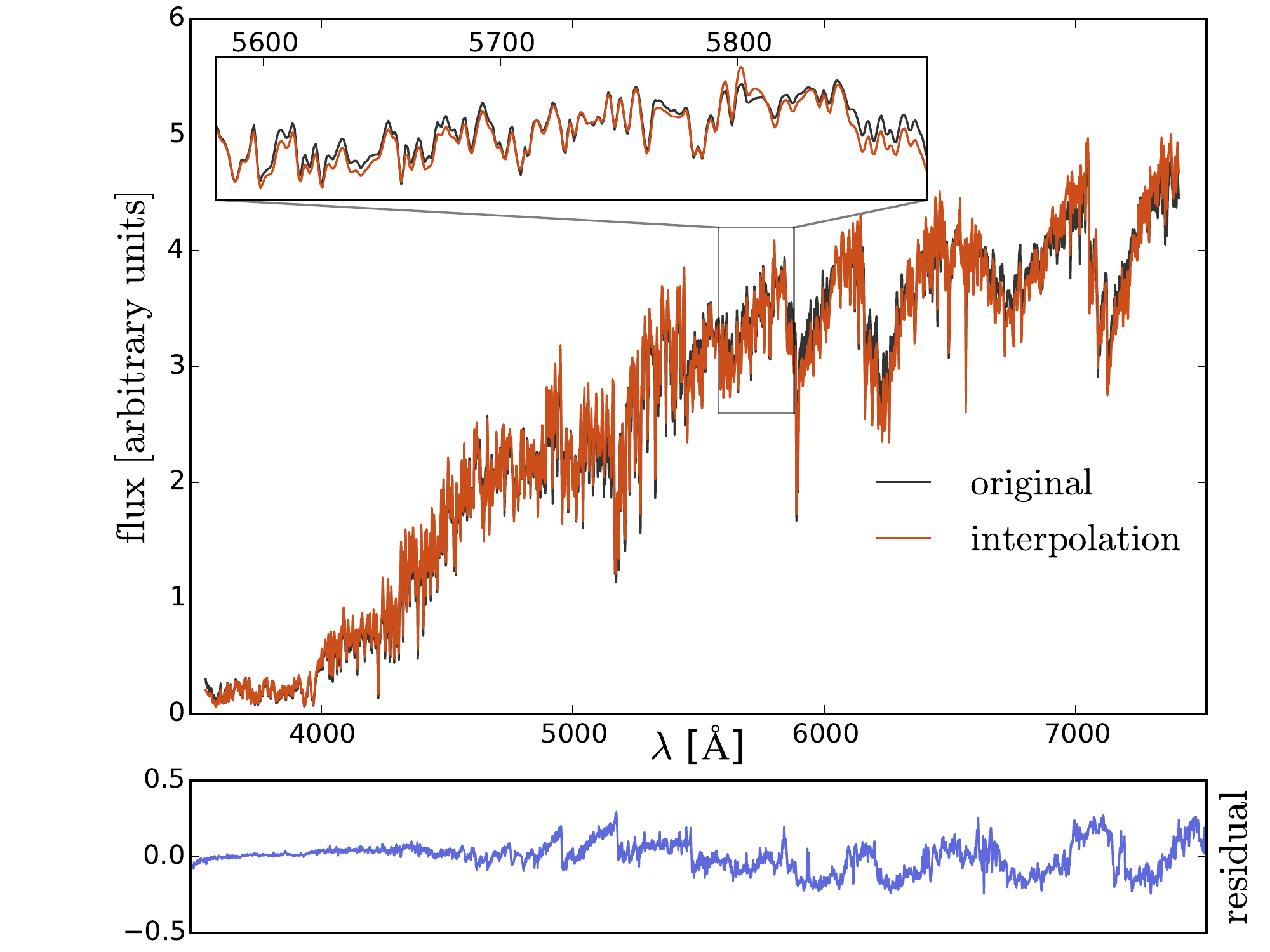}}
	\caption{Two examples where an observed MILES spectrum is compared to an interpolated spectrum with the same atmospheric parameters. The original MILES star is not part of the data set used to build the interpolator. The black lines represent the original spectra, the red lines the interpolated spectra. For both spectra, the residual between the original spectrum and the interpolated spectrum is shown in blue. \textbf{Top panel:} Interpolated spectrum MILES star with $T_{\mathrm{eff}} = 5392$ K, $\log g = 4.6$ and $\mathrm{[Fe/H]} = 0.1$. \textbf{Bottom panel:} Interpolated spectrum MILES star with $T_{\mathrm{eff}} = 3793$ K, $\log g = 1.4$ and $\mathrm{[Fe/H]} = 0.32$. The residual between the spectrum and the interpolated spectrum is on average 1.2\% of the average of the original spectrum for the first spectrum and 2.9\% for the second spectrum.}
	\label{interpolatedSpectra}
\end{figure}
To test the interpolator, we have created an interpolated spectrum $S_{\mathrm{int}}$ for each of the stars in the MILES library, in such a way that the original MILES star was excluded from the data set that we use to build the interpolator. In this way, we calculate for each of the spectra in the MILES library the average residual $R_{S}$ between the original spectrum $S_{\mathrm{or}}$ and the interpolated spectrum $S_{\mathrm{int}}$. The residual is weighted by the average of the original spectrum, such that
\begin{equation}
R_{\mathrm{S}} = \left\langle\frac{\mathrm{abs}\left( S_{\mathrm{or}} - S_{\mathrm{int}} \right)}{S_{\mathrm{or}}}\right\rangle.
\end{equation}

This allowed us to assess the quality of the interpolator and to identify stars with problems. A large mismatch between the interpolated spectrum and the original spectrum may be the result of a low SNR of the original spectrum, any form of peculiarity in the original spectrum or using incorrect atmospheric parameters for the star in question. Problematic stars with a low signal-to-noise ratio, stars with obvious problems in their spectra after visual inspection and some peculiar stars that showed a large mismatch with their interpolated spectrum were removed from the dataset. Overall we removed 46 stars from the library.

A mismatch between the continuum of the original star and the interpolated spectrum may be caused by uncertainties in both the flux calibration and the correction for extinction. To absorb this effect, we correct each of the stars in the MILES library by a first order polynomial. This polynomial correction is an iterative process. At every step, the star with the largest residual is selected and corrected by a polynomial. Then the residuals are calculated again for the new data set. This process is repeated until each star has been corrected. Each star is corrected only once. After this correction, the average residual between the stars in the MILES library and the interpolated spectra is 2.3\%. If we exclude stars with any notion of peculiarity in SIMBAD the average residual becomes 1.8\%. Fig. \ref{interpolatedSpectra} shows two examples of a comparison between a MILES spectrum and an interpolated spectrum with the same atmospheric parameters.

Having the interpolator in place, we create spectra for all the stars defined by the set of isochrones in the age-metallicity grid $\log \mathrm{age} = \{8.0-10.11\}$ yr and $[\mathrm{Fe/H}] = \{-1.0-0.4\}$. As a final step, the spectra resulting from the interpolator are scaled to match the $V$-magnitudes of the stars defined by the isochrone, for which we use the filter response defined in \cite{filterFunctions}.

\section{Results - mock single stellar populations}
\label{sec:results}
In this section we apply our model to a number of mock SSPs. We create the spectra for these mock SSPs by combining the stellar templates of one particular age and metallicity with an IMF.

To model the velocity dispersion of real stellar populations, we smooth the stellar templates to a velocity dispersion of $150 \mathrm{\:km} \mathrm{\:s}^{-1}$. Before applying the model, we smooth the stellar templates to the same velocity dispersion. In Appendix \ref{sec:velocityDispersion} we show that the results that we obtain for our mock SSPs do not depend on the velocity dispersion.

Note that although here we fix the velocity dispersion of the templates, this could in principle also be a free parameter of the model that we sample together with the other non-linear parameters. We will implement this in a future version of the code.

As a final step, we add Gaussian noise to the mock spectra to mimic an observation with a certain signal-to-noise ratio (SNR).

\subsection{The regularization scheme}
Before we apply the model to the mock SSPs we have to specify a regularization scheme $\mathbfss{C}^{-1}_{\mathrm{pr}}$. For the mock SSPs we choose to use a regularization scheme that penalizes the relative deviation of the weights from the prior and prefers smooth deviations from the prior. This is expressed through the following inverse covariance matrix for the prior
\begin{equation}
\mathbfss{C}^{-1}_{\mathrm{pr}} = \mathbfss{C}_1 + \mathbfss{C}_2,
\end{equation}
where $\mathbfss{C}_1$ is a diagonal matrix with
\begin{equation}
\mathbfss{C}_{1_{i,i}} = \frac{1}{w_{0,i}^2},
\end{equation}
and $\mathbfss{C}_2$ enforces smoothness of the deviations by using the following form of gradient regularization
\begin{equation}
\mathbfss{C}_2 = \left(
\begin{matrix}
\frac{1}{w_{0,1}^2} & -\frac{1}{w_{0,2}^2} & 0 & 0 &\ldots & 0 \\
0 & \frac{1}{w_{0,2}^2} & -\frac{1}{w_{0,3}^2} & 0 & \ldots & 0 \\
0 & 0 & \ddots & {} & {} & \vdots \\
\vdots & \vdots & {} & {} & {} & \vdots \\
0 & 0 & \ldots & \ldots & \ldots & \frac{1}{w_{0,n}^2}
\end{matrix}
\right)
\end{equation}

\subsection{First level of inference}
Within the first level of inference, we assume a certain model family and try to reconstruct its underlying parameters. Here we parametrize the IMF as a double power law with a break at $0.5$ $\mathrm{M_{\odot}}$. We split the reconstruction of the model parameters into a linear and a non-linear part. The linear part consists of the weights assigned to the individual stellar templates (i.e. the actual best-fitting IMF) whereas the non-linear part consists of determining the most probable regularization parameter, finding the age and metallicity of the SSP and sampling the parameters of the IMF prior parametrization.

\subsubsection{Linear parameters}
\label{sec:linearParameters}
\begin{figure}
	\centering
		{\includegraphics[width=0.99\columnwidth]{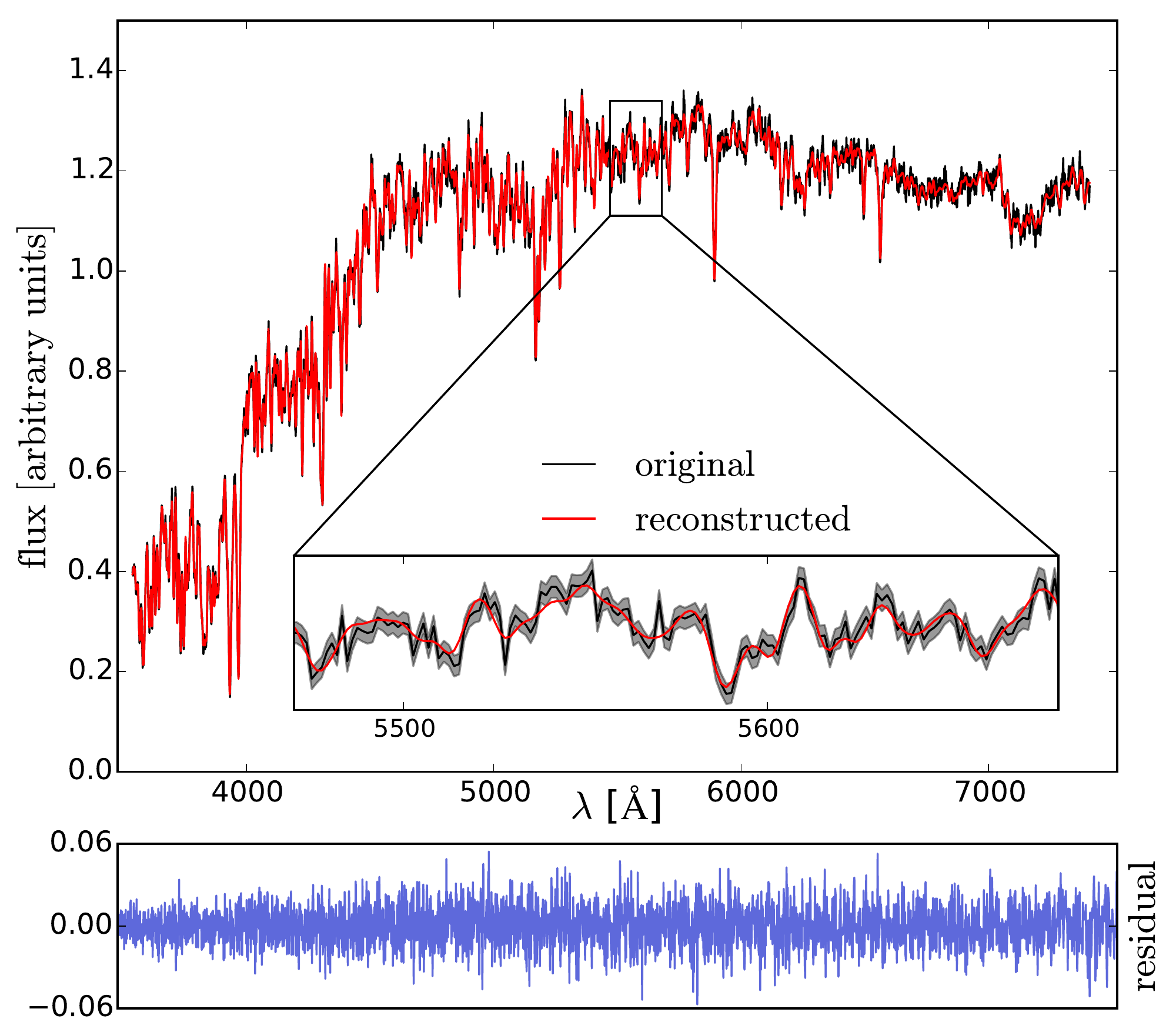}}
		{\includegraphics[width=0.99\columnwidth]{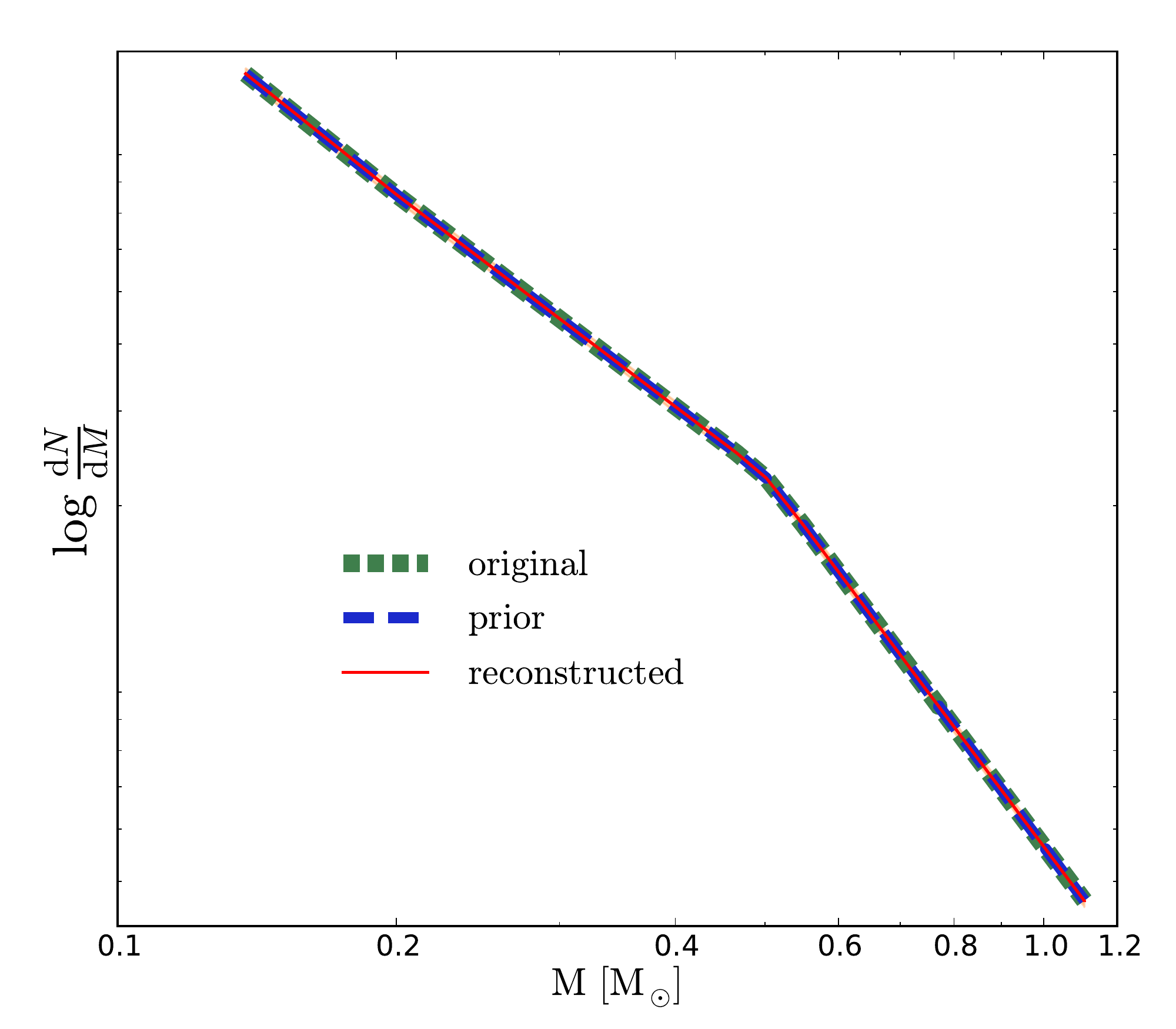}}
	\caption{Reconstructed spectrum and IMF of a mock SSP. The input spectrum is a mock SSP for which the underlying IMF is a double power law Kroupa IMF. We applied our model to the spectrum after adding random noise to this spectrum such that the SNR of the spectrum is 70. We provide the model with the correct set of stellar templates and specify the (correct) prior as a double power law Kroupa IMF. \textbf{Top panel:} Reconstructed spectrum. The \textit{black} line shows the spectrum of the mock SSP and the \textit{red} line corresponds to the spectrum reconstructed by the model. The residual between the two spectra is shown in \textit{blue}. The \textit{shaded gray} region in the zoom-in represents the one sigma uncertainty corresponding to the specified SNR. \textbf{Bottom panel:} Reconstruction of the IMF. The \textit{short-dashed green} line represents the original IMF, the \textit{long-dashed blue} line the prior IMF and the \textit{red} line represents the reconstructed IMF by the model. The \textit{shaded orange} region corresponds to the error on the weights resulting from the linear inversion, derived as described in Section \ref{sec:weightsError} (almost invisible in this plot).}
	\label{reconstructionIMF1}
\end{figure}

The linear parameters in our model are represented by the weights $\mathbfit{w}$ which correspond to the number of stars that we have for each of the templates in our model. As described in Section \ref{sec:modelDescription}, these weights allow us to reconstruct the piecewise IMF of the stellar population.

To demonstrate the reconstruction of the IMF we consider a mock SSP with an age of 8.5 Gyr and solar metallicity. This mock SSP has an average SNR of 70 and the underlying IMF of the stellar population is a double power law Kroupa IMF (i.e. a break at 0.5 $\mathrm{M_{\odot}}$, a low-mass slope $\alpha_1=1.3$ and a high-mass slope $\alpha_2 = 2.3$).

To reconstruct the piecewise IMF, we first need to find the most probable distribution of weights $\mathbfit{w}_{\!_{\mathrm{MP}}}$ given a set of stellar templates and a prior IMF. If we fix the age and metallicity of the templates to the true values (which we know a priori in this case), the only thing that we can change in the model is the prior IMF.

First we consider what happens when we use the correct prior, a double power law Kroupa IMF (but let the level of regularization be free in the optimization). The upper panel of Fig. \ref{reconstructionIMF1} shows the spectrum of the mock SSP together with the reconstructed spectrum of the model. The average of the spectrum divided by the standard deviation of the difference between the spectrum and the reconstructed spectrum is 68.6, consistent with an average SNR of 70. The lower panel of Fig. \ref{reconstructionIMF1} shows the reconstructed IMF compared to the original IMF and the prior IMF. As one might expect, in this case the original IMF, the prior IMF and the reconstructed IMF all lie on top of each other.

\begin{figure}
	\centering
		{\includegraphics[width=0.99\columnwidth]{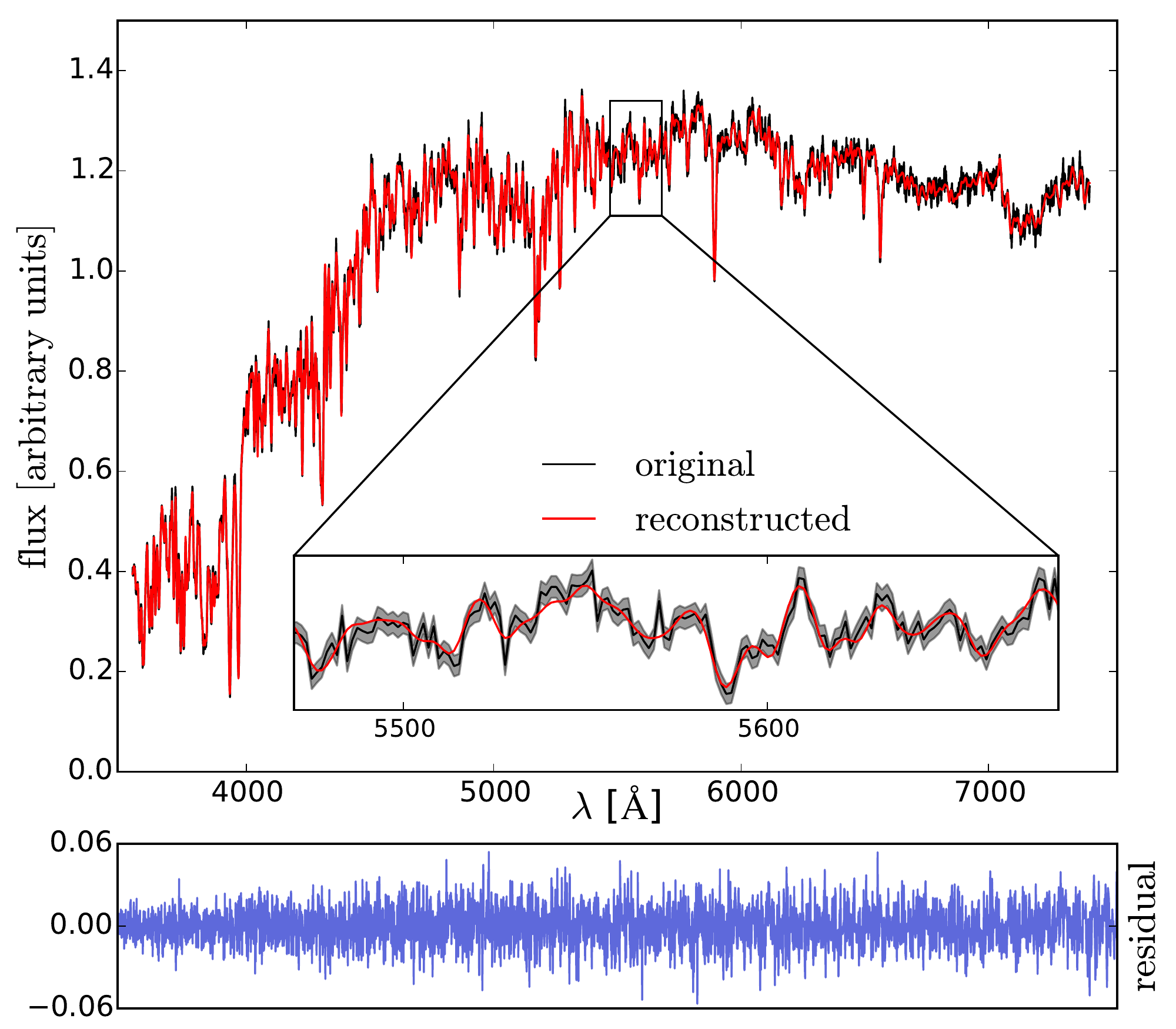}}
		{\includegraphics[width=0.99\columnwidth]{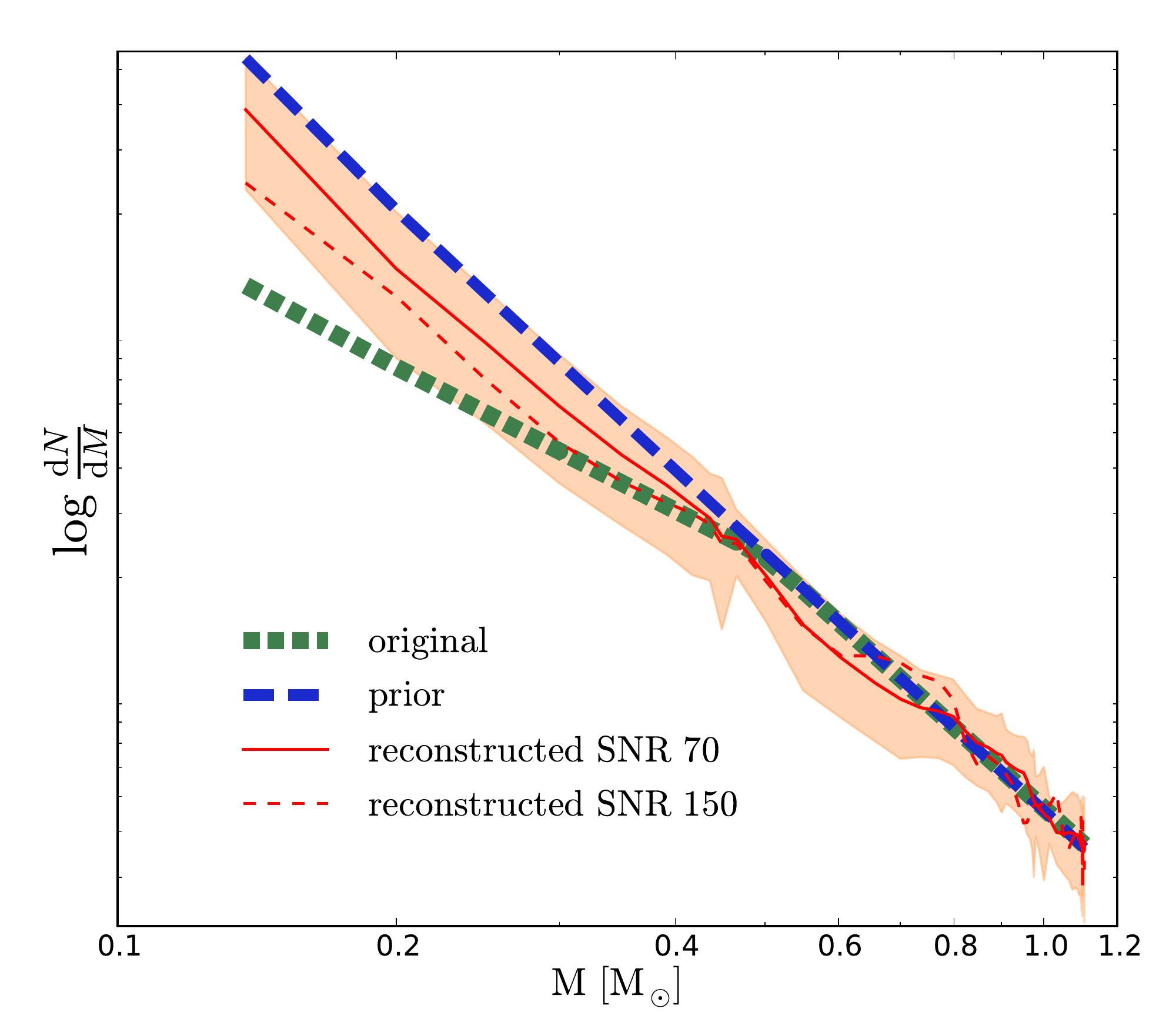}}
	\caption{As in Fig. \ref{reconstructionIMF1}, but when providing the model with an incorrect Salpeter IMF prior.}
	\label{reconstructionIMF2}
\end{figure}

\begin{figure}
	\resizebox{\hsize}{!}{\includegraphics[width=\columnwidth]{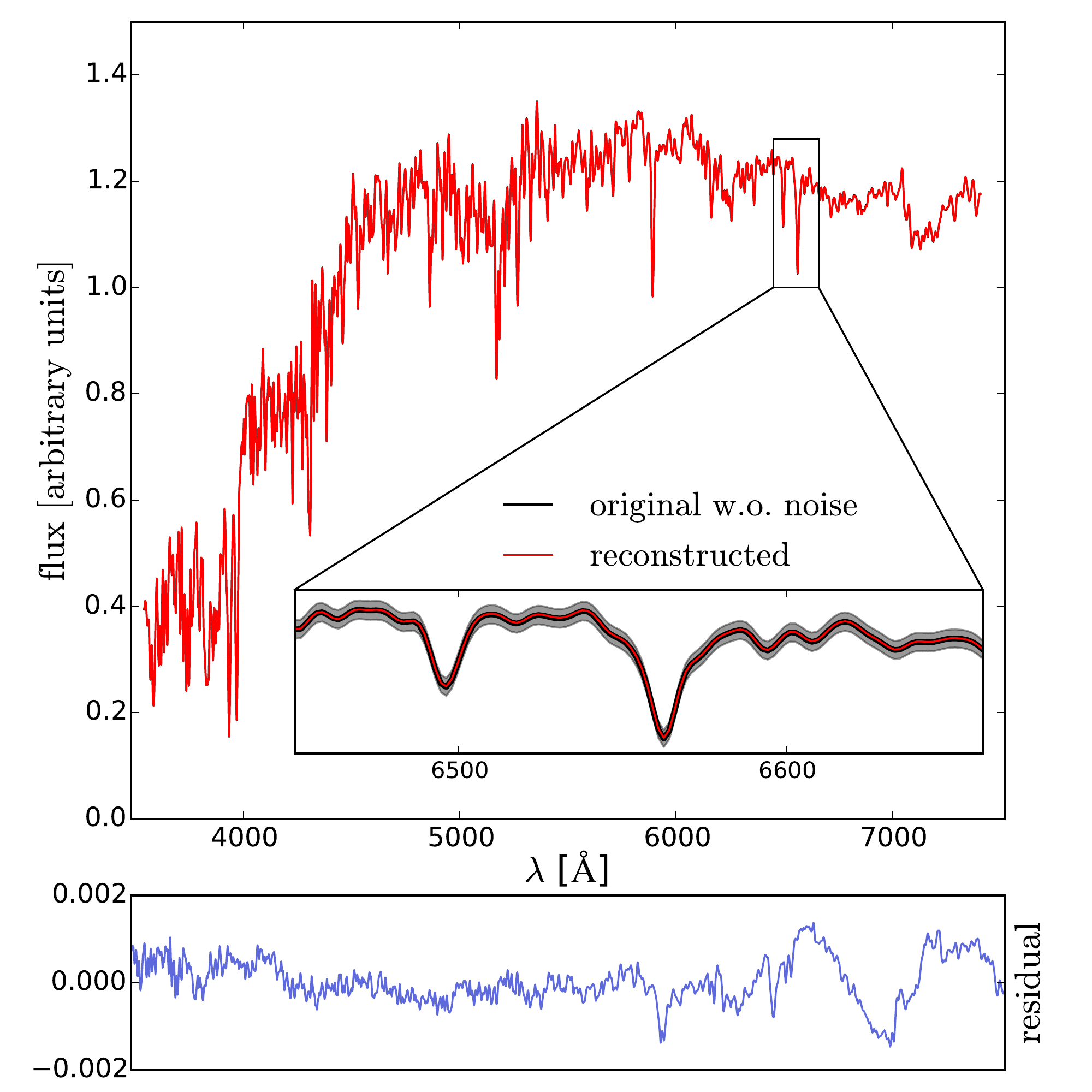}}
	\caption{Reconstructed spectrum of mock SSP from figure \ref{reconstructionIMF2}. We compare the reconstructed spectrum against the noise-free input spectrum. The figure shows us that the incorrect solution that we obtain in figure \ref{reconstructionIMF2} provides an excellent fit to the data. This illustrates that there exists a degeneracy between the stellar templates in the model.}
	\label{reconstructionIMF2-residual}
\end{figure}

\begin{figure}
	\centering
		{\includegraphics[width=0.99\columnwidth]{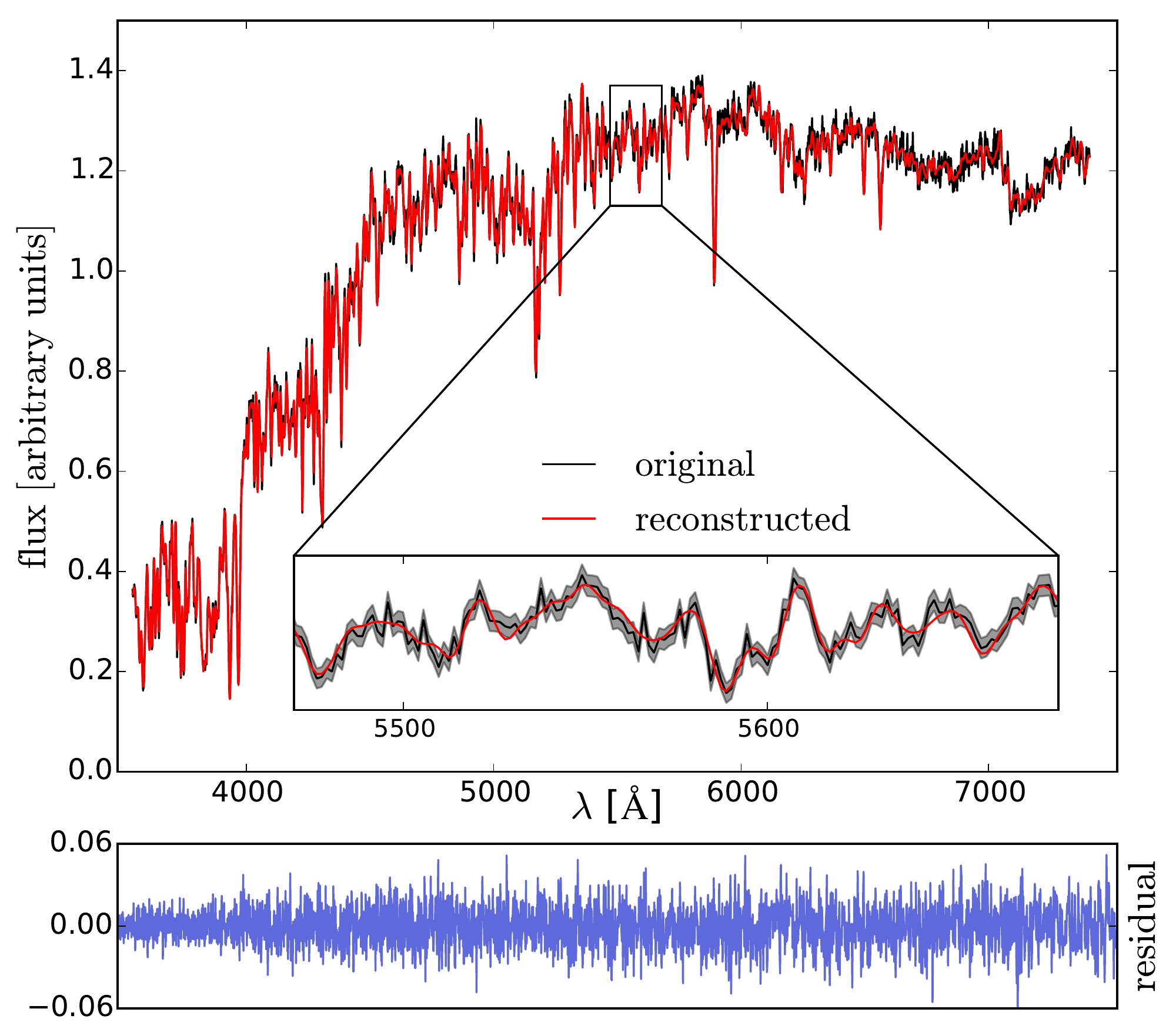}}
		{\includegraphics[width=0.99\columnwidth]{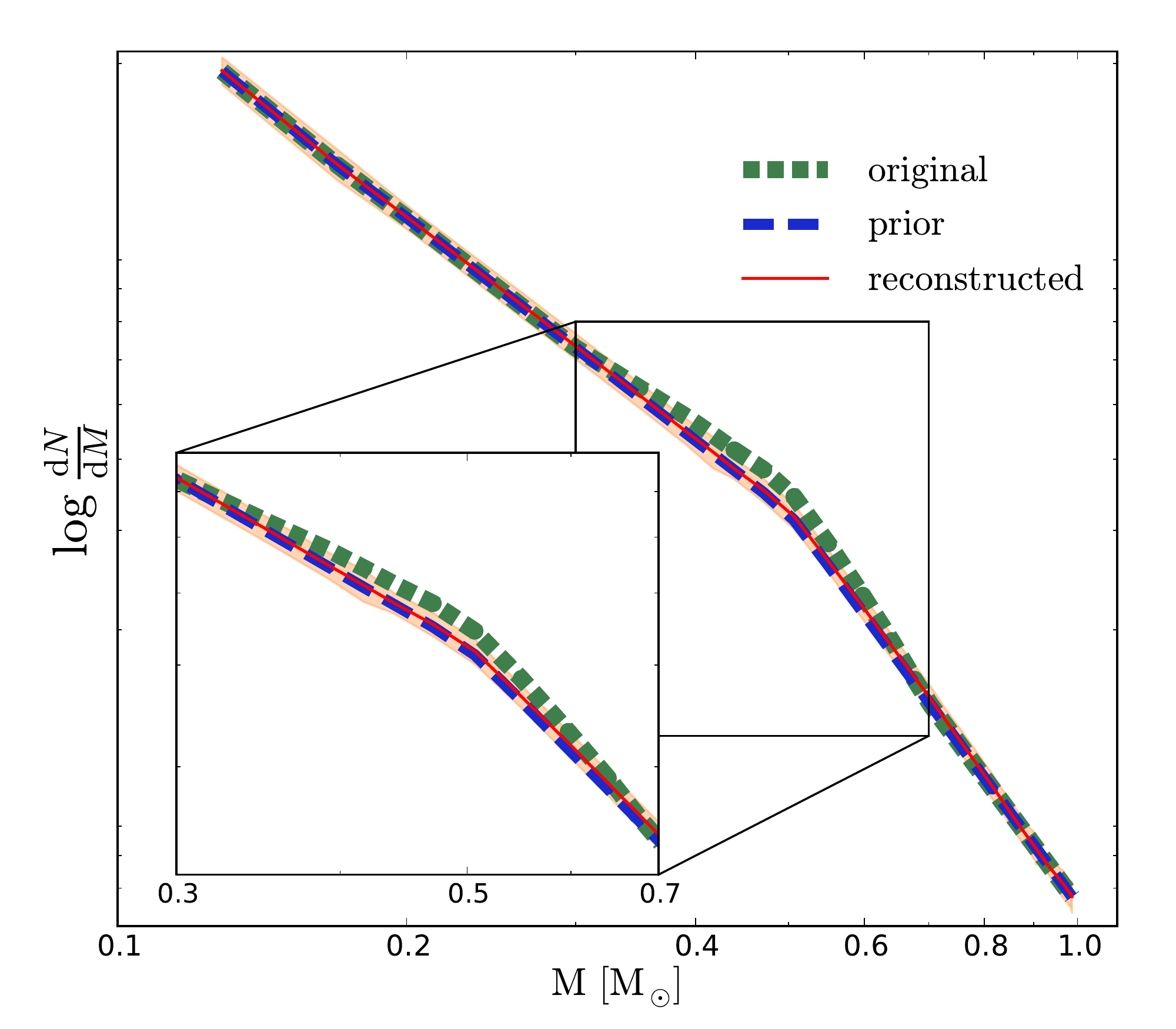}}
	\caption{As in Fig. \ref{reconstructionIMF1}, but now the underlying IMF has a bump between $M = 0.3$ $\mathrm{M_{\odot}}$ and $M = 0.7$ $\mathrm{M_{\odot}}$, as described in the text.}
	\label{reconstructionIMF3}
\end{figure}

As a next step, we provide the model with an (incorrect) Salpeter IMF prior (i.e. a constant slope of $\alpha_1 = \alpha_2 = 2.35$). The regularization parameter is again free and optimized by the model. The reconstructed spectrum and reconstructed IMF that we obtain are shown in the upper and lower panel of Fig. \ref{reconstructionIMF2}. In this case the reconstructed IMF converges between the original IMF and the prior IMF, reflecting the fact that the minimization routine on the one hand wants to provide a good fit to the data and on the other hand wants to stay as close as possible to the prior.

Moreover, the obtained solution tends more and more towards the prior as we go to lower masses. This is probably related to the fact that the lowest mass stars contribute very little light to the spectrum. Therefore, it becomes much harder for the model to derive the abundance of these lower mass stars and hence the solution tends more and more towards the incorrect prior.

There also appears to be a degeneracy between the lower mass templates ($M \lesssim 0.4$ $\mathrm{M_{\odot}}$) and the higher mass templates ($M \gtrsim 0.4$ $\mathrm{M_{\odot}}$). This degeneracy is such that the model corrects the over-abundance of the lower mass templates with respect to the original weights (enforced by the prior) by decreasing the abundance of the templates that have slightly higher masses. In Fig. \ref{reconstructionIMF2-residual} we show the reconstructed spectrum from the most probable weights as compared to the original input spectrum without noise. This figure shows us that the obtained solution provides a very good fit to the data and it appears that this fit is as good as that obtained for the real solution within the uncertainty computed from the covariance matrix. Since this solution lies closer to the prior IMF, it is preferred over the actual solution. Except for the lowest mass bin, the actual solution lies within the one sigma errors of our solution.

One might expect that this degeneracy disappears if we increase the SNR. However, repeating this analysis for a number of (higher) SNRs shows that the degeneracy does not completely disappear. As an example, we show in the lower panel of Fig. \ref{reconstructionIMF2} the reconstructed IMF for a spectrum with SNR=150. Although the obtained solution is now closer to the true one, apparently the data is not informative enough to break the degeneracy.

The final test that we discuss here is that of an IMF which is not part of the prior space allowed by the IMF parametrization. For this test we consider a 13 Gyr old mock SSP with solar metallicity. We assign weights to the different stellar templates based on a Kroupa IMF ($w_{\!_{\mathrm{Kroupa}}}$) and then add to these weights a sinusoidal bump. This bump is added to the templates in the mass range 0.3 through 0.7 $\mathrm{M_{\odot}}$ with a maximum of 10\% for $M=0.5$ $\mathrm{M_{\odot}}$, so that we have
\begin{equation}
w(M) = w_{\!_{\mathrm{Kroupa}}}\left(1 + 0.1 \cos{\left( \frac{\pi}{0.4}M -0.5 \right)}\right),
\end{equation}
for templates in the range $M = [0.3,0.7]$ $\mathrm{M_{\odot}}$. 

To reconstruct the IMF for this mock SSP we use a Kroupa IMF prior. The reconstructed spectrum and IMF are shown in Fig. \ref{reconstructionIMF3}. This figure shows that, although not by much, the obtained solution is different from the prior in the sense that it deviates slightly towards the true solution. And although this may not allow us to find the true solution, it provides a clear indication that the prior we used does not completely fit the data.

\begin{table*}
	\centering
	\caption{Overview of the 12 mock SSPs considered in Section \ref{sec:nonlinearParameters}. The table provides both the input and reconstructed values of each mock SSP. In addition, we present the difference in log evidence with the second best set of stellar templates. The reconstructed IMF slopes are the median values of the distribution. The errors on the reconstructed slopes correspond to the distance between the median and the 16th and the 84th percentile. In the last two columns we show the maximum a posteriori (MAP) values of the sample for the two IMF slopes.}
	\label{tab:table_mockSSPs}
	\begin{tabular}{ccccccccccc} 
		\hline
		name & age [Gyr] & [Fe/H] & IMF & reconstructed & reconstructed & $\Delta$ evidence & \multicolumn{4}{c}{reconstruction IMF prior parameters}\\
		     &			&		&	& age [Gyr] & [Fe/H] & 2nd best & $\alpha_{1,\mathrm{med}}$ & $\alpha_{2,\mathrm{med}}$ & $\alpha_{1,\mathrm{map}}$ & $\alpha_{2,\mathrm{map}}$\\
		\hline
		\rule{0pt}{3ex}
		\texttt{mock1} & 3.1 & -0.2 & Kroupa & 3.1 & -0.2 & 15.8 & $1.39^{+0.40}_{-0.73}$ & $2.31^{+0.07}_{-0.07}$ & 1.50 & 2.32\\
		\rule{0pt}{3ex}
		\texttt{mock2} & 3.1 & 0.0 & Kroupa & 3.1 & 0.0 & 16.0 & $0.95^{+0.58}_{-1.40}$ & $2.32^{+0.08}_{-0.08}$ & 1.41 & 2.34\\
		\rule{0pt}{3ex}
		\texttt{mock3} & 3.1 & 0.2 & Kroupa & 3.1 & 0.2 & 7.5 & $0.93^{+0.51}_{-0.97}$ & $2.32^{+0.10}_{-0.08}$ & 1.20 & 2.27\\
		\rule{0pt}{3ex}
		\texttt{mock4} & 8.5 & -0.2 & Kroupa & 8.5 & -0.2 & 9.7 & $1.17^{+0.22}_{-0.40}$ & $2.35^{+0.10}_{-0.10}$ & 1.34 & 2.32\\
		\rule{0pt}{3ex}
		\texttt{mock5} & 8.5 & 0.0 & Kroupa & 8.5 & 0.0 & 8.0 & $1.30^{+0.23}_{-0.30}$ & $2.33^{+0.09}_{-0.09}$ & 1.29 & 2.34\\
		\rule{0pt}{3ex}
		\texttt{mock6} & 8.5 & 0.2 & Kroupa & 8.5 & 0.2 & 11.9 & $1.33^{+0.21}_{-0.27}$ & $2.24^{+0.09}_{-0.09}$ & 1.37 & 2.23\\
		\rule{0pt}{3ex}
		\texttt{mock7} & 13.0 & -0.2 & Kroupa & 13.0 & -0.2 & 7.8 & $1.33^{+0.16}_{-0.39}$ & $2.43^{+0.12}_{-0.10}$ & 1.40 & 2.41\\
		\rule{0pt}{3ex}
		\texttt{mock8} & 13.0 & 0.0 & Kroupa & 13.0 & 0.0 & 10.0 & $1.35^{+0.24}_{-0.29}$ & $2.14^{+0.12}_{-0.10}$ & 1.41 & 2.12\\
		\rule{0pt}{3ex}
		\texttt{mock9} & 13.0 & 0.2 & Kroupa & 13.0 & 0.2 & 4.6 & $1.46^{+0.21}_{-0.21}$ & $2.20^{+0.10}_{-0.10}$ & 1.38 & 2.22\\
		\rule{0pt}{3ex}
		\texttt{mock10} & 3.1 & 0.2 & bottom-heavy & 3.1 & 0.2 & 3.9 & $2.95^{+0.07}_{-0.10}$ & $3.00^{+0.07}_{-0.07}$ & 2.99 & 2.97\\
		\rule{0pt}{3ex}
		\texttt{mock11} & 8.5 & 0.0 & bottom-heavy & 8.5 & 0.0 & 9.1 & $3.03^{+0.08}_{-0.09}$ & $2.91^{+0.08}_{-0.08}$ & 3.06 & 2.90\\
		\rule{0pt}{3ex}
		\texttt{mock12} & 13.0 & -0.2 & bottom-heavy & 13.0 & -0.2 & 5.5 & $3.00^{+0.13}_{-0.07}$ & $3.03^{+0.09}_{-0.12}$ & 2.98 & 3.06\\
		\hline
	\end{tabular}
\end{table*}

\subsubsection{Non-linear parameters}
\label{sec:nonlinearParameters}
The non-linear parameters of our model are represented by the age and metallicity of the stellar templates and by the parameters of the IMF prior model. We split the sampling of these parameters into two parts. 

First we determine the evidence for a grid of ages and metallicities. Each point in this grid is associated with a set of stellar templates with corresponding age and metallicity. We select the templates with the highest evidence. Secondly, we use these templates to refine the sampling of the parameters of the IMF prior model. Note that in principle \texttt{Multinest} already provides us with a sample of the parameters of the IMF prior and that the sampling of these parameters with \texttt{emcee} for the stellar templates with the highest evidence is only required to get a more precise sampling. Inside this loop, for every sample of the IMF prior model parameters $p_i$, the model continuously determines the most probable regularization parameter and solves for the most probable weights $\mathbfit{w}_{\!_\mathrm{MP}}$.

To demonstrate the reconstruction of the non-linear parameters we consider a set of twelve mock SSPs. The first nine mock SSPs have a Kroupa IMF ($\alpha_1=1.3$, $\alpha_2=2.3$) and the last three mock SSPs have a bottom-heavy IMF with $\alpha_1=\alpha_2=3.0$ (the two low-mass slopes limit the range of low-mass slopes that are currently being considered in the literature as reasonable for early-type galaxies). All of the mock SSPs have a SNR of 70. The parameters of the different mock SSPs are summarized in Table \ref{tab:table_mockSSPs}. 

\begin{figure*}
	\includegraphics[width=0.8\textwidth]{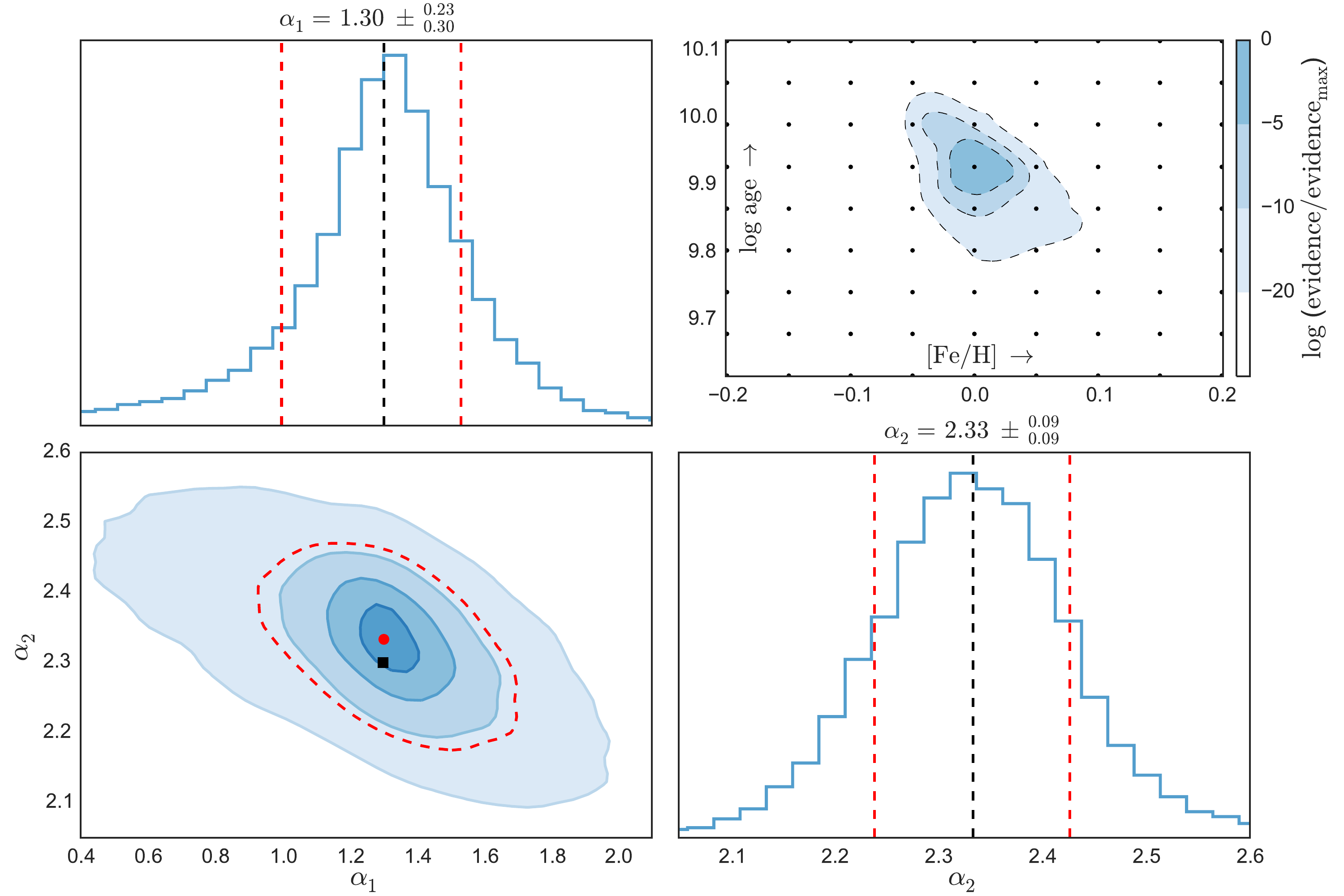}
	\caption{Reconstructed parameters for \texttt{mock5}. \textbf{Upper-right panel:} log evidence obtained for the different stellar templates in the age-metallicity grid. The evidences are re-scaled such that the log evidence of the templates with the highest evidence is zero. \textbf{Lower-left panel:} two-dimensional plot of the probability density distribution resulting from sampling the IMF slopes $\alpha_1$ and $\alpha_2$. For the sampling we used the stellar templates with the highest evidence. The different colors contain 10\%, 33\%, 60\% and 90\% of the points in the sample. The red-dashed line corresponds to the $1\sigma$ confidence interval. The red dot corresponds to the median of the sample whereas the black square corresponds to the input Kroupa IMF. \textbf{Upper-left panel:} marginalized distribution for the low-mass slope $\alpha_1$. \textbf{Lower-right panel:} marginalized distribution for the high-mass slope $\alpha_2$. The black-dashed lines in the histograms correspond to the median values and the red-dashed lines to the 16th and 84th percentiles of the marginalized distribution.}
	\label{reconstruction-mock5}
\end{figure*}

Fig. \ref{reconstruction-mock5} shows the results for the reconstruction of the non-linear parameters of \texttt{mock5}. The age-metallicity grid in this plot shows us that the stellar templates that we used as an input (i.e., $\log t[\mathrm{Gyr}] = 9.93$ and $\mathrm{[Fe/H]=0.0}$) give us the highest evidence. Although the grid shows an age-metallicity degeneracy, the evidence difference with the other grid points is at least more than 8. This implies that there is substantial evidence in favour of the correct stellar templates and that we are able to reconstruct the age and metallicity of the mock SSP. Note that in the future we plan to further expand the sampling to cover the full space of age, metallicity and IMF prior parameters.

In addition to the age-metallicity grid, Fig. \ref{reconstruction-mock5} also shows the reconstruction of the two IMF slopes. These slopes are part of the parameters that define the assumed double power law IMF prior parametrization. The values presented for $\alpha_1$ and $\alpha_2$ in Fig. \ref{reconstruction-mock5} correspond to the median values of the marginalized distributions whereas the errors represent the difference between the median and the 16th and 84th percentile. Within these confidence limits, the reconstructed IMF prior parameters for \texttt{mock5} agree well with the input parameters. 

\begin{figure}
	\centering
		{\includegraphics[width=0.99\columnwidth]{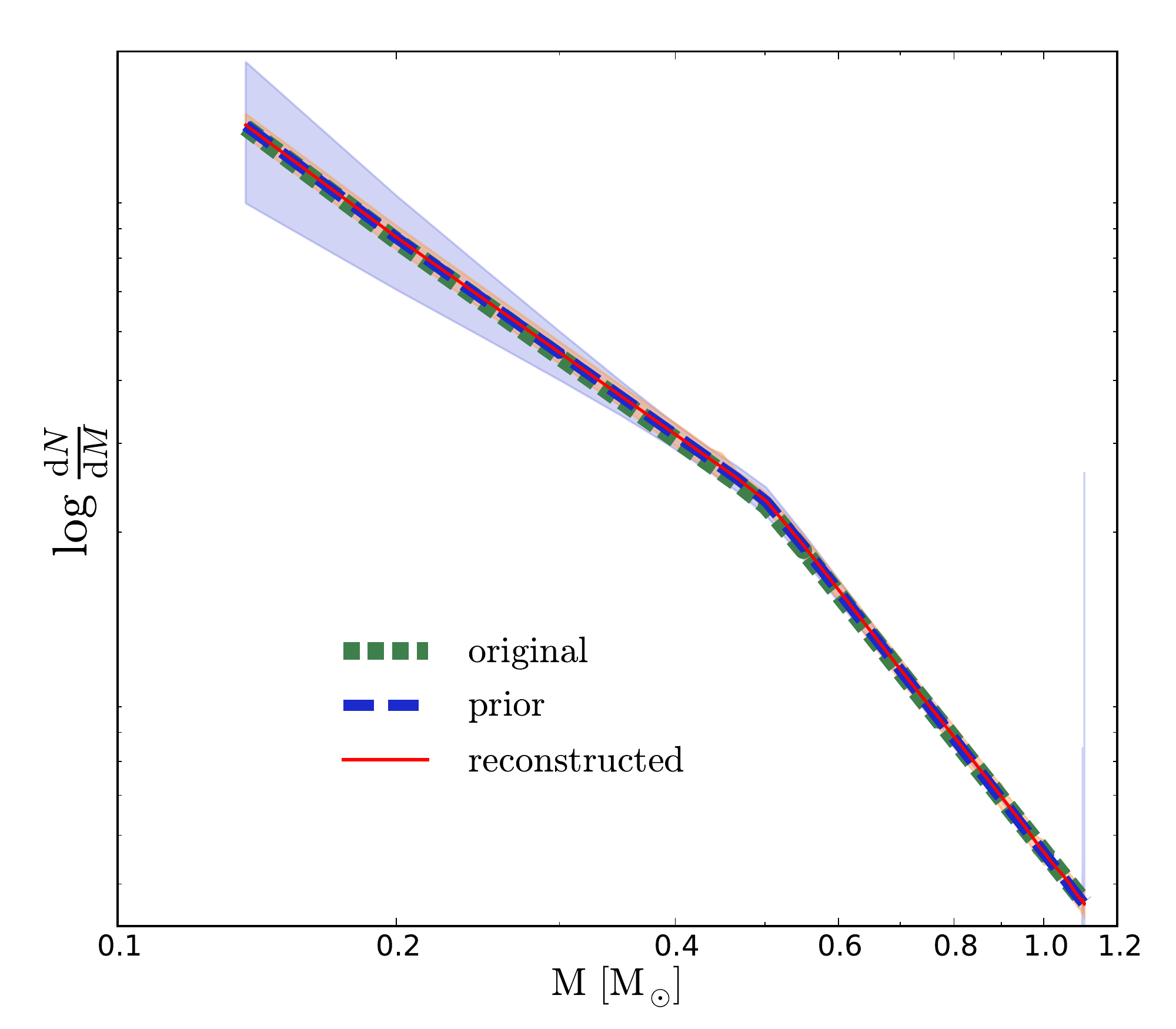}}
		{\includegraphics[width=0.99\columnwidth]{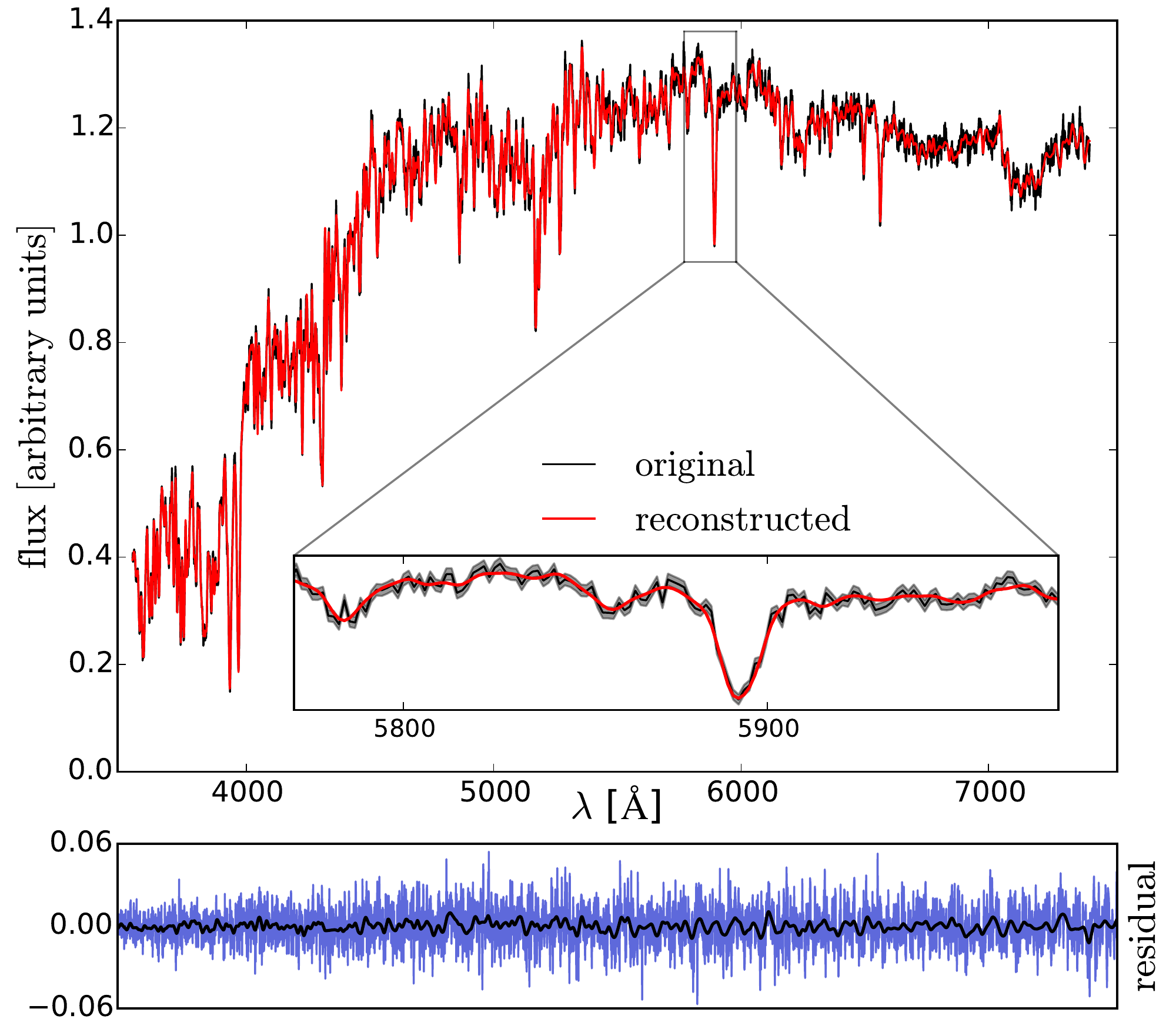}}
	\caption{Reconstruction of the piecewise IMF and the spectrum for \texttt{mock5}. \textbf{Top panel:} Reconstructed piecewise IMF for \texttt{mock5}. The \textit{short-dashed green} line represents the original IMF, the \textit{long-dashed blue} line represents the prior IMF, and the \textit{red} line represents the IMF reconstructed by the model. The \textit{shaded orange} region  corresponds to the error on the weights from the linear inversion. The \textit{shaded blue} region represents the area between the 16th and 84th percentile in the distribution of most probable weights. \textbf{Middle panel:} The \textit{black} line corresponds to the spectrum of the mock SSP and the \textit{red} line corresponds to the spectrum reconstructed by the model. The \textit{shaded gray} region in the zoom-in represents the one-sigma uncertainty corresponding to the SNR of the spectrum. \textbf{Bottom panel:}, the \textit{blue} line represent the residual between the original spectrum and the reconstructed spectrum. In the same panel, the \textit{black} line represents the residual smoothed with a Gaussian kernel with $\sigma=5$ pixels.}
	\label{reconstructionIMF4}
\end{figure}

Once we have a set of best fit values\footnote{We use the maximum a posteriori values of the sampled IMF prior parameters.} for the parameters of the IMF prior model, we use these parameters to construct a prior for the IMF. This prior may then be used to reconstruct the piecewise IMF, similar to what we did in Section \ref{sec:linearParameters}. The reconstructed IMF for \texttt{mock5} is shown in Fig. \ref{reconstructionIMF4}. Also shown in this figure is the reconstructed spectrum obtained by using this IMF together with the most probable stellar templates.

The plots corresponding to the other mock SSPs in Table \ref{tab:table_mockSSPs} are shown in Appendix \ref{sec:resultsMockSSPs}. For all of the mock SSPs, the reconstructed parameters are summarized in Table \ref{tab:table_mockSSPs}. We are able to select the true age and metallicity for all of our mock SSPs. As a measure of the robustness of this reconstruction, we present in Table \ref{tab:table_mockSSPs} the difference in log evidence with the second best set of stellar templates. These numbers show us that for \texttt{mock3}, \texttt{mock4}, \texttt{mock5}, \texttt{mock7}, \texttt{mock9}, \texttt{mock10}, \texttt{mock11} and \texttt{mock12} there is substantial evidence in favour of the true set of stellar templates. For \texttt{mock1}, \texttt{mock2}, \texttt{mock6} and \texttt{mock8} there is strong evidence in favour of the true stellar templates. See also Section \ref{sec:reconstruction_IMFmodelparameters} for the interpretation of the difference in log evidence.

The reconstructed IMF slopes for the twelve mock SSPs are listed in Table \ref{tab:table_mockSSPs}. Except for \texttt{mock7}, \texttt{mock8} and \texttt{mock11}, we are able to reconstruct the input slopes of the IMF within the confidence limits resulting from the sampling procedure. For \texttt{mock7}, \texttt{mock8} and \texttt{mock11}, the true high-mass slope $\alpha_2$ is just outside the confidence limits. Since the confidence limits correspond to the 16th and 84th percentile of the distribution, one would expect that indeed about one third of the test cases will be outside of these limits.

For the intermediate and older populations with a Kroupa IMF, the errors on the low-mass slope $\alpha_1$ are significantly smaller than those of the younger populations with a Kroupa IMF. Although the absolute signal of the low-mass stars in an old and a young population may be the same, the additional light of the young stars that are still present in the younger population effectively reduces the SNR of the low-mass stars. So the younger an SSP, the lower the SNR of the low-mass stars (for a spectrum with a given SNR) and the more difficult it becomes to constrain the low-mass slope $\alpha_1$. 

Driven by the increasing error on $\alpha_1$ for the mock SSPs with a Kroupa IMF, we also consider the fractional contribution $L_{0.5}$ of low-mass stars (i.e. $M < 0.5$ M$_{\odot}$) to the integrated spectrum across the MILES wavelength range. Table \ref{tab:lightFractions} provides an overview of this fraction for the different ages and IMF prior models that we consider in Table \ref{tab:table_mockSSPs} (for solar metallicity). We conclude that the signal from the low-mass stars in the youngest populations with a Kroupa IMF becomes comparable to or even lower than the intrinsic noise of the spectrum. Hence, this explains why there is a sudden increase in the error on $\alpha_1$ if we go from the intermediate to the youngest populations and why the effect is much smaller if we go from the oldest populations to the intermediate populations.

\begin{table}
	\centering
	\caption{Fractional contribution $L_{0.5}$ of low-mass stars ($M < 0.5$ M$_{\odot}$) to the integrated spectrum across the MILES wavelength range. These fractions are derived for the ages and IMF prior models considered in Table \ref{tab:table_mockSSPs} for solar metallicity using our models.}
	\label{tab:lightFractions}
	\begin{tabular}{ccc} 
		\hline
		age [Gyr] & IMF & $L_{0.5}$ \\
		\hline
		3.1 & Kroupa & 0.9\%\\
		8.5  & Kroupa & 2.2\%\\
		13.0 & Kroupa & 3.2\%\\
		3.1 & bottom-heavy & 4.2\%\\
		8.5 & bottom-heavy & 8.6\%\\
		13.0 & bottom-heavy & 11.3\%\\
		\hline
	\end{tabular}
\end{table}

By increasing the age of an SSP, one reduces light from the more massive stars in the SSP and effectively increases the SNR of the low-mass stars. A different way to increase the SNR of the low-mass stars is to increase the number of low-mass stars. To realize this, we consider a set of three bottom-heavy SSPs (\texttt{mock10}, \texttt{mock11} and \texttt{mock12}). Table \ref{tab:lightFractions} shows that the relative contribution of low-mass stars to the integrated spectrum is much higher than it is for a Kroupa IMF. This is reflected in the smaller errors on the low-mass slope $\alpha_1$ in Table \ref{tab:table_mockSSPs}, even for the youngest population where the signal of the low-mass stars is still well above the intrinsic noise of the spectrum.

The high-mass slope $\alpha_2$ is, independently of age, determined by the stars that emit most of the light. Therefore we expect that the model accurately reconstructs $\alpha_2$ for all mock SSPs. This is confirmed by the relatively small errors on $\alpha_2$ in Table \ref{tab:table_mockSSPs}, which are more or less constant as a function of age.

\begin{figure}
	\centering
	\includegraphics[width=0.99\columnwidth]{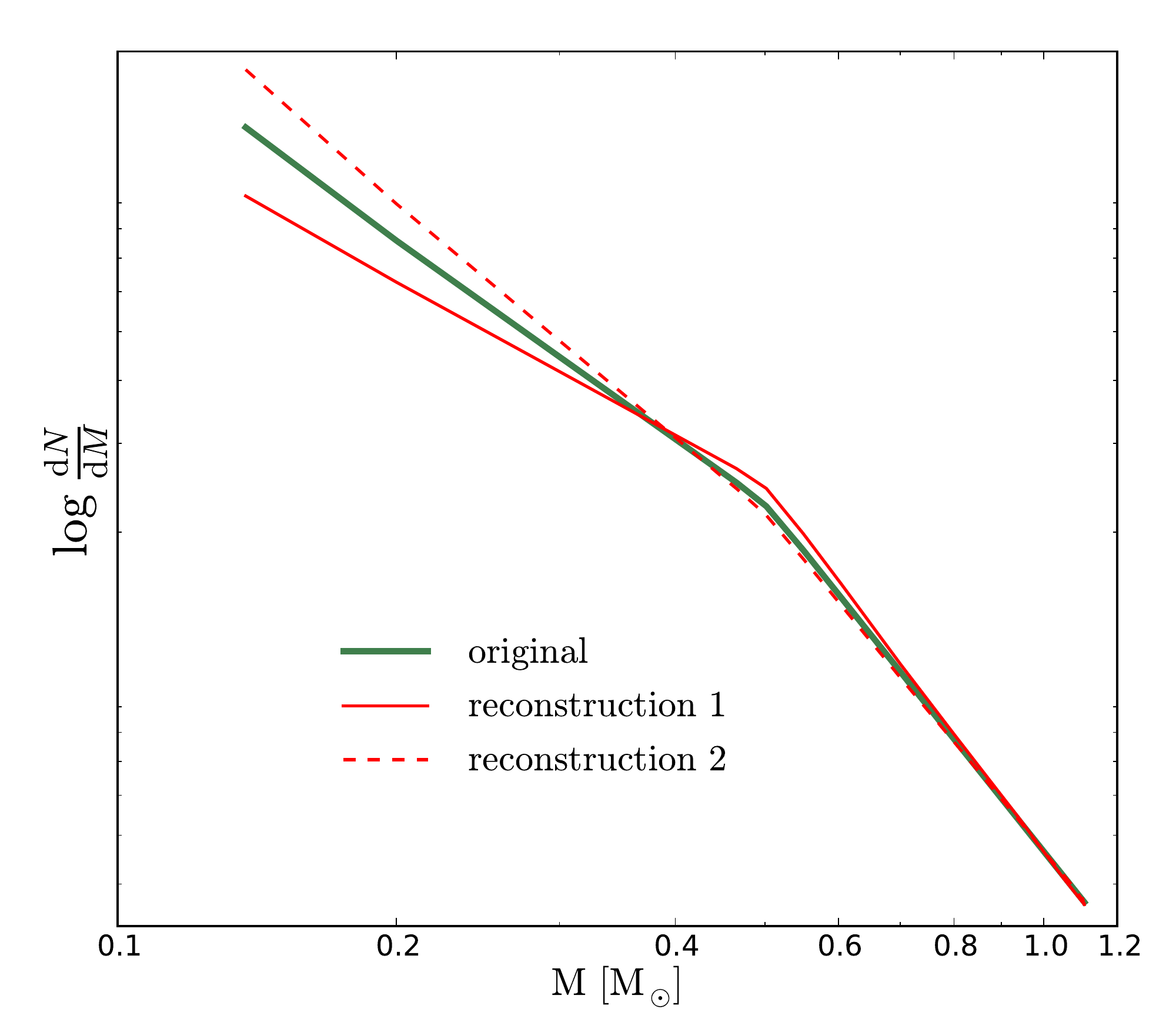}
	\caption{Two different IMF reconstructions for \texttt{mock5}. The first reconstruction combines a lower low-mass slope with a higher high-mass slope and the second reconstruction combines a higher low-mass slope with a lower high-mass slope. These reconstructions correspond to the one sigma values around the median of the sampled IMF slopes and reflect the degeneracy between the IMF slopes in Fig. \ref{reconstruction-mock5}.}
	\label{plotSlopeDegeneracy}
\end{figure}

All of the two-dimensional probability density plots in Fig. \ref{reconstruction-mock5} and Appendix \ref{sec:resultsMockSSPs} show a clear degeneracy between the low-mass slope $\alpha_1$ and high-mass slope $\alpha_2$. This degeneracy is such that an increasing low-mass slope seems to prefer a decreasing high-mass slope. To interpret this result we show in Fig. \ref{plotSlopeDegeneracy} two reconstructed IMFs corresponding to the one sigma values around the medians of the sampled values. For the first reconstruction we combine a lower low-mass slope with a higher high-mass slope (i.e. $\alpha_1 = 1.0$, $\alpha_2 = 2.42$) and for the second reconstruction we combine a higher low-mass slope with a lower high-mass slope (i.e. $\alpha_1 = 1.53$, $\alpha_2 = 2.24$). The parameters of these two reconstructions reflect the degeneracy visible in Fig. \ref{reconstruction-mock5}. The normalization of the IMF is lower for the second reconstruction. For both reconstructions the model appears to correct an over-abundance (under-abundance) of the lowest mass templates with respect to the real IMF by an under-abundance (over-abundance) of the templates around 0.5 $\mathrm{M}_{\odot}$ (luminosity conservation). In fact, the observed degeneracy between the IMF slopes may therefore be closely related to the degeneracy observed in Fig. \ref{reconstructionIMF2}.

\subsubsection{Mass fraction of low-mass stars}
Stellar templates in the same stellar mass range can have spectra that look very similar. As a consequence, there may be degeneracies between stellar templates with similar masses. Our regularization scheme ensures that these degeneracies do not cause problems when we solve for the most probable weights. However, if such a degeneracy exists it is basically impossible to find the exact contribution of the degenerate templates and hence the balance between these templates is set by the parametrization of the IMF prior. Although in that case we may not be able to completely reconstruct the piecewise IMF, we may still be able to constrain broader IMF-related quantities, for example the dwarf-to-giant ratio. 

\label{sec:massFractions}
\begin{table}
	\centering
	\caption{Mass fraction of stars with $M < 0.5$ $\mathrm{M_{\odot}}$ for the twelve mock SSPs in Table \ref{tab:table_mockSSPs}. The mass fractions and corresponding errors are derived from the most probable distribution of weights. As a reference, for every mock SSP we also provide $F_{0.5,\mathrm{original}}$, the value of $F_{0.5}$ that the corresponding SSP would have for a the input IMF (i.e. Kroupa or bottom-heavy).}
	\label{tab:massfractions}
	\begin{tabular}{ccc} 
		\hline
		name & $F_{0.5,\mathrm{original}}$ & $F_{0.5}$ \\
		\hline
		\rule{0pt}{3ex}
		\texttt{mock1} & $0.52$ & $0.55^{+0.05}_{-0.13}$\\
		\rule{0pt}{3ex}
		\texttt{mock2}  & $0.51$ & $0.53^{+0.02}_{-0.23}$\\
		\rule{0pt}{3ex}
		\texttt{mock3} & $0.51$ & $0.49^{+0.05}_{-0.15}$\\
		\rule{0pt}{3ex}
		\texttt{mock4} & $0.58$ & $0.59^{+0.01}_{-0.08}$\\
		\rule{0pt}{3ex}
		\texttt{mock5} & $0.58$ & $0.58^{+0.03}_{-0.04}$\\
		\rule{0pt}{3ex}
		\texttt{mock6} & $0.58$ & $0.58^{+0.03}_{-0.04}$\\
		\rule{0pt}{3ex}
		\texttt{mock7} & $0.62$ & $0.64^{+0.02}_{-0.06}$\\
		\rule{0pt}{3ex}
		\texttt{mock8} & $0.61$ & $0.61^{+0.02}_{-0.05}$\\
		\rule{0pt}{3ex}
		\texttt{mock9} & $0.61$ & $0.61^{+0.04}_{-0.02}$\\
		\rule{0pt}{3ex}
		\texttt{mock10} & $0.81$ & $0.80^{+0.01}_{-0.01}$\\
		\rule{0pt}{3ex}
		\texttt{mock10} & $0.83$ & $0.83^{+0.01}_{-0.01}$\\
		\rule{0pt}{3ex}
		\texttt{mock12} & $0.85$ & $0.85^{+0.01}_{-0.01}$\\
		\hline
	\end{tabular}
\end{table}

One of the important questions that we try to answer with our model is what the relative importance of low-mass stars is to the total stellar mass. In that context, transforming the most probable weights determined by our model into a fractional contribution of low-mass stars to the total stellar mass allows for a simple interpretation of the results.

\cite{LaBarbera} define the fraction of the total initial stellar mass in stars with $M < 0.5 \mathrm{\:M_{\odot}}$ as 
\begin{equation}
\mathrm{Fraction}(<0.5 \mathrm{M_{\odot}}) \equiv \frac{\int_{0.1\mathrm{\: M_{\odot}}}^{0.5 \mathrm{\:M_{\odot}}} \xi(M) M \mathrm{d}M}{\int_{0.1\mathrm{\:M_{\odot}}}^{100 \mathrm{\:M_{\odot}}} \xi(M) M \mathrm{d}M}.
\end{equation}

However, the SSPs that we consider here do not have stars more massive than $\sim 1.5$ $\mathrm{M_{\odot}}$. Everything beyond the high-mass-cut-off (HMCO) of the current mass function (i.e. the highest mass star in an SSP of a given age and metallicity that is still present) is more sensitive to the parametrization of the IMF than to the actual distribution of stellar masses. This is particularly true if we extrapolate our reconstructed IMFs, which can possibly be irregular. Therefore, we define the quantity $F_{0.5}$ as the fraction of the total current stellar mass that is in stars with $M < 0.5$ $\mathrm{M_{\odot}}$
\begin{equation}
\label{eq:DGratio}
F_{0.5}(<0.5 \mathrm{M_{\odot}}) \equiv \frac{\int_{0.1\mathrm{\: M_{\odot}}}^{0.5 \mathrm{\:M_{\odot}}} M\xi(M) \mathrm{d}M}{\int_{0.1\mathrm{\:M_{\odot}}}^{m_{\mathrm{HMCO}}} M\xi(M) \mathrm{d}M}.
\end{equation}

In Table \ref{tab:massfractions} we summarize the mass fractions $F_{0.5}$ for our twelve mock SSPs. As a reference, for every SSP we report $F_{0.5,\mathrm{original}}$: the value of $F_{0.5}$ for an SSP with the same age and metallicity and an IMF equal to the input IMF. The results in Table \ref{tab:massfractions} show that for all of our mock SSPs the value of $F_{0.5,\mathrm{original}}$ lies within one sigma of the reconstructed $F_{0.5}$ and these results are therefore consistent with the input data.

\subsubsection{The signal-to-noise ratio}
\label{sec:SNRratio}
\begin{table}
	\centering
	\caption{The effect of the SNR on the reconstructed parameters. For two of the mock SSPs in Table \ref{tab:table_mockSSPs} we compare the reconstructed parameters for spectra with a SNR of 70 and spectra with a SNR of 150.}
	\label{tab:table_mockSSPs_SNR}
	\begin{tabular}{ccccccc} 
		\hline
		name & age [Gyr] & [Fe/H] & SNR & $\alpha_1$ & $\alpha_2$\\
		\hline
		\texttt{mock3-70} & 3.1 & 0.2 & 70 & $0.93^{+0.51}_{-0.97}$ & $2.32^{+0.10}_{-0.08}$\\
		\rule{0pt}{3ex}
		\texttt{mock3-150} & 3.1 & 0.2 & 150 & $1.38^{+0.23}_{-0.39}$ & $2.29^{+0.04}_{-0.05}$\\
		\rule{0pt}{3ex}
		\texttt{mock7-70} & 13.0 & -0.2 & 70 & $1.33^{+0.16}_{-0.39}$ & $2.43^{+0.12}_{-0.10}$\\
		\rule{0pt}{3ex}
		\texttt{mock7-150} & 13.0 & -0.2 & 150 & $1.28^{+0.08}_{-0.09}$ & $2.29^{+0.04}_{-0.05}$\\
		\hline
	\end{tabular}
\end{table}

The spread that we find in the reconstructed parameters for the mock SSPs is related to the SNR of the input spectra. To demonstrate this, we compare the reconstructed parameters for four mock SSPs with different SNRs. We will consider a young, metal-rich population with the same parameters as \texttt{mock3} and an old, metal-poor population with the same parameters as \texttt{mock7}. For both populations we have a spectrum with a SNR=70 and a spectrum with a SNR=150. The spectra are summarized in Table \ref{tab:table_mockSSPs_SNR}.

\begin{figure}
	\centering
		{\includegraphics[width=0.99\columnwidth]{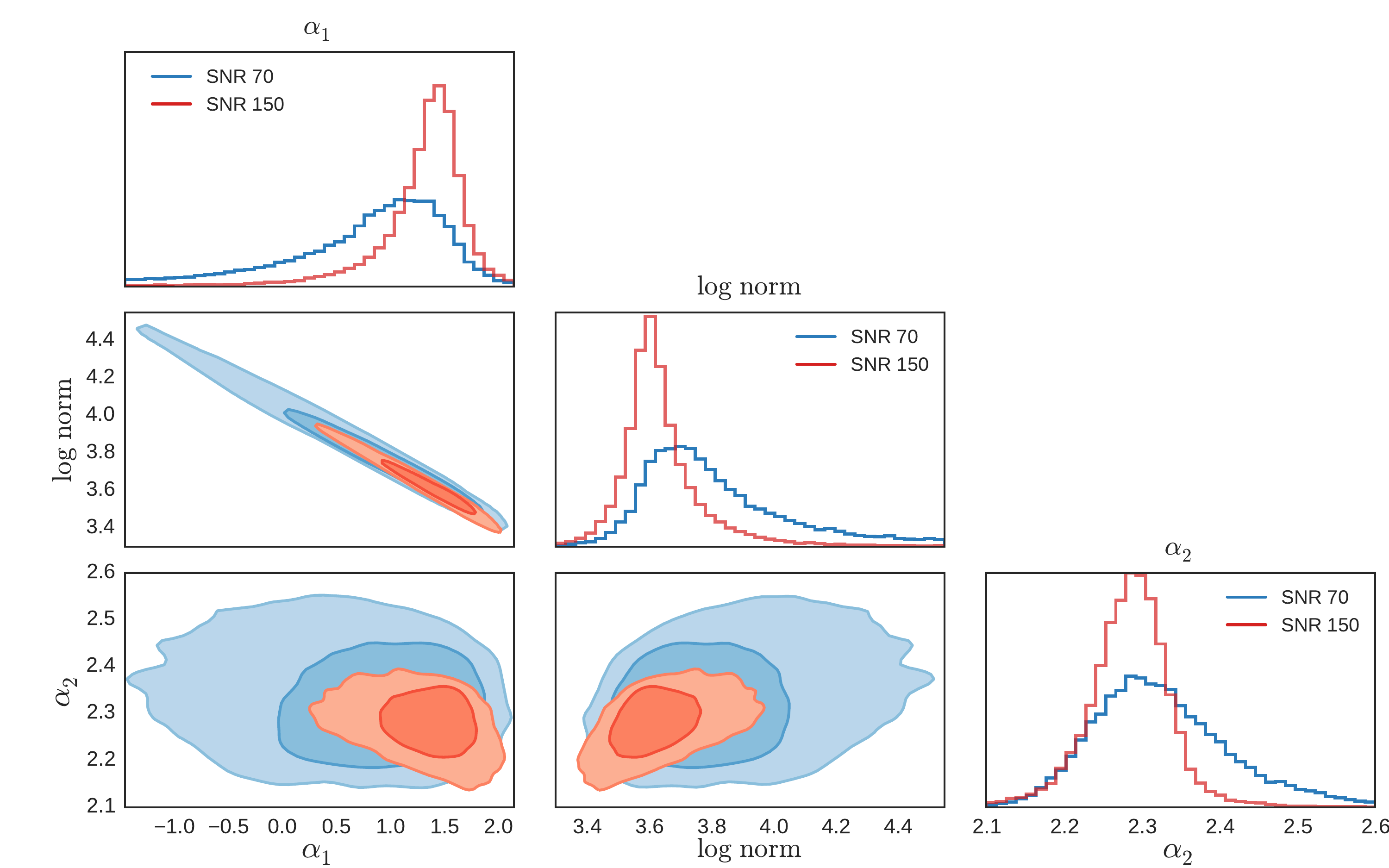}}
		{\includegraphics[width=0.99\columnwidth]{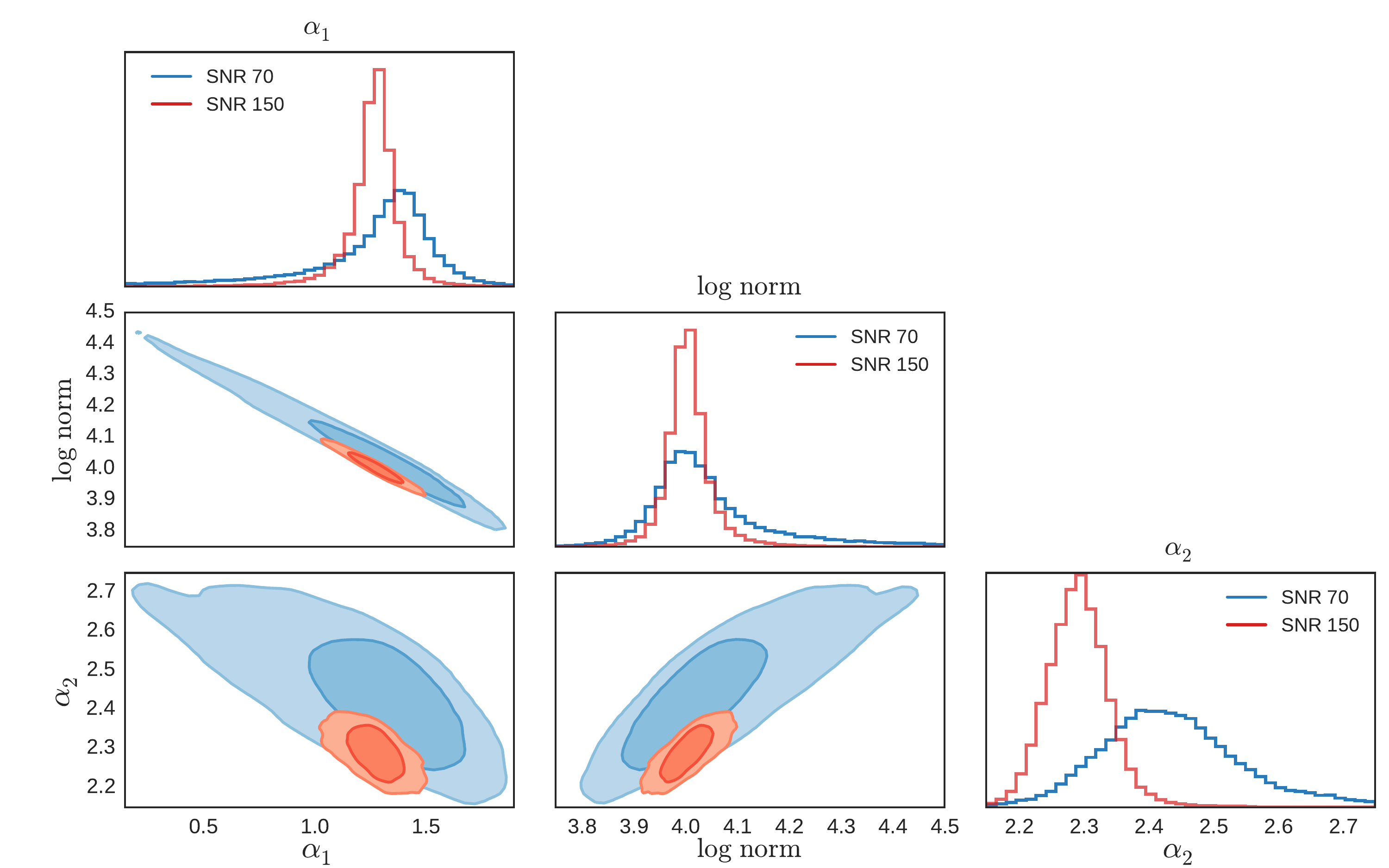}}
	\caption{The effect of the SNR on the reconstructed IMF parameters for the mock SSPs in Table \ref{tab:table_mockSSPs_SNR}. In the two-dimensional probability density plots, the \textit{dark} and \textit{light blue} contours contain 68\% and 90\% of the sampled points for the spectra with SNR=70. The \textit{dark} and \textit{light red} contours correspond to 68\% and 90\% of the sampled points for the spectra with SNR=150. Also shown are the marginalized distributions for the IMF parameters: \textit{blue} corresponds to a SNR=70 and \textit{red} to a SNR=150. \textbf{Top panel:} Reconstructed IMF parameters for \texttt{mock3-70} and \texttt{mock3-150}. \textbf{Bottom panel:} Reconstructed IMF parameters for \texttt{mock7-70} and \texttt{mock7-150}.}
	\label{reconstruction-effectSNR}
\end{figure}

The reconstructed IMF prior parameters ($\alpha_1$, $\alpha_2$ and the normalization of the IMF) for the mock SSPs in Table \ref{tab:table_mockSSPs_SNR} are shown in Fig. \ref{reconstruction-effectSNR}. This figure shows that increasing the SNR from 70 to 150 decreases the spread in the reconstructed IMF prior parameters and results in a much sharper peak in the marginalized distributions. In this case, the effect of increasing the SNR is very clear. However, for these mock SSPs the only source of uncertainty that we have is the Gaussian noise that we add a priori to the spectra of the SSPs. When we consider real data, there are additional problems and uncertainties that play a role, including flux-calibration issues, telluric residuals, abundance ratios and systematic uncertainties between the interpolated stellar templates and the true stellar templates. These errors are non-negligible and therefore increasing the SNR will not always improve the reconstruction of the IMF model parameters.

\subsubsection{Wavelength dependence}
The residuals between the original spectrum and the reconstructed spectrum for \texttt{mock5} in Fig. \ref{reconstructionIMF4} do not show a strong dependence on wavelength. To investigate a possible wavelength dependence of the residuals in more detail, we smooth the residuals with a Gaussian kernel with $\sigma=5$ pixels. The smoothed version of the residual is also shown in Fig. \ref{reconstructionIMF4}. The smoothed residuals seem to be slightly smaller at the bluest wavelengths, but this is mainly related to the lower flux in that region (the residuals are absolute).

Although we do not see a wavelength dependence in the residuals of our results, this does not tell us which wavelength regions are better suited to constrain the IMF. To investigate the relation between the reconstruction of the IMF prior parameters and the considered wavelength region we repeat our analysis for \texttt{mock5} using only the blue half of the spectrum and using only the red half of the spectrum. The results of this analysis are presented in Table \ref{tab:wavelengthDependence}.

\begin{table}
	\centering
	\caption{Effect of wavelength region on the reconstruction of the IMF slopes $\alpha_1$ and $\alpha_2$ for \texttt{mock5}.}
	\label{tab:wavelengthDependence}
	\begin{tabular}{ccc} 
		\hline
		wavelength region & $\alpha_1$ & $\alpha_2$ \\
		\hline
		\rule{0pt}{3ex}
		3540-7410 \AA & $1.30^{+0.23}_{-0.30}$ & $2.33^{+0.09}_{-0.09}$\\
		\rule{0pt}{3ex}
		3540-5122 \AA  & $-0.59^{+1.15}_{-0.96}$ & $2.54^{+0.14}_{-0.16}$\\
		\rule{0pt}{3ex}
		5122-7410 \AA & $0.99^{+0.53}_{-0.98}$ & $2.23^{+0.17}_{-0.18}$\\
		\hline
	\end{tabular}
\end{table}

Firstly, the results in Table \ref{tab:wavelengthDependence} show that ignoring half of the data points significantly increases the scatter in the reconstructed parameters. Secondly, the reconstructed parameters that we obtain using the reddest wavelength region are consistent with the input data whereas the reconstructed parameters derived from the bluest wavelength region are not. Since the overall flux in the blue part of the spectrum is lower, the SNR of the blue part is also slightly lower. Moreover, according to our models in the blue part of the spectrum the low-mass stars contribute only 1.5\% to the integrated spectrum whereas in the red part they contribute 2.6\%. Taking into account that these numbers have the same order of magnitude as the intrinsic noise in the spectrum this represents an important difference. This difference and the lower SNR in the blue part of the spectrum might explain why we are able to constrain $\alpha_1$ correctly in the red part of the spectrum and not in the blue part of the spectrum. The offset that we find for $\alpha_2$ in the blue part of the spectrum may be related to the model being unable to correctly break the degeneracy between the IMF slopes.

The results in this section show that in order to constrain the contribution of low-mass stars to the spectrum it is important to consider a broad wavelength region. Furthermore, it is essential to consider wavelength regions and features that are sensitive to changes in the IMF. Since low-mass stars emit most of their light in the (infra)red part of the spectrum and most of the IMF sensitive features are also found in this part of the spectrum \citep{Spiniello_2013}, (infra)red wavelength regions will be more useful to constrain the low-mass IMF. In future work, we therefore plan to combine our model with the X-Shooter Spectral Library which provides a much broader and redder wavelength coverage.

\subsection{Second level of inference}
The second level of inference allows us to compare different model families based on the evidence. Here we consider two model families: one in which the IMF is parametrized as a double power law and one in which it is parametrized as a single power law. For simplicity, although there could in principle also be a degeneracy between age/metallicity and IMF model family, for now we fix the age and metallicity in both model families to the true values, such that the stellar templates are the same for the two model families.

To test the second level of inference we consider two mock SSPs with $t = 8.5$ Gyr and $\mathrm{[Fe/H]=0.0}$. The SNR of the spectra of these SSPs is 70. One of the SSPs has a Kroupa IMF and the other a Salpeter IMF.

We then use \texttt{Multinest} to calculate the evidence for both model families by applying them to the two mock SSPs. The results are presented in Table \ref{tab:table2ndLevelOfInference}. 

\begin{table}
	\centering
	\caption{Evidence for two different model families obtained by applying it to two different mock SSPs. The mock SSPs are created with the same stellar templates. One of the mock SSPs has a Kroupa IMF and the other has a Salpeter IMF. We compare a single power law IMF model family against a double power law IMF model family.}
	\label{tab:table2ndLevelOfInference}
	\begin{tabular}{ccc} 
		\hline
		true IMF & IMF model & log evidence \\
		     &	family		&		\\
		\hline
		Kroupa & single power law & $-22085.4 \pm 0.2$ \\
		Kroupa & double power law & $-22083.0\pm 0.2$ \\
		Salpeter & single power law & $-22049.3 \pm 0.2$  \\
		Salpeter & double power law & $-22051.7 \pm 0.2$ \\
		\hline
	\end{tabular}
\end{table}

First consider the Kroupa mock SSP. The difference in log evidence is 2.4 in favour of the double power law model family. Since a broken power law IMF is not part of the model family of single power laws, this is what we expect. However, the difference in log evidence between the two model families is not substantial. Apparently there is a single power law that, in combination with the allowed deviations from the prior IMF, is still able to provide a reasonable fit to the data. If this would not have been the case, the difference in the evidence would have been much more significant.

For the Salpeter mock SSP, the difference in log evidence is 2.4 in favour of of the single power law model family. In this case the single power law input IMF is part of both model families. Therefore we expect the two model families to be able to fit the data equally well. Nevertheless, the model prefers the simpler single power law model. This is the result of Occam's razor: the model is set up in such a way that it tries to find the simplest model that fits the data.

Considering the errors on the evidences in Table \ref{tab:table2ndLevelOfInference}, we conclude that for these specific mock SSPs we are able to discriminate between the single- and double-power-law model families. The difference in log evidence between the two model families is however not substantial.

\section{Results - model versus model}
\label{sec:modelvsmodel}
\begin{table*}
	\centering
	\caption{Overview of the three MILES SSPs considered in Section \ref{sec:modelvsmodel}. All of these mock SSPs have been downloaded from the MILES website and have a SNR of 70. The table provides both the input parameters and reconstructed values for each of the mock SSPs. In addition, we present the difference in log evidence with the second best set of stellar templates. The reconstructed IMF slopes are the median values of the distribution. Errors on the reconstructed slopes correspond to the 16th and the 84th percentile. The last two columns represent the reconstructed IMF slopes using a set of stellar templates from the MILES website.}
	\label{tab:table_MILES-SSPs}
	\begin{tabular}{ccccccccccc} 
		\hline
		 &  &  &  & \multicolumn{2}{c}{reconstructed} & $\Delta$ evidence & \multicolumn{4}{c}{reconstructed IMF prior parameters}\\
		   name  &	age [Gyr]		&	[Fe/H]	& IMF &	age [Gyr] & [Fe/H] & 2nd best & $\alpha_1$ & $\alpha_2$ & $\alpha_{1,\mathrm{MILES}}$ & $\alpha_{2,\mathrm{MILES}}$\\
		\hline
		\rule{0pt}{3ex}
		\texttt{MILES1} & 3.2 & 0.22 & Kroupa & 3.6 & 0.30 & 4.0 & $3.87^{+0.36}_{-0.44}$ & $0.89^{+0.27}_{-0.30}$ &  $-0.23^{+1.23}_{-1.17}$ & $2.44^{+0.14}_{-0.14}$ \\
		\rule{0pt}{3ex}
		\texttt{MILES2} & 8.9 & 0.0 & Kroupa & 9.8 & -0.05 & 4.2 & $2.08^{+0.44}_{-0.61}$ & $2.16^{+0.20}_{-0.20}$ & $1.01^{+0.54}_{-0.69}$ & $2.34^{+0.12}_{-0.15}$ \\
		\rule{0pt}{3ex}
		\texttt{MILES3} & 12.6 & -0.40 & Kroupa & 13.0 & -0.40 & 4.3 & $1.57^{+0.58}_{-0.71}$ & $1.97^{+0.30}_{-0.29}$ & $1.42^{+0.43}_{-0.52}$ & $2.21^{+0.22}_{-0.24}$ \\
		\hline
	\end{tabular}
\end{table*}

As a next step, we apply our model to a set of SSPs that have been constructed with a different SPS code. Because the current version of our model is based on the MILES library, the obvious choice is to compare our model against the MILES SPS models \citep{MILES-SPS-models, MILES_v10}. Here we consider three SSPs. The parameters of these SSPs are given in Table \ref{tab:table_MILES-SSPs}. We downloaded the spectra of these SSPs from the MILES website 

Before we discuss the reconstruction of the IMF for these three SSPs we first specify the regularization scheme that we used for the analysis of the MILES spectra.

\subsection{The regularization scheme}
For the MILES mock SSPs, there are also systematic uncertainties between the model and the input spectra in addition to the Gaussian noise that we add to the spectra. Since we do not take these uncertainties into account in the covariance matrix, sometimes this may result in unrealistic IMFs. This is a consequence of the model trying to provide a fit to the data that is too good with respect to the real uncertainties and at the cost of large deviations from the prior IMF (i.e. if the regularization parameter is low; this is also related to the use of NNLS). We solve this problem for now by using a different regularization scheme. For the MILES SSPs, we use $\mathbfss{C}^{-1}_{\mathrm{pr}} = \mathbfss{I}$ (i.e. the identity matrix). This regularization scheme penalizes the absolute deviation of the weights from the prior whereas the regularization scheme that we used in Section \ref{sec:results} penalizes both the relative deviation of the weights from the prior and the gradient of the relative deviations. In general there are much more low-mass stars than high-mass stars. Therefore the regularization scheme that we use in this section makes it harder for the model to deviate from the prior for the low-mass templates and helps to prevent strange solutions that are not regulated strongly enough. However, it comes at the cost of a less flexible model for the low-mass end.

\subsection{IMF reconstruction of MILES SSPs}
\label{sec:MILESresults}
We analyse the spectra of the three MILES mock SSPs in the same way as in Section \ref{sec:results}, except for the different regularization scheme. We first calculate the evidence for every set of stellar templates in the age-metallicity grid. Then we refine the sampling of the non-linear IMF prior parameters for the stellar templates with the highest evidence.

\begin{figure}
	\centering
		{\includegraphics[width=\columnwidth]{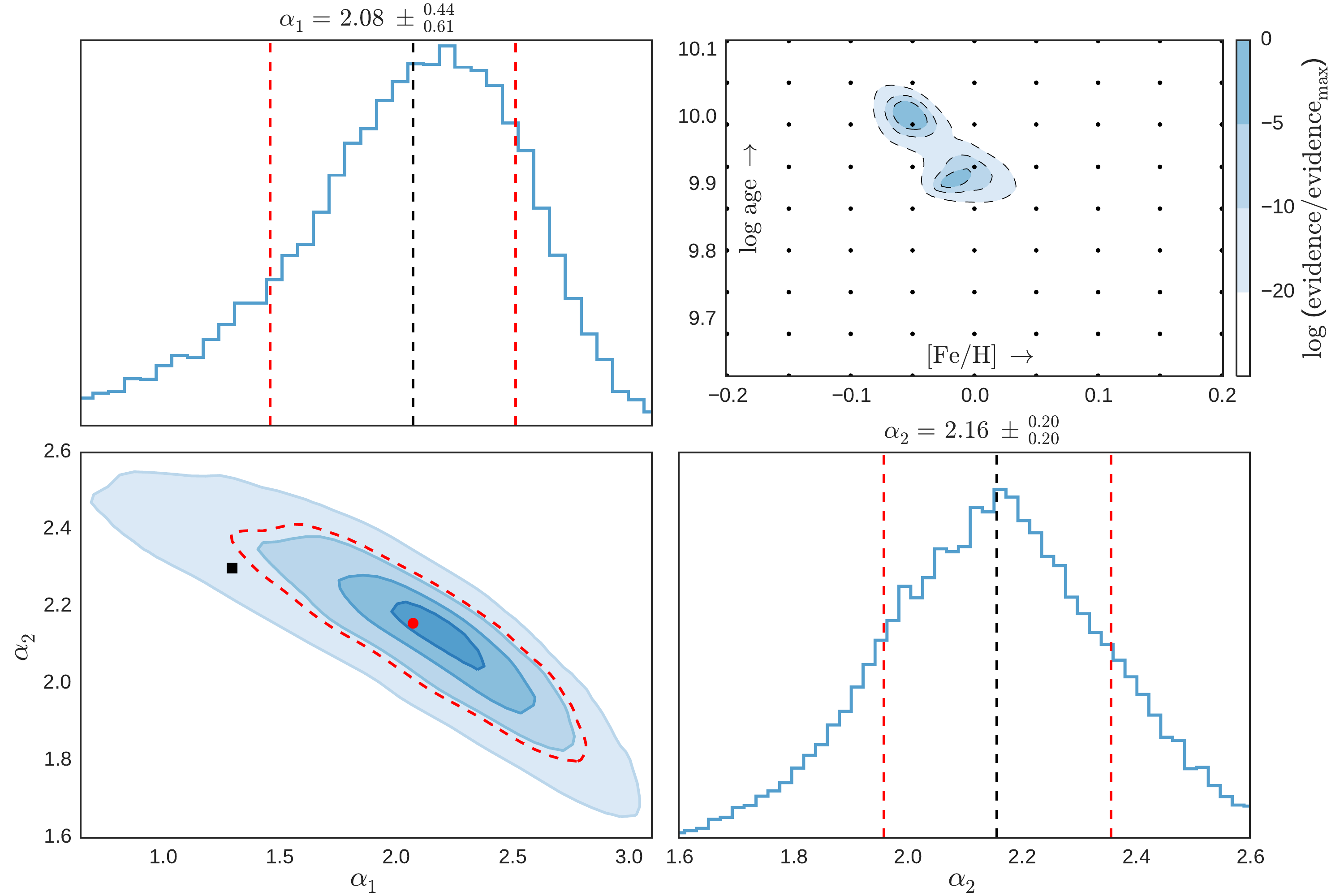}}
		{\includegraphics[width=\columnwidth]{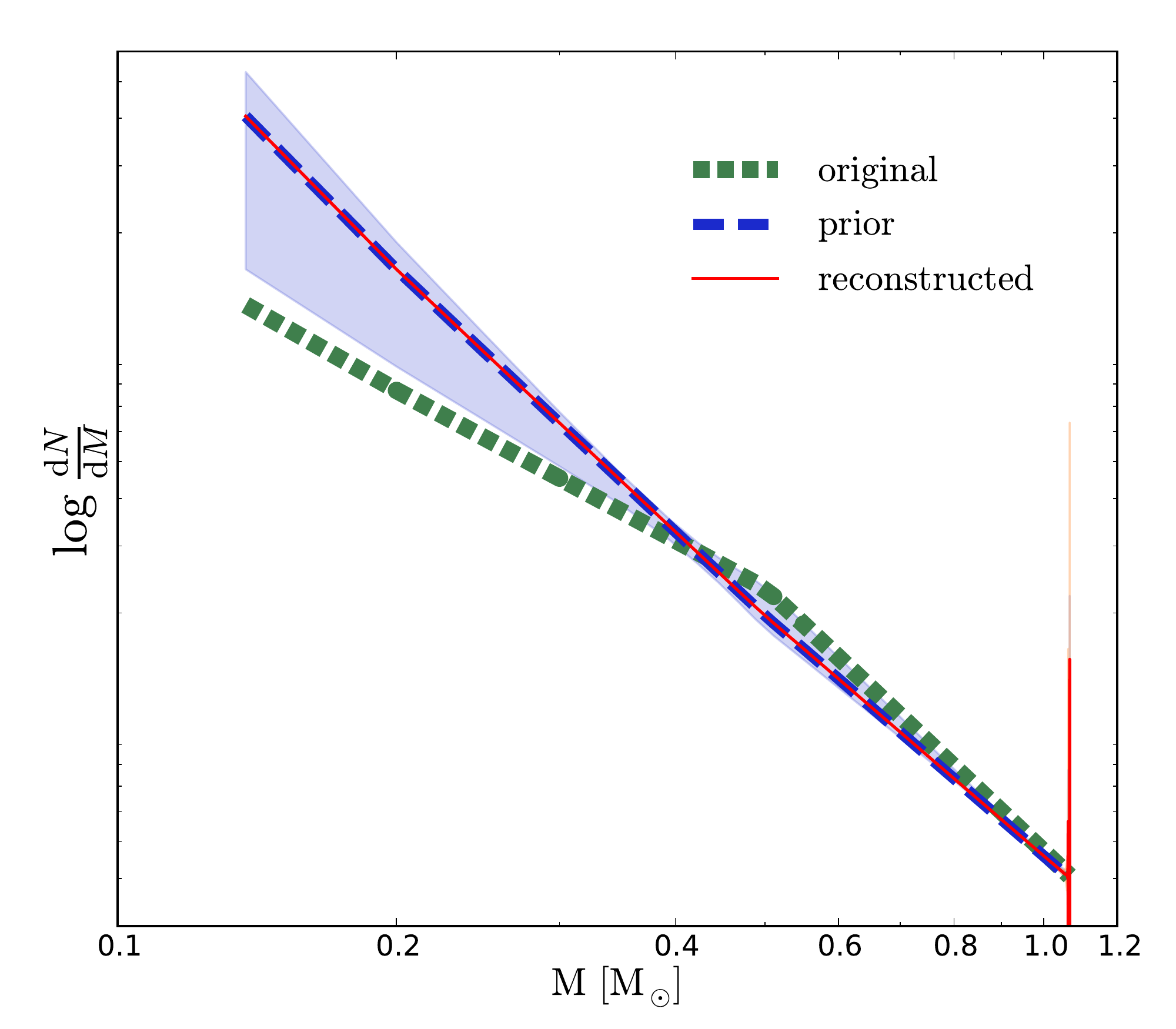}}
	\caption{Sampling of the IMF prior model parameters and reconstruction of the IMF for \texttt{MILES2}. \textbf{Top panel:} As in Fig. \ref{reconstruction-mock5} only now for \texttt{MILES2}. \textbf{Bottom panel:} As in the top panel of Fig. \ref{reconstructionIMF4} only now for \texttt{MILES2}.}
	\label{MILES_results2}
\end{figure}

The top panel of Fig. \ref{MILES_results2} shows the reconstructed parameters for the SSP with $t = 8.9$ Gyr and $\mathrm{[Fe/H] = 0.0}$ (\texttt{MILES2}). The preferred set of stellar templates has an age of 9.8 Gyr and a metallicity $\mathrm{[Fe/H] = -0.05}$. Note that the age-metallicity grid in our model does not contain templates with $t = 8.9$ Gyr. The stellar templates in our age-metallicity grid that are closest in age have $t = 8.5$ Gyr and $t = 9.8$ Gyr. Although the selected stellar templates are slightly too metal-poor, the reconstructed age and metallicity agree reasonably well with the input values.

The reconstructed high-mass slope for \texttt{MILES2} is consistent with the input values of a Kroupa IMF. The value obtained for the low-mass slope is, on the other hand, too high and just outside the one sigma confidence limits given by the model. Notice, however, we only take into account the noise in the data in the covariance matrix $\mathbfss{C}_{\mathrm{D}}$ and that systematic uncertainties are not taken into account. Therefore we expect the errors returned by the model to be a lower limit.

The bottom panel of Fig. \ref{MILES_results2} shows the reconstructed IMF for \texttt{MILES2}. The reconstructed IMF and the input IMF are consistent at the high-mass end. However, the original IMF for the low-mass end is just outside the one sigma confidence limits of the model. Note that the spike in the reconstructed IMF at the high-mass end corresponds to a template with a very small mass bin and that this spike is not visible in the reconstructed distribution of weights. For more details see also Appendix \ref{sec:MILES_results}.

For the other two MILES SSPs, the results are shown in Appendix \ref{sec:MILES_results}. The reconstructed parameters are summarized in Table \ref{tab:table_MILES-SSPs}. For all MILES SSPs, the reconstructed age and metallicity are relatively close to the input values.

Although the high-mass slope is slightly too low, the reconstructed IMF slopes for \texttt{MILES3} are consistent with a Kroupa IMF. The reconstructed slopes for \texttt{MILES1}, on the other hand, are inconsistent with the input values. There are a number of differences between the stellar templates in our model and the stellar templates in the MILES models that might explain the inconsistent reconstructions. First of all, we use a different set of isochrones. Second, although the interpolation mechanisms in the two models are similar they are not the same. In the MILES models there is an additional correction term that minimizes the difference between the requested stellar parameters and the stellar parameters of the interpolated spectrum whereas in our model there is a polynomial correction as described in Section \ref{sec:polCorrections}. Finally, the MILES models use empirical color-temperature relations to scale the stellar templates, whereas we use the colors provided by the isochrones.

\begin{figure}
	\centering
		{\includegraphics[width=\columnwidth]{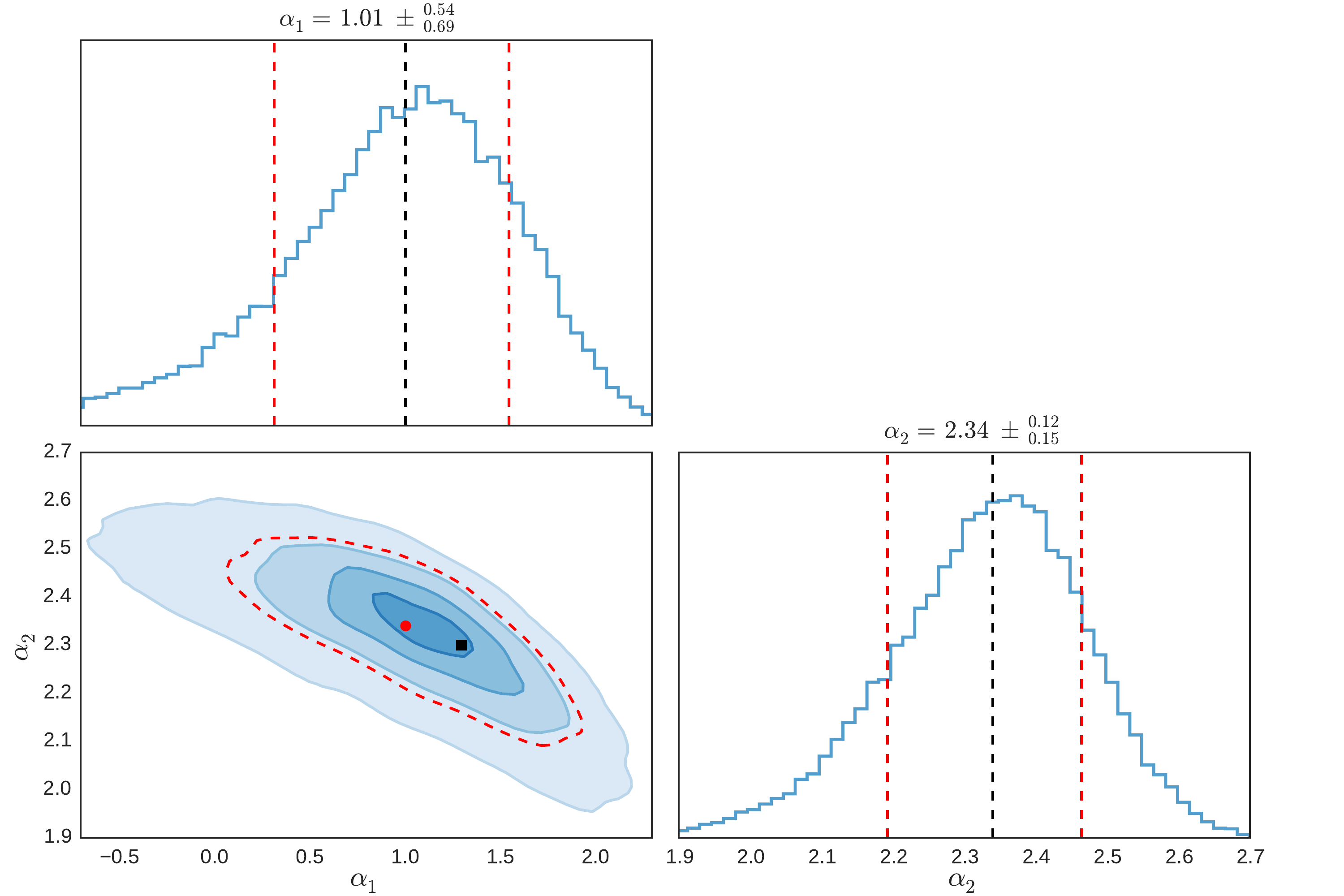}}
		{\includegraphics[width=\columnwidth]{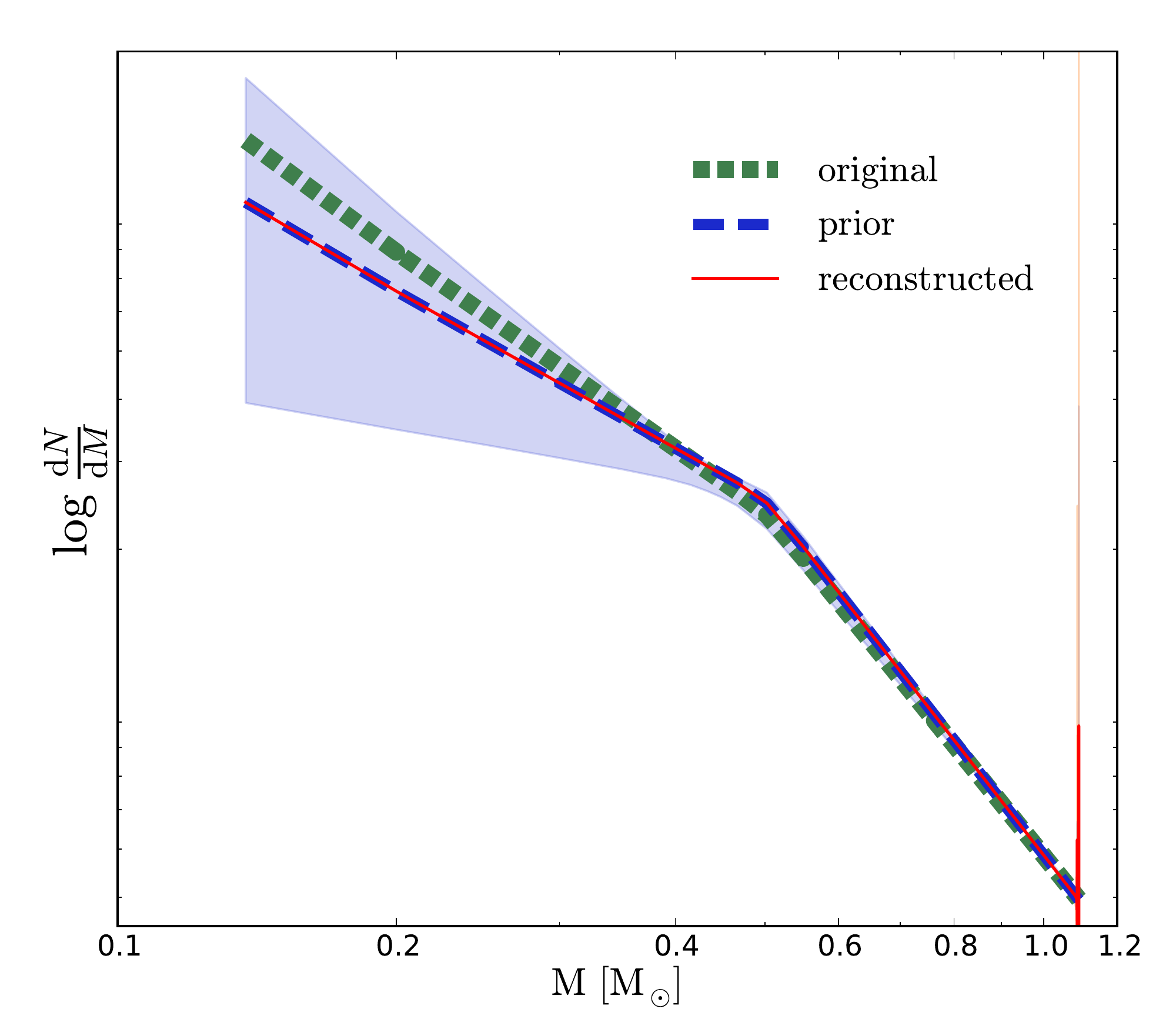}}
	\caption{As in Fig. \ref{MILES_results2}, except for the age-metallicity plot in the top panel, but now using the stellar templates created with the interpolator on the MILES website.}
	\label{MILES_results2_MILES_templates}
\end{figure}

We suspect that the stellar interpolator is one of the most important sources of uncertainty responsible for the discrepancy between the input IMF slopes for the MILES SSPs and the slopes reconstructed by our model. To test this hypothesis, we select an isochrone with $t=8.9$ Gyr and $\mathrm{[Fe/H]} = 0.0$. For all of the stars in this isochrone, we create a spectrum with the MILES interpolator by using the webtool on the MILES website. Using this set of stellar templates we once again apply our model to \texttt{MILES2}. The results are shown in Fig. \ref{MILES_results2_MILES_templates}. This figure clearly shows that, using the same interpolator, the reconstructed IMF slopes agree well with the input parameters.

For \texttt{MILES1} and \texttt{MILES3} we did a similar test. The results that we obtain are shown in Appendix \ref{sec:MILES_results}. The slopes that we reconstruct by applying our model using the MILES templates of the same age and metallicity as the input SSP are also given in Table \ref{tab:table_MILES-SSPs}. For \texttt{MILES3} the reconstructed IMF slopes are now completely consistent with a Kroupa IMF. For \texttt{MILES1} we are able constrain the high-mass slope but the data does not allow us to reconstruct the low-mass slope. As one might expect, the results in Table \ref{tab:table_MILES-SSPs} clearly show that it is easier to constrain the low-mass end of the IMF in older populations.

Our results show that the quality of the stellar templates that we provide to the model is a very important factor in our ability to reconstruct the IMF. If we use a set of stellar templates for which the systematic uncertainties are larger than the typical uncertainties in the data, we may find incorrect IMF parameters. The strength of our model lies in the fact that we can easily compare different sets of stellar templates based on their evidence. For example, if we apply our model to the MILES SSPs using the MILES templates, the log evidence is 28.3, 31.2, and 75.8 higher for, respectively, \texttt{MILES1}, \texttt{MILES2}, and \texttt{MILES3} than when we apply the model with our own stellar templates.

\subsection{Mass fraction of low-mass stars}
As in Section \ref{sec:massFractions}, we summarize our results by calculating the reconstructed mass fractions $F_{0.5}$. The reconstructed mass fractions for the MILES SSPs are given in Table \ref{tab:massfractionsMILES}. As a reference, this table also provides the mass fraction $F_{0.5}$ of the original IMF (Kroupa) for the same age and metallicity as the SSP. The results in Table \ref{tab:massfractionsMILES} show that we are only able to reconstruct the true value of $F_{0.5}$ for the oldest SSP.

\begin{table}
	\centering
	\caption{Mass fraction of stars with $M < 0.5$ $\mathrm{M_{\odot}}$ for the three MILES SSPs in Table \ref{tab:table_MILES-SSPs}. For each of the SSPs, the table provides both the mass fraction $F_{0.5}$ derived from the stellar templates described in Section \ref{sec:stellarTemplates} and the mass fraction $F_{0.5,\mathrm{MILES\:templates}}$ derived from the MILES templates. As a reference, for every mock SSP we also provide $F_{0.5,\mathrm{original}}$, the value of $F_{0.5}$ that the corresponding SSP would have for the input IMF (Kroupa).}
	\label{tab:massfractionsMILES}
	\begin{tabular}{cccc} 
		\hline
		name & $F_{0.5,\mathrm{original}}$ & $F_{0.5}$ & $F_{0.5,\mathrm{MILES\:templates}}$ \\
		\hline
		\rule{0pt}{3ex}
		\texttt{MILES1} & $0.51$ & $0.76^{+0.01}_{-0.04}$ & $0.41^{+0.06}_{-0.15}$\\
		\rule{0pt}{3ex}
		\texttt{MILES2} & $0.58$ & $0.72^{+0.02}_{-0.07}$ & $0.55^{+0.06}_{-0.09}$\\
		\rule{0pt}{3ex}
		\texttt{MILES3} & $0.63$ & $0.67^{+0.03}_{-0.16}$ & $0.67^{+0.02}_{-0.10}$\\
		\hline
	\end{tabular}
\end{table}

If we apply the MILES templates to the MILES SSPs, the reconstructed mass fractions $F_{0.5,\mathrm{MILES\:templates}}$ agree better with the input data. Except for \texttt{MILES1}, for which the obtained mass fraction is slightly too low, the reconstructed mass fractions lie within one sigma of the mass fraction for a Kroupa IMF. Since we expect that approximately one third of the results is outside the one sigma contours, these results are consistent with the data.

The discrepancy between the results that we obtain for the two sets of stellar templates may be explained by the systematic differences between the stellar templates used to create the MILES SSPs and the stellar templates described in Section \ref{sec:stellarTemplates}. These systematic differences have been discussed in Section \ref{sec:MILESresults} and include a different interpolator, different sets of isochrones and a different normalization scheme for the stellar templates. By using the stellar templates from the MILES website, we remove the interpolator as a source of systematic uncertainty. The remaining systematic uncertainties and the fact that the SNR of the low-mass stars is lower for younger populations (for a spectrum with constant SNR) most likely explains why the reconstructed value $F_{0.5,\mathrm{MILES\:templates}}$ for \texttt{MILES1} is inconsistent with the input IMF.

\section{Summary and discussion}
We have designed a new SPS code to reconstruct the shape of the IMF. The model that we have developed consists of a Bayesian framework with a number of different layers.

At the innermost level, the spectrum of a stellar population is represented as a linear combination of a set of stellar templates. For an SSP, these templates are defined by an isochrone. Combining an isochrone with a stellar library and an interpolator allows us to create a spectrum for each of these templates. The contribution of each of these spectra to the spectrum of the SSP (weights) is obtained through a linear inversion. We regularize this linear inversion by using a prior IMF, which translates into a prior on the weights.

The prior IMF that we use to regularize the linear inversion of the weights is chosen as being part of an IMF prior model family. This IMF prior model family is characterized by  a set of (non-linear) parameters $p_i$. For every combination of $p_i$, we are able to construct a prior on the IMF which may be transformed into a prior on the weights. Given the input spectrum, the stellar templates, and the prior on the weights, our model determines the Bayesian evidence for that particular set of parameters. This allows us to sample the parameters of the IMF prior model family using MCMC techniques.

We then applied our model to a number of mock SSPs to demonstrate its validity. What we have shown is that we are able to reconstruct the input parameters of these mock SSPs. The quality of the reconstruction for these SSPs is mostly determined by the SNR of the input spectra and by the relative contribution of low-mass stars to the integrated spectrum. We have shown that the latter depends on both the age of the SSP and on the slope of the IMF. For younger SSPs, more light is emitted by stars more massive than $0.5$ M$_{\odot}$. This effectively decreases the SNR of the stars less massive than $0.5$ M$_{\odot}$. By increasing the low-mass slope of the IMF, we increase the number of low-mass stars which in turn increases the SNR of the low-mass stars. Constraining the low-mass IMF is therefore easier for older SSPs and for IMFs that are more bottom-heavy.

As a next step, we applied our model to three (mock) SSPs created with the MILES models. The age and metallicity reconstructed by our model are consistent with the input parameters. We are not able to correctly reconstruct the input IMF of the youngest MILES SSP. For the intermediate and oldest MILES SSPs there is also an offset between the input IMF and the reconstructed IMF. Nevertheless, for these two SSPs the input IMF is around the one sigma contour of the reconstructed IMF. The offsets that we find for the MILES SSPs may be explained by systematic differences between the MILES models and our models.

The application of our model to a set of mock SSPs shows that if the SNR of a spectrum is high enough to reveal the signal of the low-mass stars, in principle we are able to reconstruct the IMF of these SSPs. However, the application of the model to three MILES SSPs demonstrates that if there are systematic uncertainties this may introduce a bias on the obtained results. At the moment we do not take these systematic uncertainties into account but the results for the MILES SSPs show that it is crucial to model these uncertainties as well.

One of the most important sources of systematic uncertainty between the MILES models and our model is the interpolator that is used to create the stellar templates. To demonstrate this, we have downloaded a set of interpolated spectra from the MILES website using their interpolator. Using these spectra, we are able to reconstruct the input IMF of the MILES SSPs with the exception of the low-mass slope of the youngest SSP. This demonstrates very clearly that our model is only able to reconstruct the IMF if we provide it with a representative set of stellar templates. In that respect, the Bayesian framework of our model is very important as it allows us to objectively compare different model ingredients with each other in light of the evidence.

The application of our model to the MILES models shows us that the reconstructed IMF slopes are very sensitive to the interpolation method used. Therefore, reconstructing the IMF requires a reliable interpolator. In practice, interpolation between stellar spectra can be difficult. The stellar libraries on which interpolators are based, in general, do not provide complete coverage of the parameter space, and there is uncertainty in the parameters of the stars in the library. Moreover, one has to define the parameters that are relevant for interpolating between stellar spectra. For now, we use the effective temperature, surface gravity and $\mathrm{[Fe/H]}$ ratio. However, for some stellar populations it may be necessary to also include, for example, the $\mathrm{[\alpha/Fe]}$ ratio. In addition to that, there are variable stars and stars with peculiarities. These stars are very hard to model and most often found in the low-mass end of the main sequence and the upper giant and asymptotic giant branch.

One of the main questions that we are trying to answer by reconstructing the IMF is the ratio of dwarf to giant stars. This question is difficult to answer in the spectral window of the MILES library because the great spectral similarity between low-mass stars and K and M giants makes these objects difficult to differentiate. In the future we plan to use the X-shooter Spectral Library (XSL) as an input to our model. This stellar library offers much broader wavelength coverage, extending from the UV to the NIR. Compared to MILES, the spectral window of XSL contains many more IMF-sensitive features that will help to better constrain the IMF of distant galaxies.

\section*{Acknowledgements}
This research is partially based on data from the MILES project. We thank Philippe Prugniel and Alexandre Vazdekis for helpful discussions. We thank the anonymous referee for useful comments which improved the quality of this paper.




\bibliographystyle{mnras}
\bibliography{references} 



\appendix

\section{Velocity dispersion mock SSPs}
\label{sec:velocityDispersion}
If we increase the velocity dispersion of the mock SSPs, more and more spectral features will be washed out of the spectrum. For very high velocity dispersion, one can therefore imagine that this becomes problematic for reconstructing the IMF. However, if the velocity dispersion is not too high we do not expect that our results depend on the velocity dispersion of the mock SSP.

To test the robustness of our results as a function of velocity dispersion we reconsider mock SSP \texttt{mock5}. We create five different versions of this mock SSP, each of them smoothed to a different velocity dispersion (but using the same noise spectrum). For each of these mock SSPs we reconstruct the low-mass slope $\alpha_1$ and the high-mass slope $\alpha_2$. The different velocity dispersions and the results for the reconstruction of $\alpha_1$ and $\alpha_2$ are given in Table \ref{tab:velocityDispersion}. Over a range in velocity dispersion of $0$ -- $300\,\mathrm{km\,s^{-1}}$, the inferred IMF parameters show no dependence on velocity dispersion. Hence, we do not expect that our choice to smooth our spectra to a velocity dispersion of 150 km s$^{-1}$ has implications for the reconstructed IMFs that we obtain.

\begin{table}
	\centering
	\caption{Reconstruction of low-mass slope $\alpha_1$ and high-mass slope $\alpha_2$ for \texttt{mock5} as a function of velocity dispersion.}
	\label{tab:velocityDispersion}
	\begin{tabular}{ccc} 
		\hline
		velocity & $\alpha_1$ & $\alpha_2$ \\
		dispersion [km s$^{-1}$]\\
		\hline
		 0 & $1.26 \pm 0.51$ & $2.23 \pm 0.16$ \\
		\rule{0pt}{3ex}
		75 & $1.26 \pm 0.53$ & $2.23 \pm 0.16$\\
		\rule{0pt}{3ex}
		150 & $1.26 \pm 0.51$ & $2.23 \pm 0.17$\\
		\rule{0pt}{3ex}
		225 & $1.27 \pm 0.56$ & $2.23 \pm 0.18$\\
		\rule{0pt}{3ex}
		300 & $1.29 \pm 0.48$ & $2.21 \pm 0.16$\\		
		\hline
	\end{tabular}
\end{table}

\section{Results for mock SSPs}
\label{sec:resultsMockSSPs}
In Section \ref{sec:nonlinearParameters} we showed the results for the reconstruction of the IMF for a mock SSP with an age of 8.5 Gyr and metallicity $\mathrm{[Fe/H] = 0.0}$. The plots in this appendix show the results that we obtain for the remaining mock SSPs discussed in Section \ref{sec:nonlinearParameters}. Table \ref{tab:table_mockSSPs} gives an overview of the twelve mock SSPs that we consider. 

For each of the remaining mock SSPs we show the distribution of the evidence in the age-metallicity grid, the reconstruction of the (non-linear) IMF slopes and the (linear) reconstruction of the piecewise IMF by using the best-fitting non-linear parameters as a prior.

\begin{figure}
	\centering
		{\includegraphics[width=0.99\columnwidth]{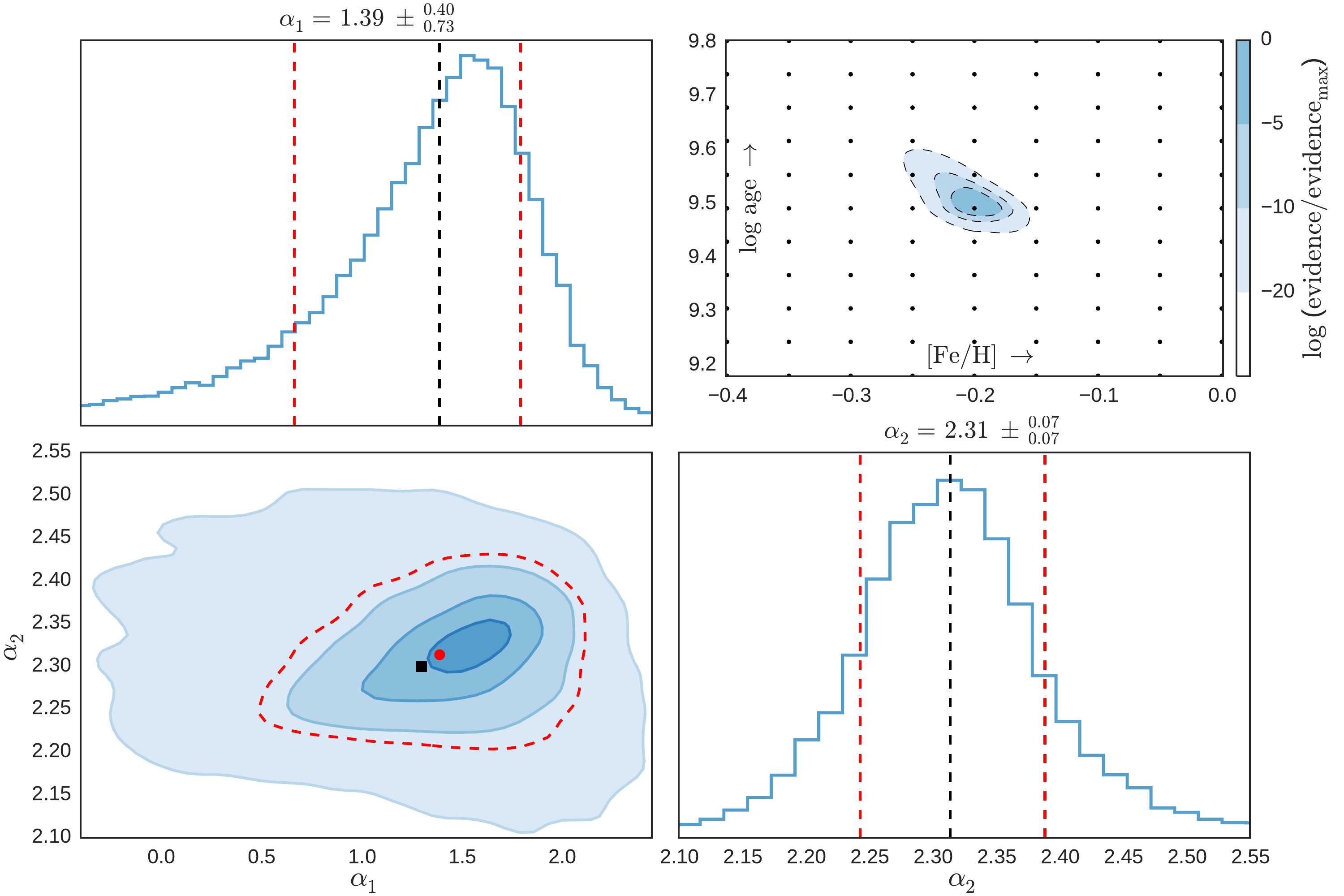}}
		{\includegraphics[width=0.9\columnwidth]{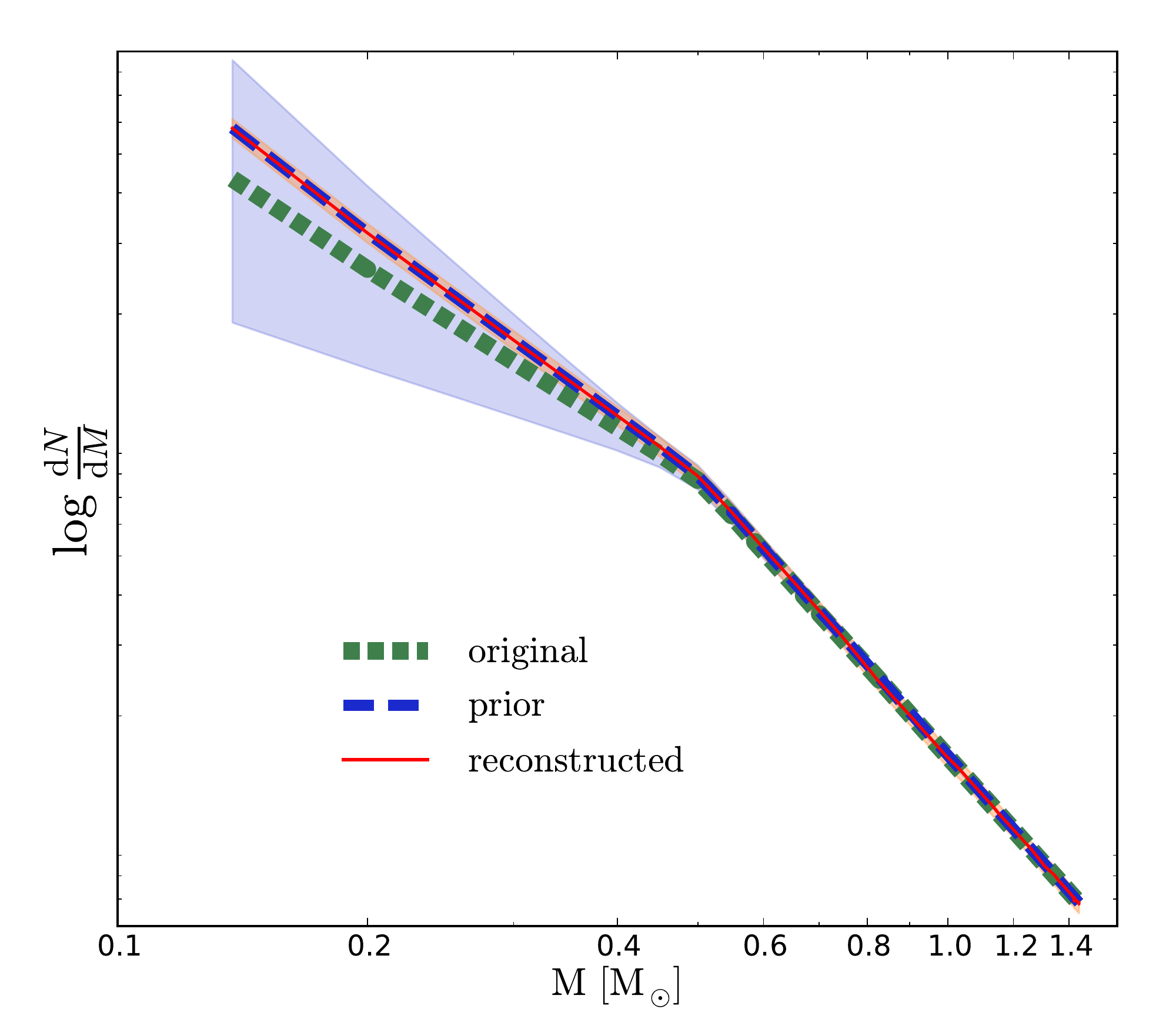}}
			\newpage
	\caption{Results for \texttt{mock1} with $t = 3.1$ Gyr, $\mathrm{[Fe/H]=-0.2}$ and a Kroupa IMF. \textbf{Top panels:} As in Fig. \ref{reconstruction-mock5} only now for \texttt{mock1}. \textbf{Bottom panel:} As in the top panel of Fig. \ref{reconstructionIMF4} only now for \texttt{mock1}}
	\label{summary1}
\end{figure}

\begin{figure}
	\centering
		{\includegraphics[width=0.99\columnwidth]{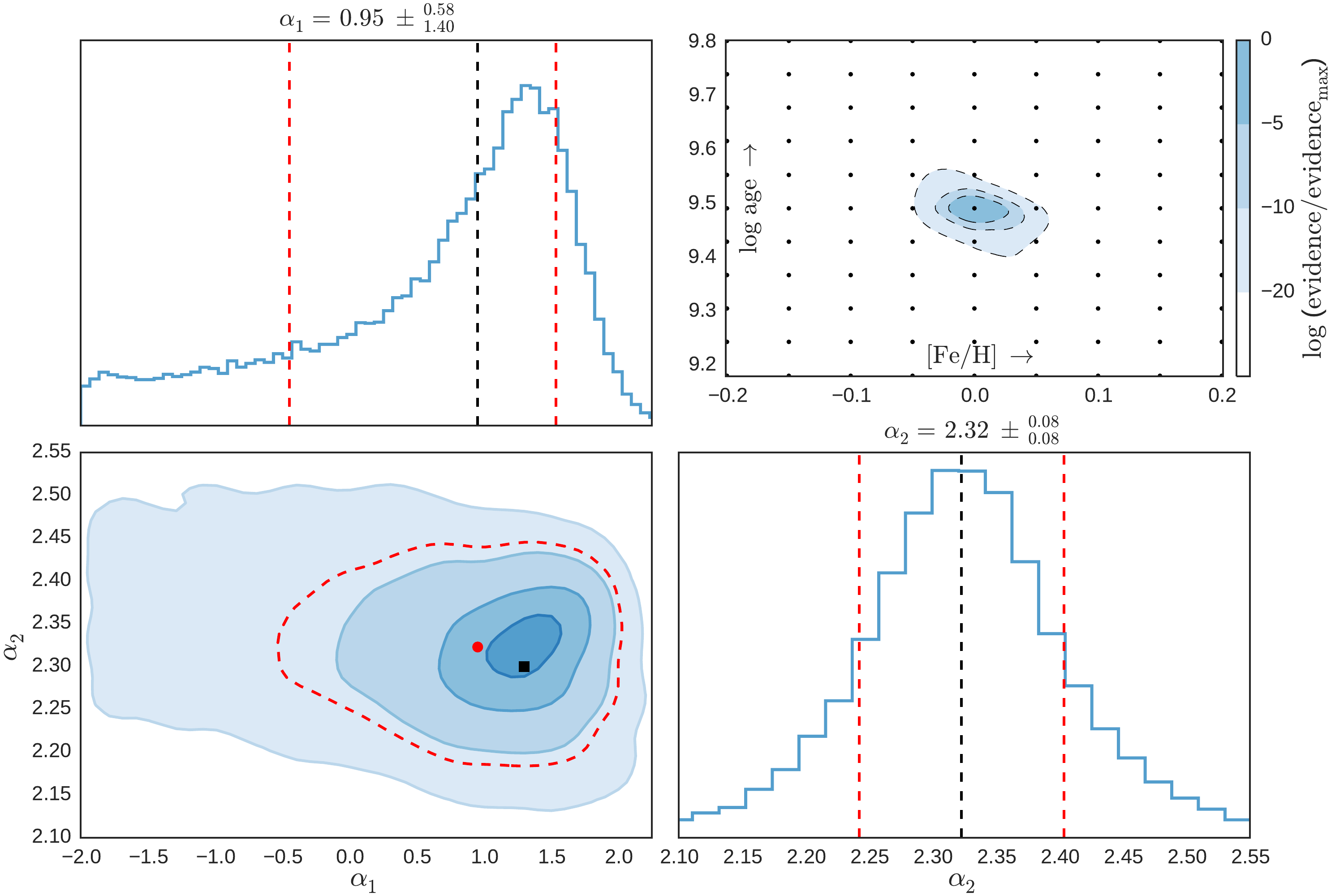}}
		{\includegraphics[width=0.9\columnwidth]{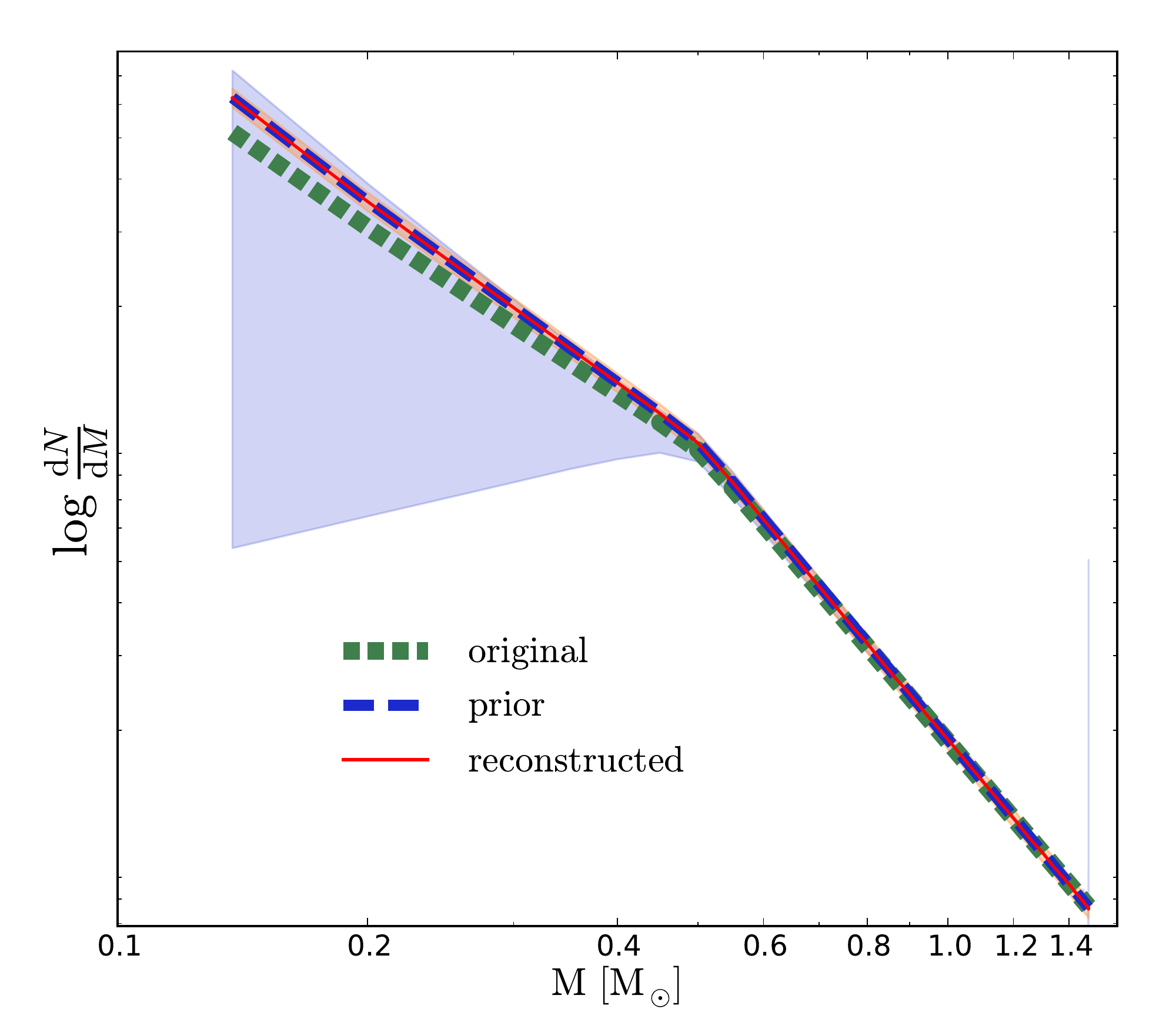}}
	\caption{As in Fig. \ref{summary1} for \texttt{mock2} with $t = 3.1$ Gyr, $\mathrm{[Fe/H]=0.0}$ and a Kroupa IMF.}
	\label{summary2}
\end{figure}

\begin{figure}
	\centering
		{\includegraphics[width=0.99\columnwidth]{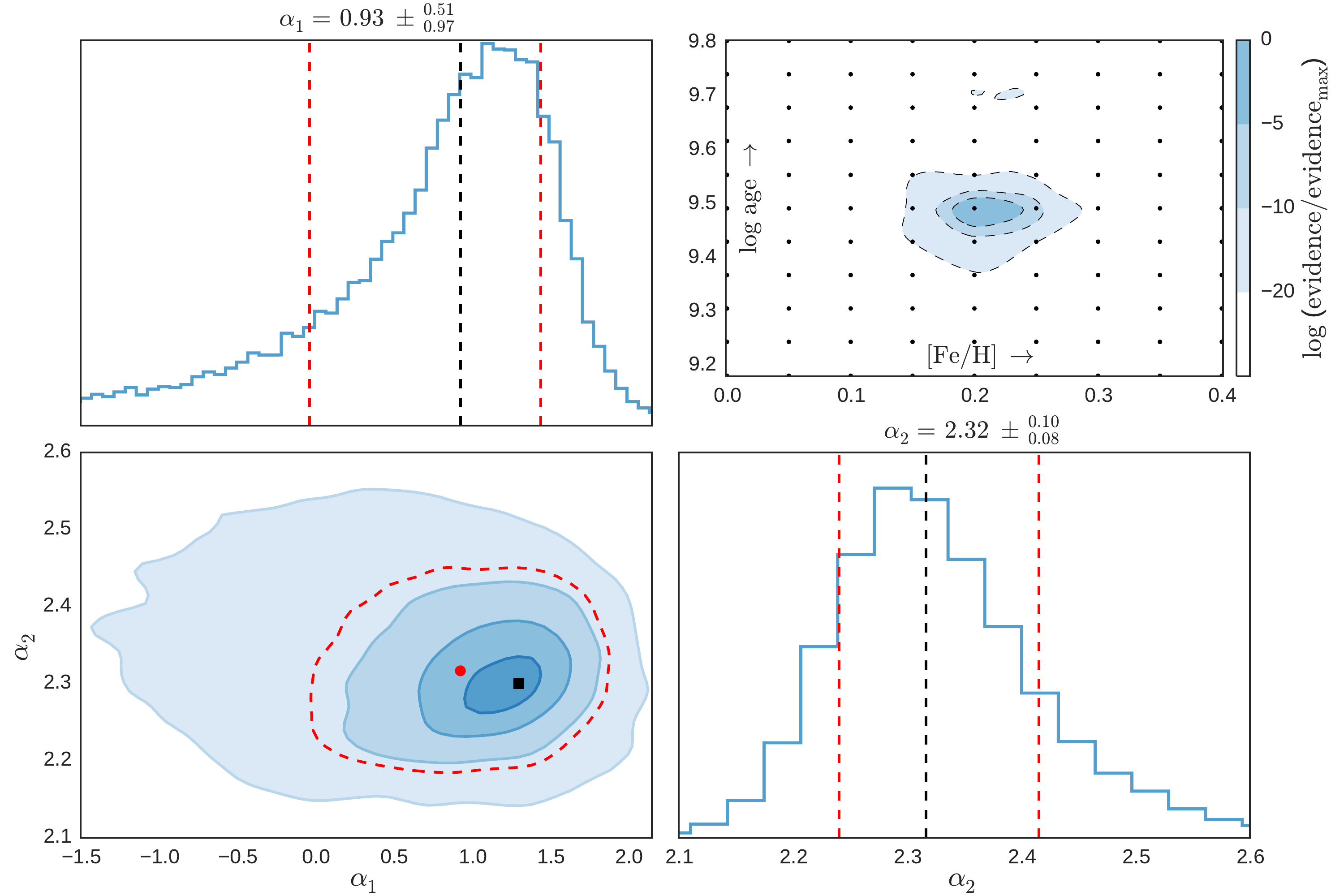}}
		{\includegraphics[width=0.9\columnwidth]{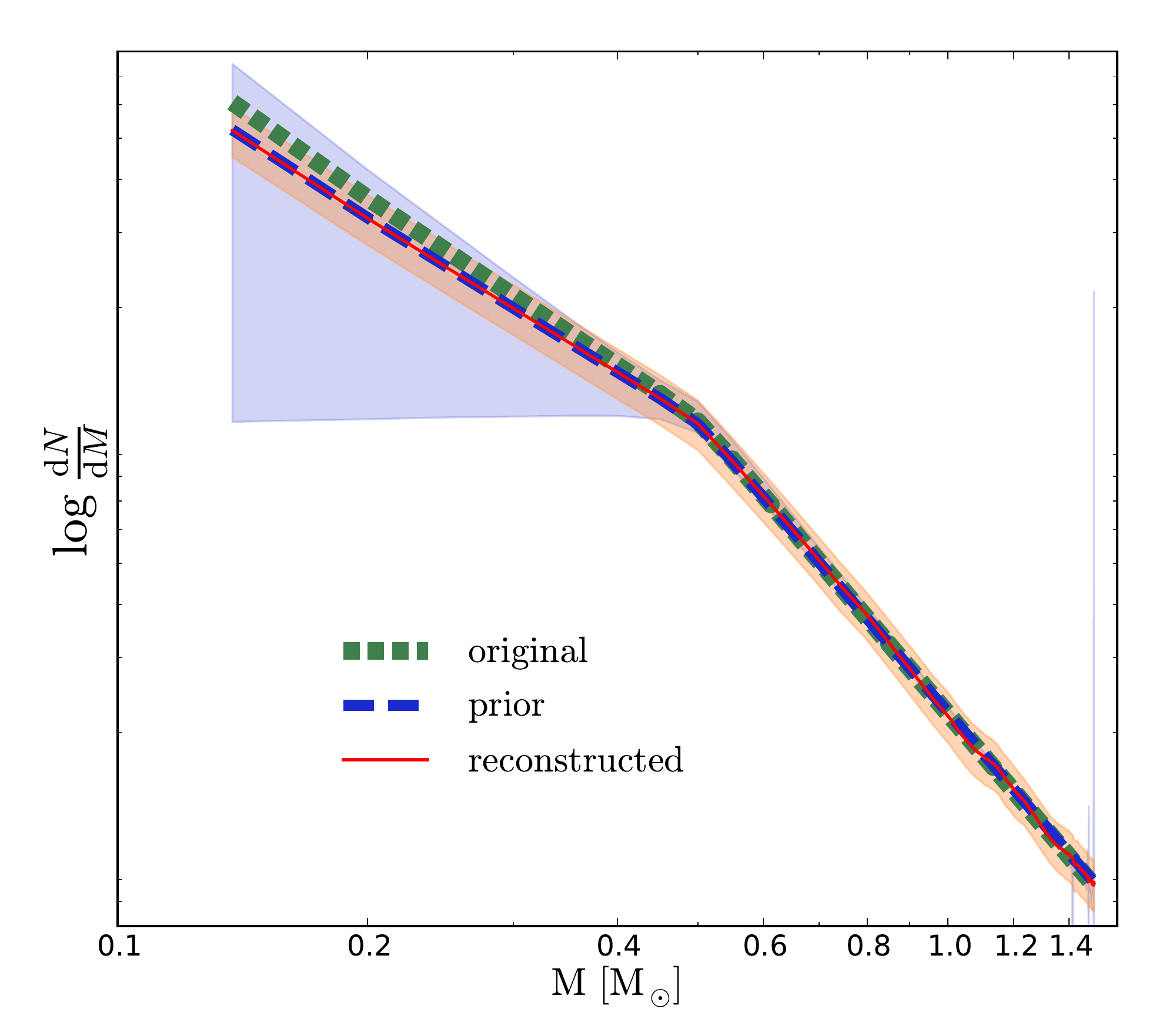}}
	\caption{As in Fig. \ref{summary1} for \texttt{mock3} with $t = 3.1$ Gyr, $\mathrm{[Fe/H]=0.2}$ and a Kroupa IMF.}
	\label{summary3}
\end{figure}

\begin{figure}
	\centering
		{\includegraphics[width=0.99\columnwidth]{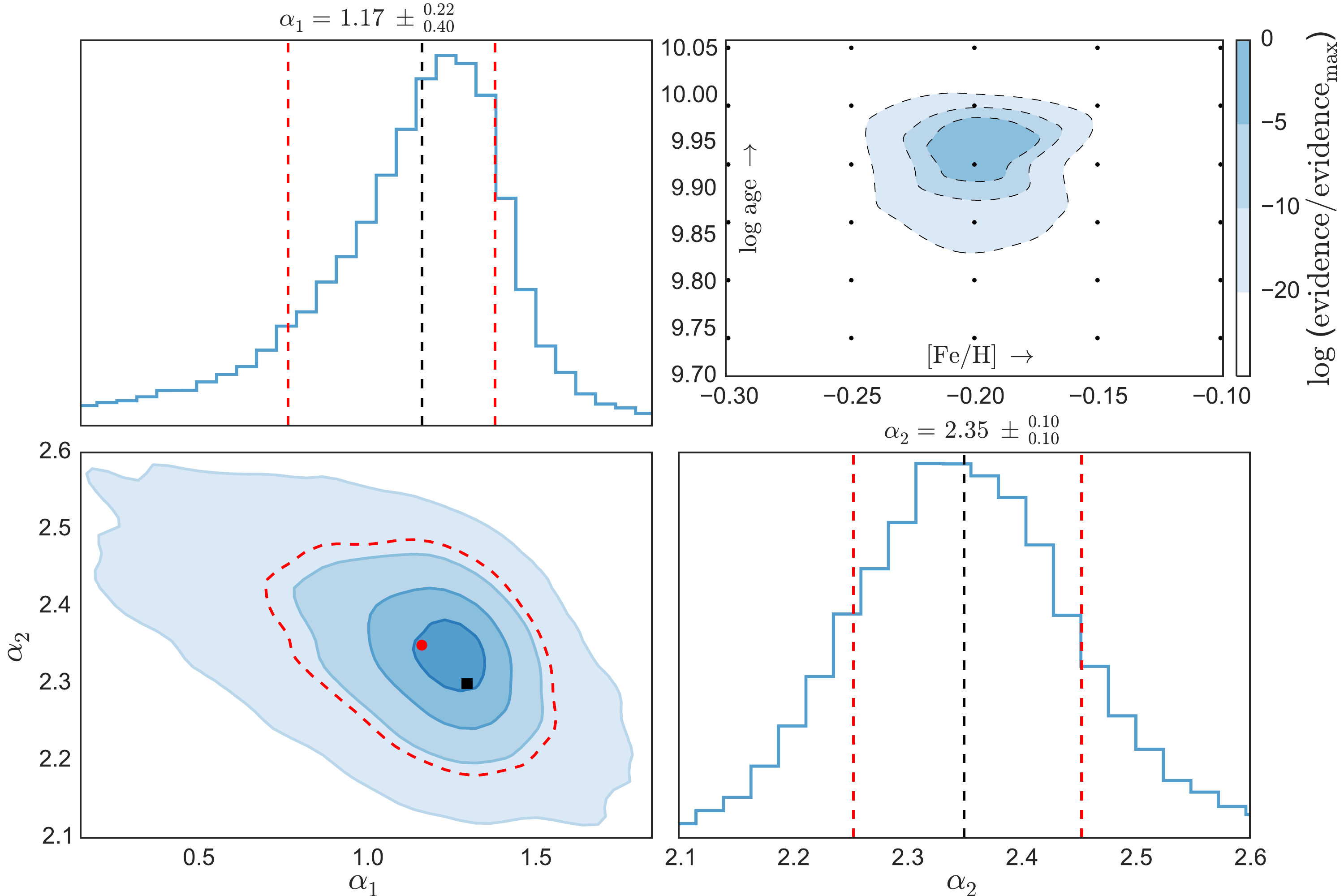}}
		{\includegraphics[width=0.9\columnwidth]{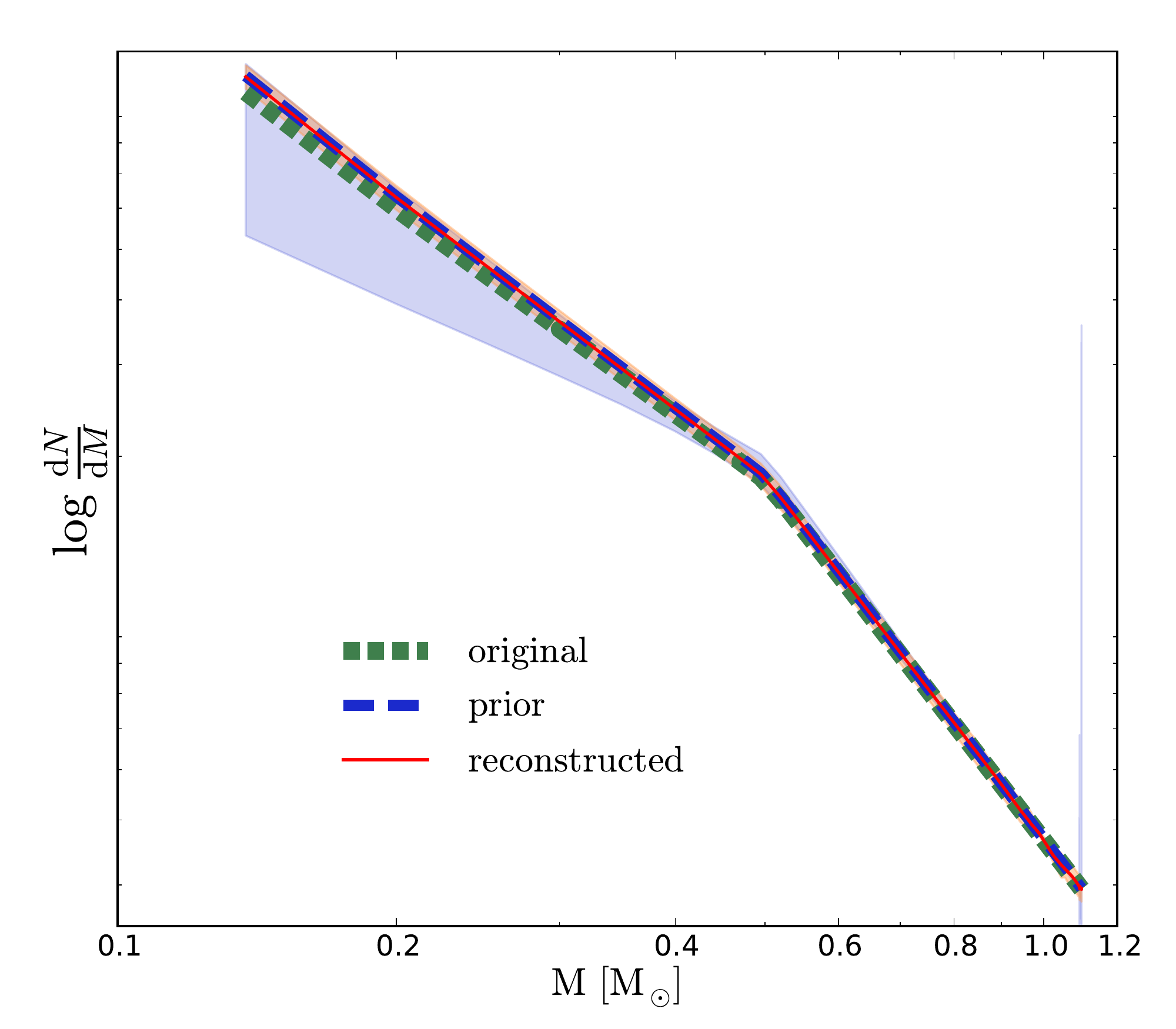}}
	\caption{As in Fig. \ref{summary1} for \texttt{mock4} with $t = 8.5$ Gyr, $\mathrm{[Fe/H]=-0.2}$ and a Kroupa IMF.}
	\label{summary4}
\end{figure}

\begin{figure}
	\centering
		{\includegraphics[width=0.99\columnwidth]{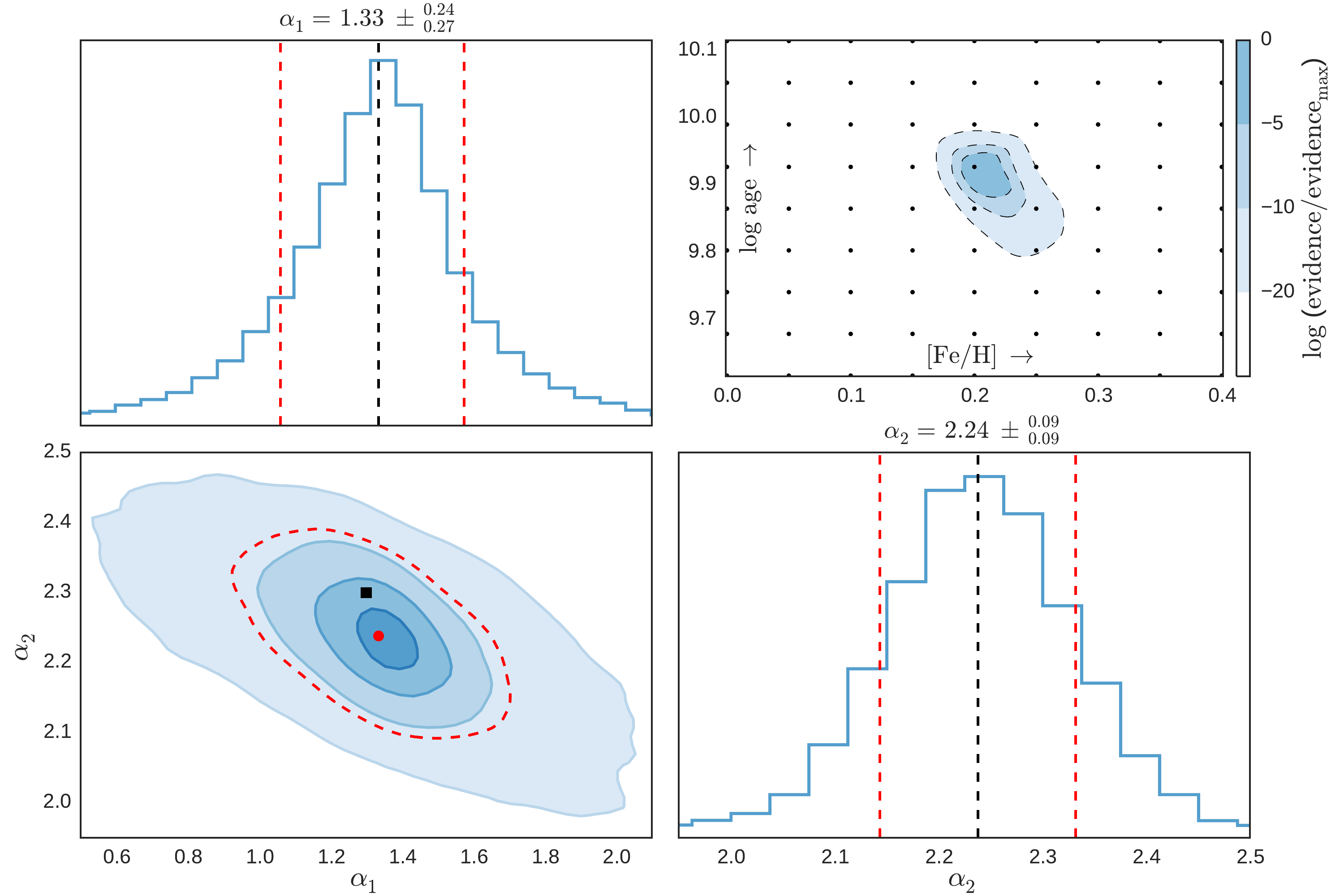}}
		{\includegraphics[width=0.9\columnwidth]{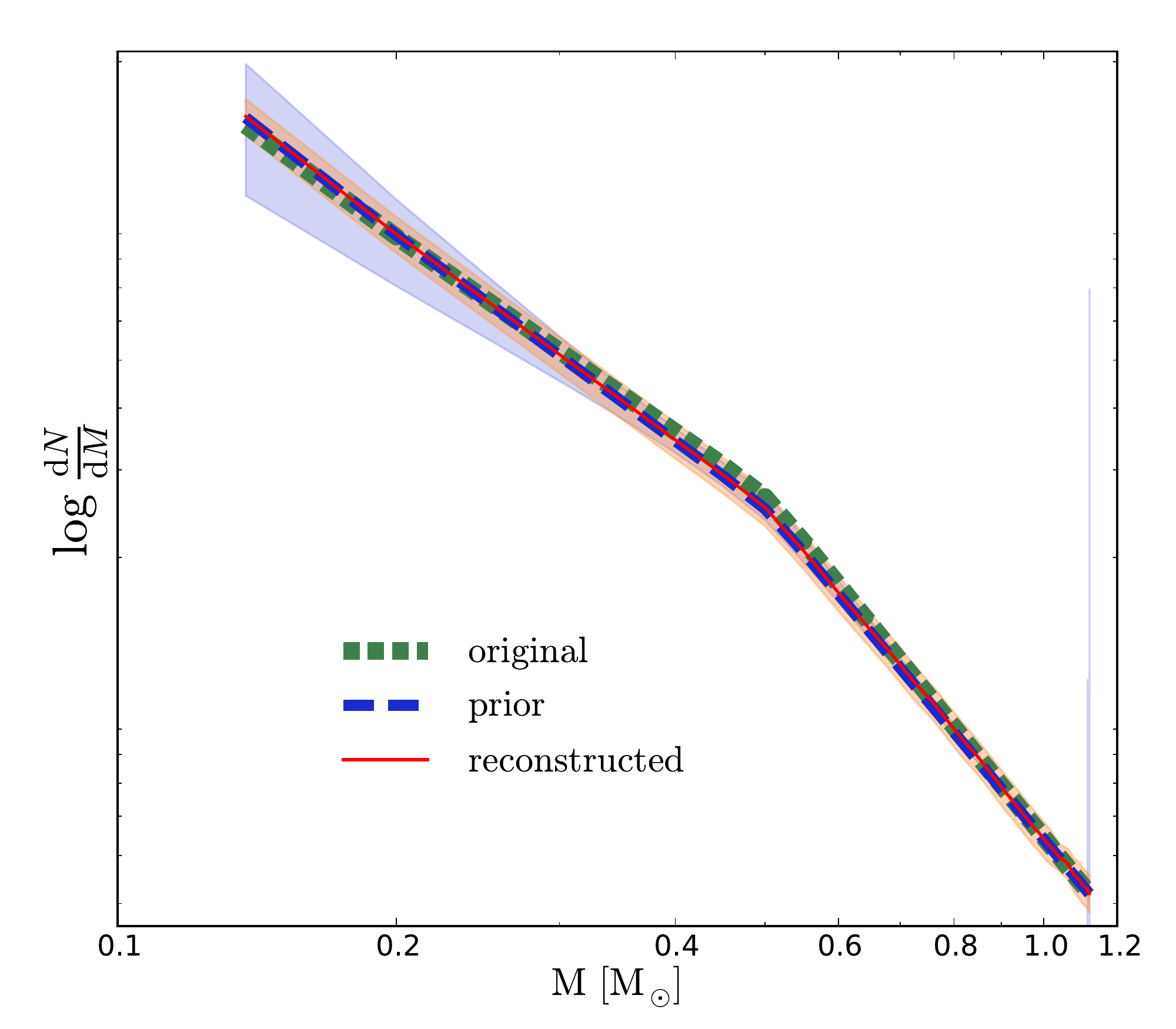}}
	\caption{As in Fig. \ref{summary1} for \texttt{mock6} with $t = 8.5$ Gyr, $\mathrm{[Fe/H]=0.2}$ and a Kroupa IMF.}
	\label{summary5}
\end{figure}

\begin{figure}
	\centering
		{\includegraphics[width=0.99\columnwidth]{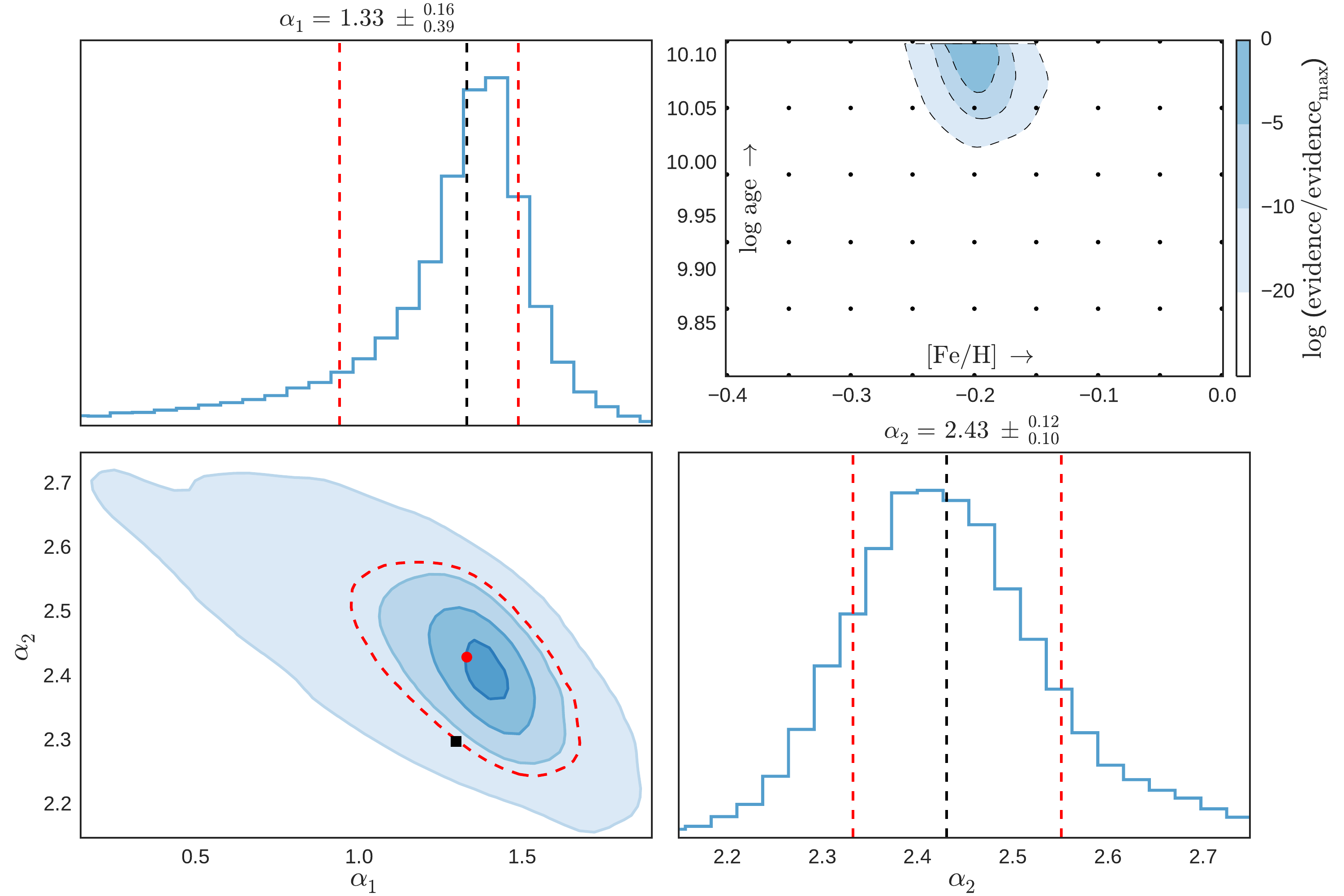}}
		{\includegraphics[width=0.9\columnwidth]{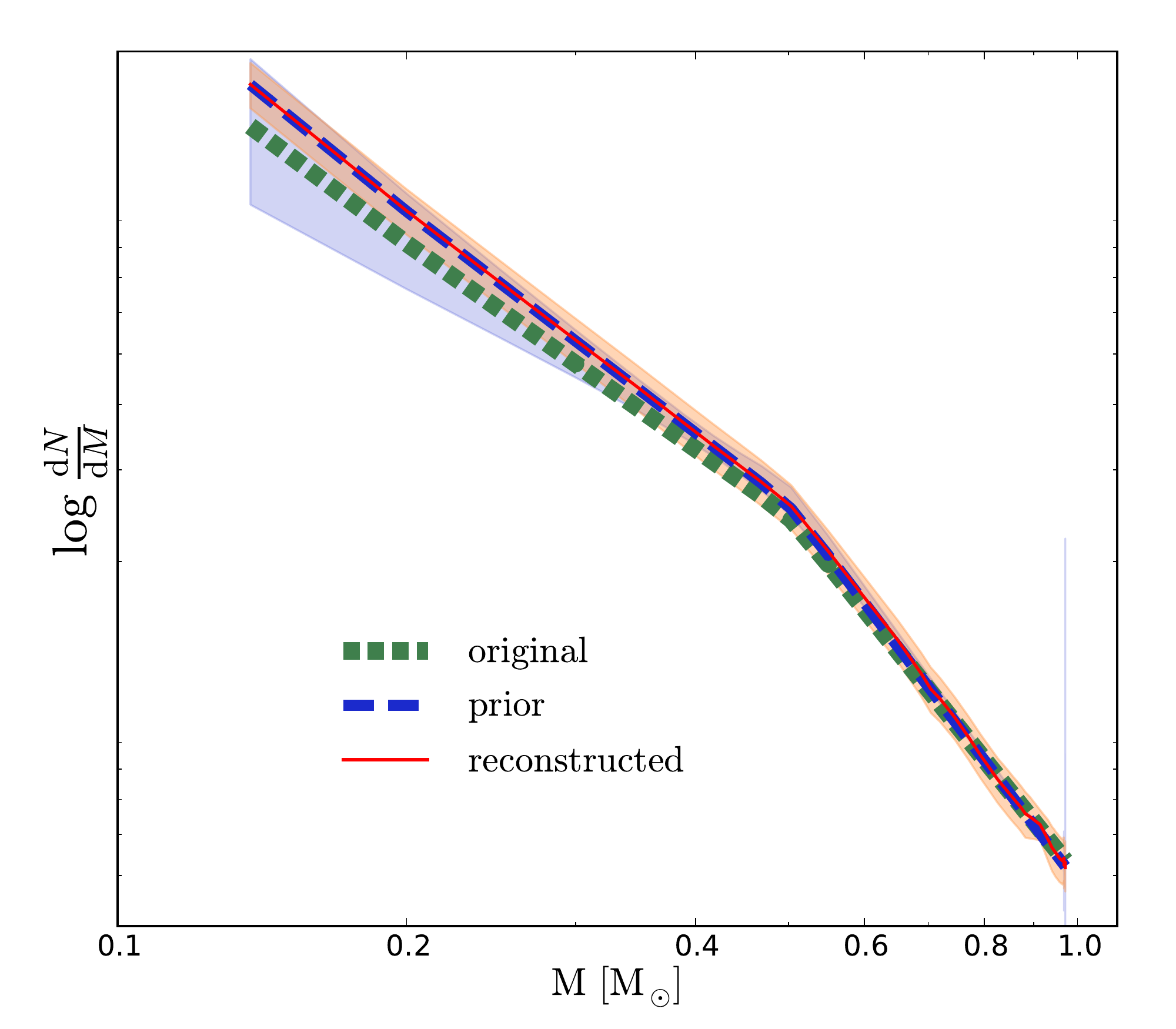}}
	\caption{As in Fig. \ref{summary1} for \texttt{mock7} with $t = 13.0$ Gyr, $\mathrm{[Fe/H]=-0.2}$ and a Kroupa IMF.}
	\label{summary6}
\end{figure}

\begin{figure}
	\centering
		{\includegraphics[width=0.99\columnwidth]{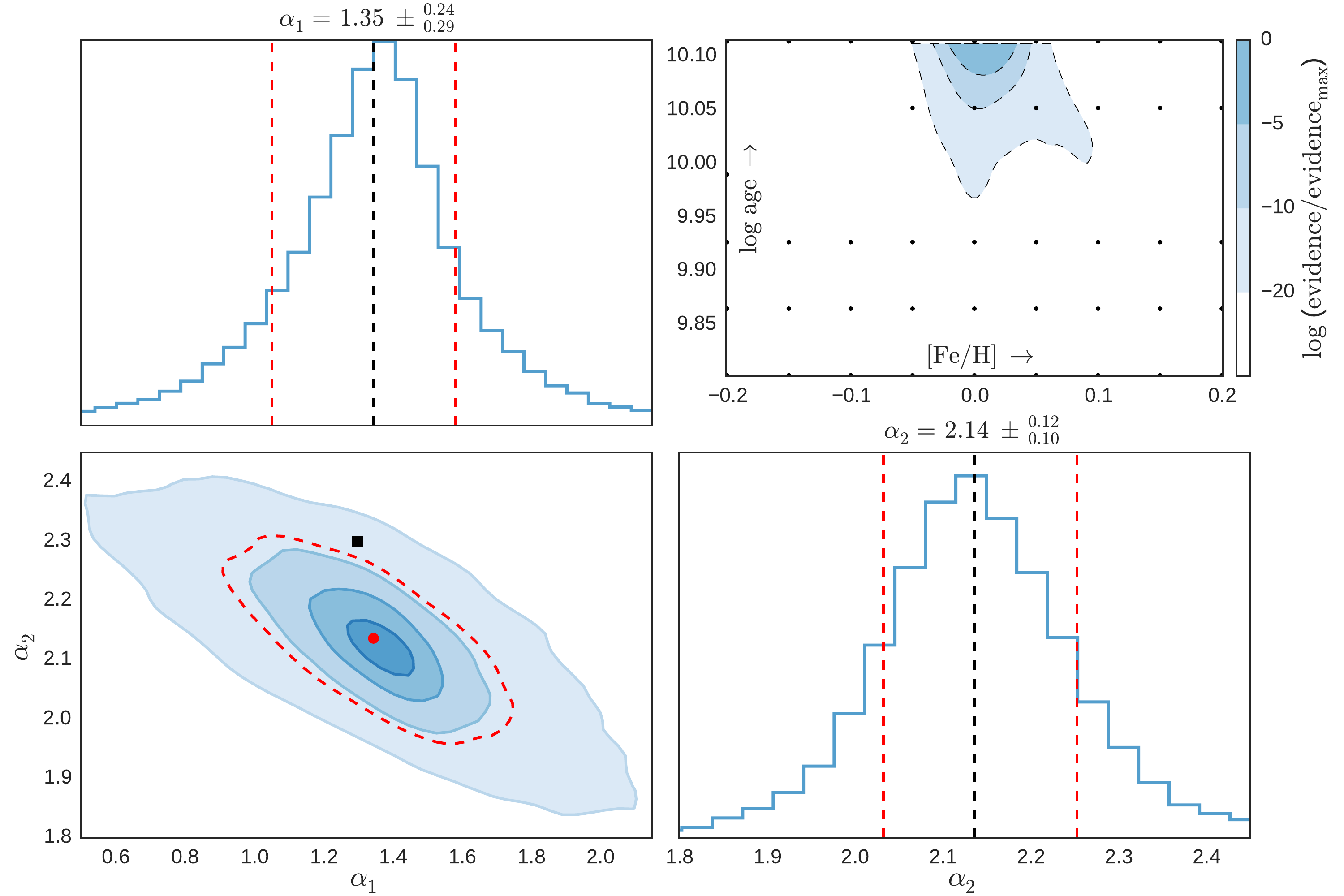}}
		{\includegraphics[width=0.9\columnwidth]{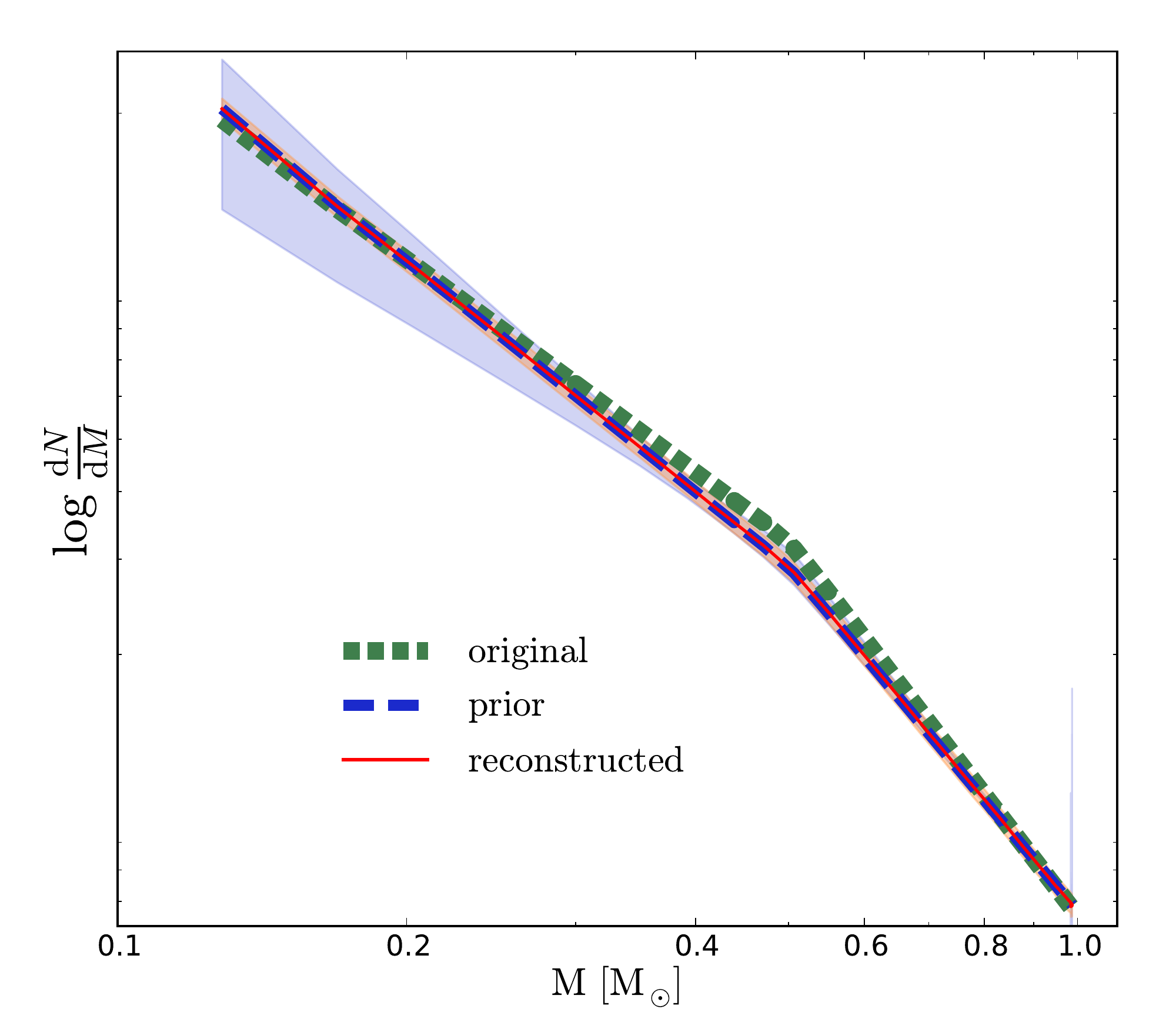}}
	\caption{As in Fig. \ref{summary1} for \texttt{mock8} with $t = 13.0$ Gyr, $\mathrm{[Fe/H]=0.0}$ and a Kroupa IMF.}
	\label{summary7}
\end{figure}

\begin{figure}
	\centering
		{\includegraphics[width=0.99\columnwidth]{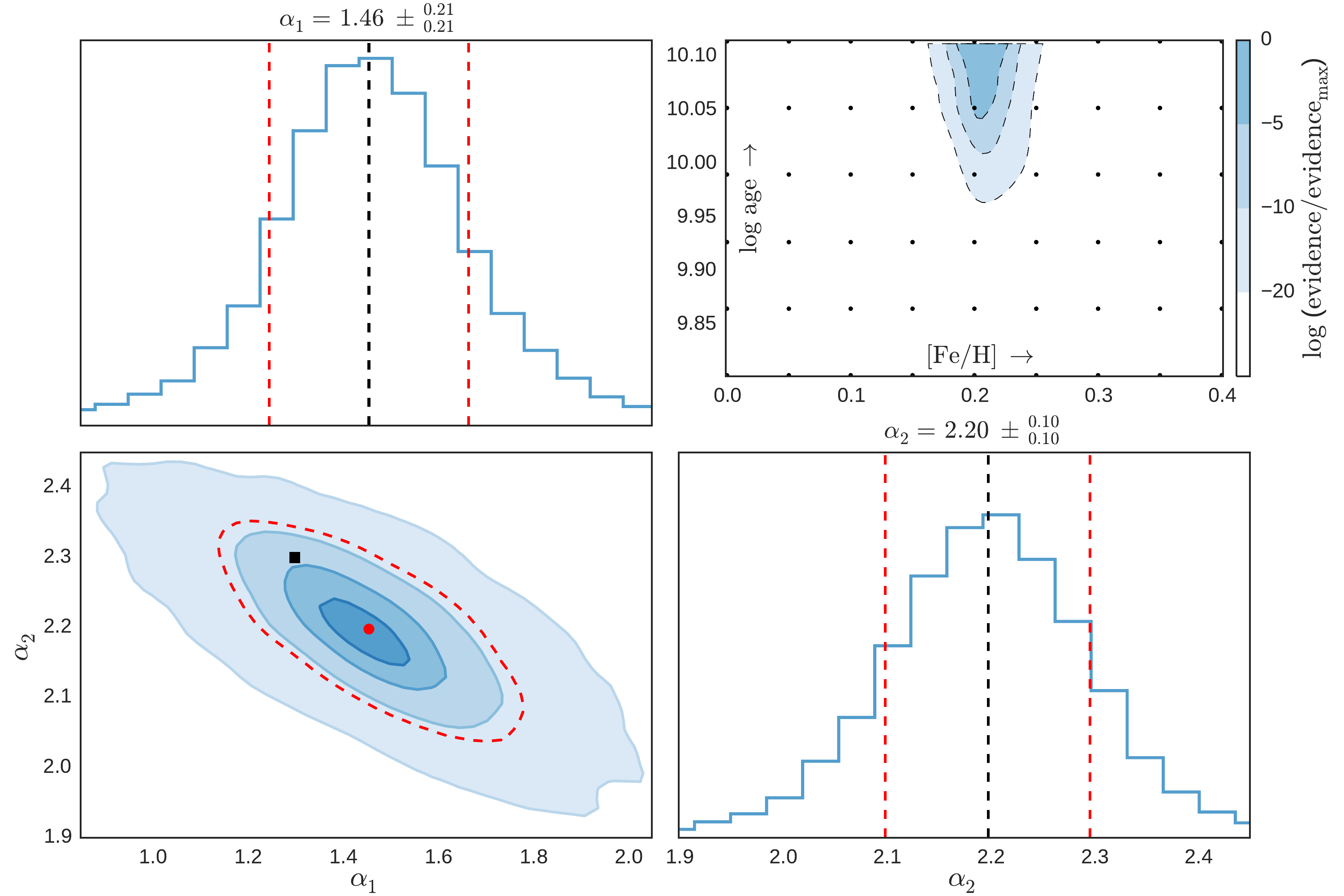}}
		{\includegraphics[width=0.9\columnwidth]{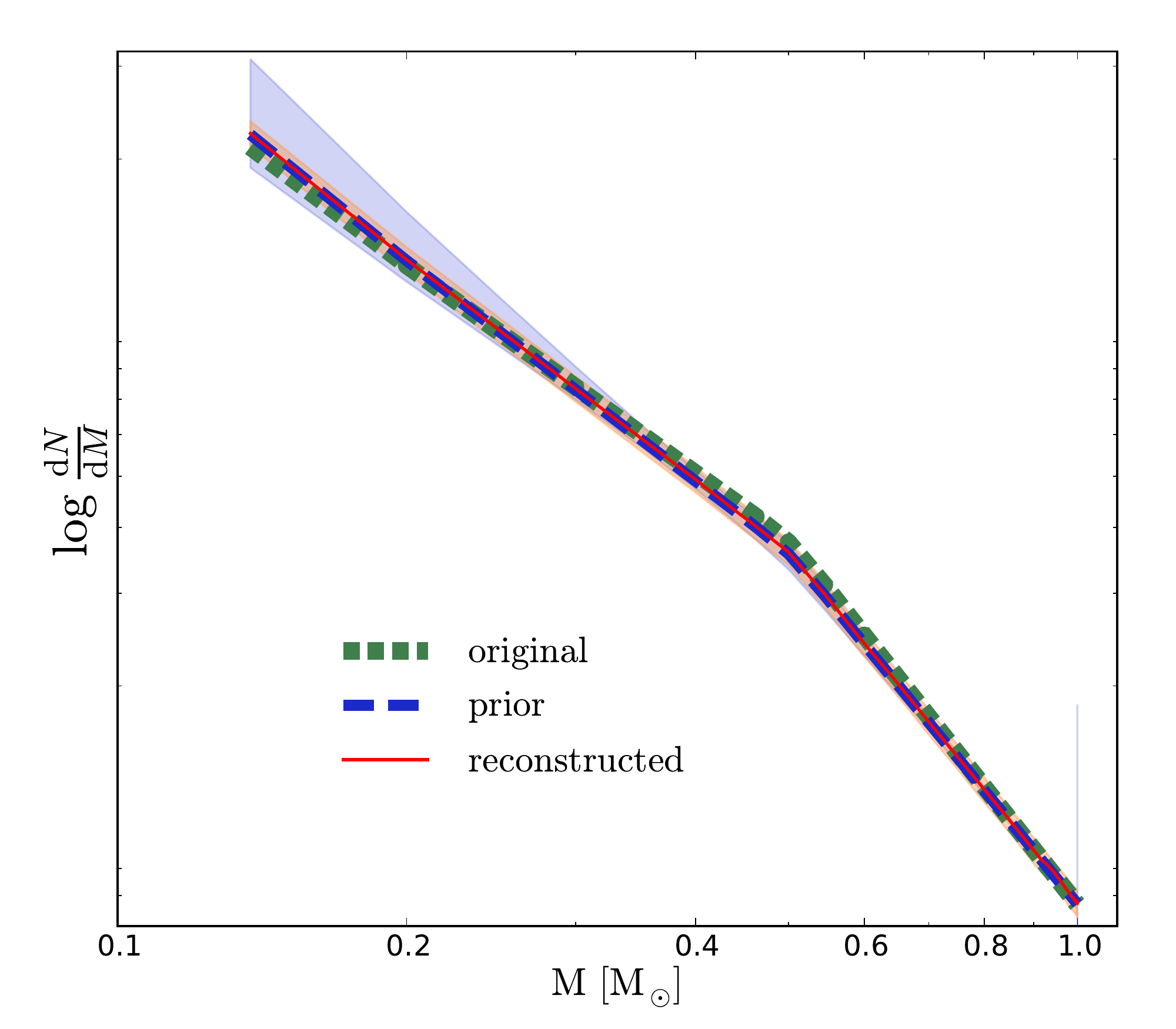}}
	\caption{As in Fig. \ref{summary1} for \texttt{mock9} with $t = 13.0$ Gyr, $\mathrm{[Fe/H]=0.2}$ and a Kroupa IMF.}
	\label{summary8}
\end{figure}

\begin{figure}
	\centering
		{\includegraphics[width=0.99\columnwidth]{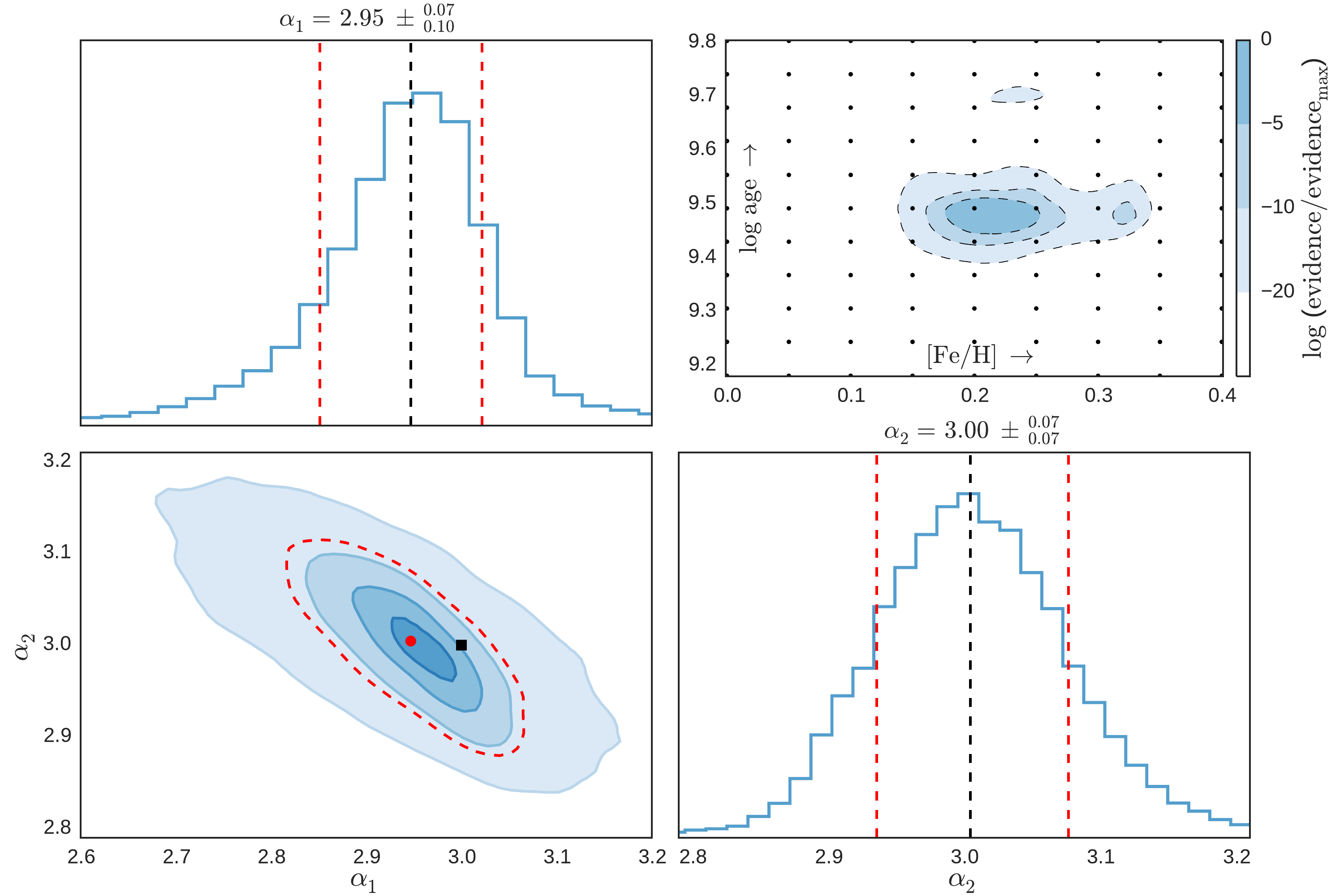}}
		{\includegraphics[width=0.9\columnwidth]{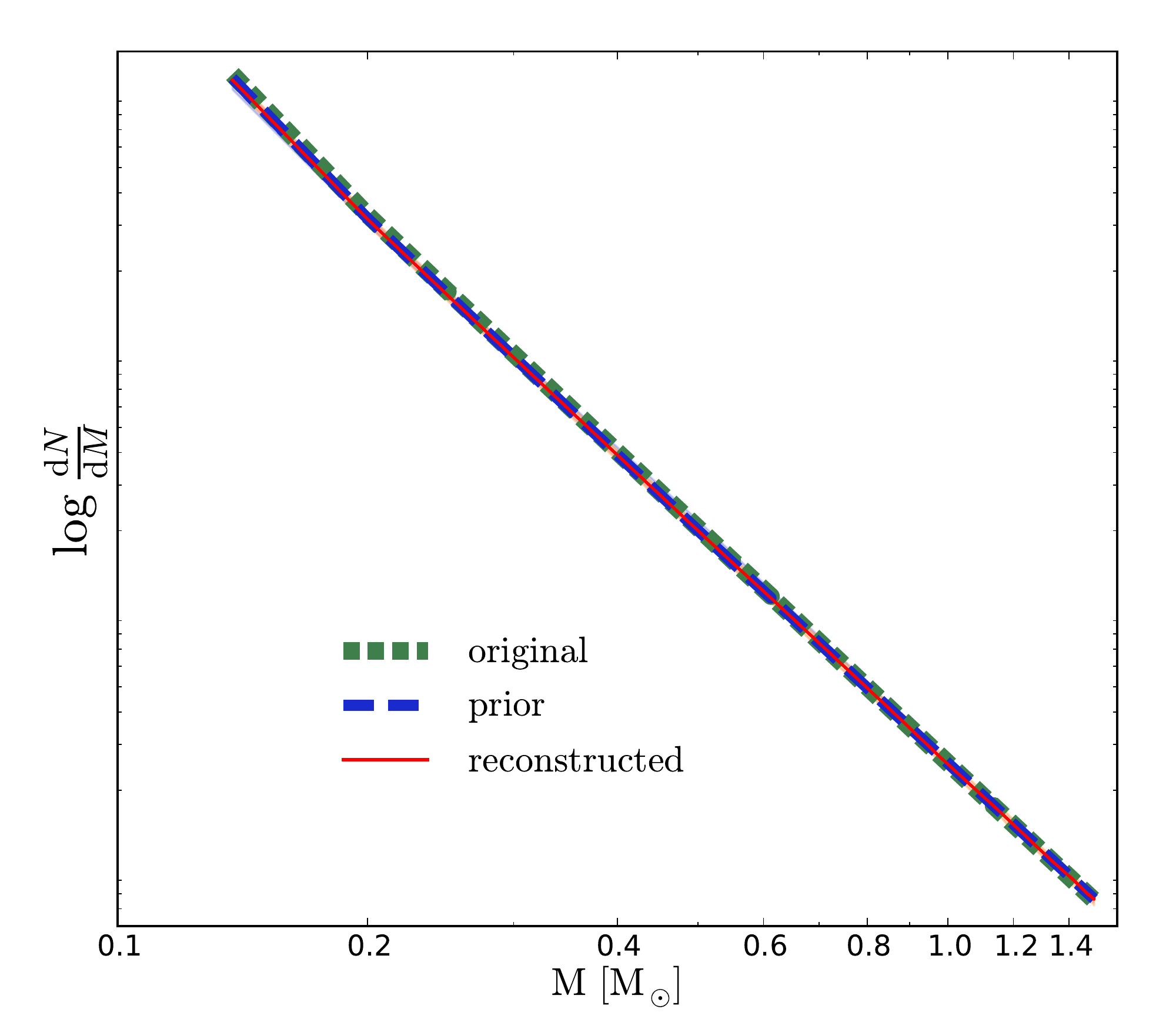}}
	\caption{As in Fig. \ref{summary1} for \texttt{mock10} with $t = 3.1$ Gyr, $\mathrm{[Fe/H]=0.2}$ and a bottom-heavy IMF.}
	\label{summary9}
\end{figure}

\begin{figure}
	\centering
		{\includegraphics[width=0.99\columnwidth]{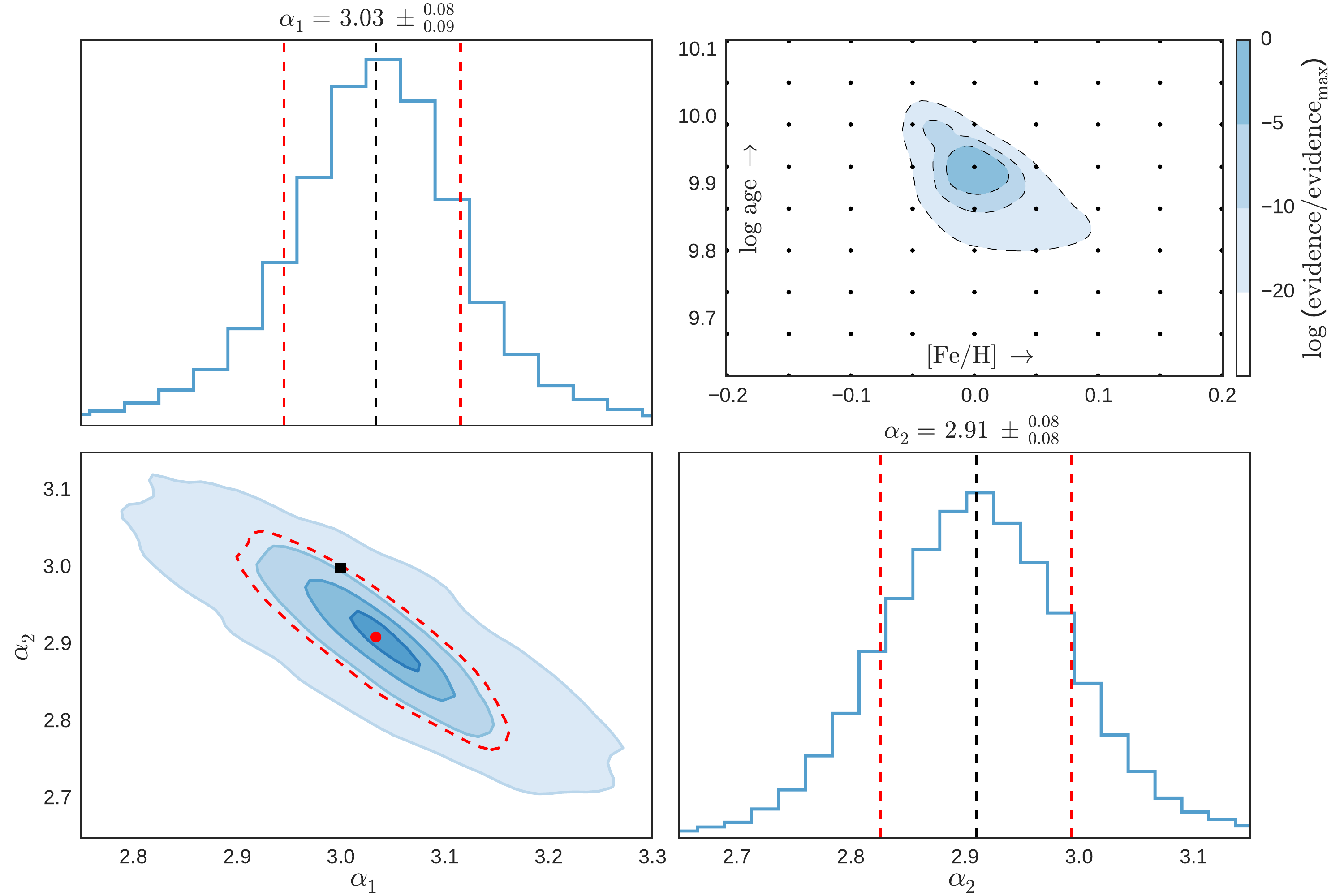}}
		{\includegraphics[width=0.9\columnwidth]{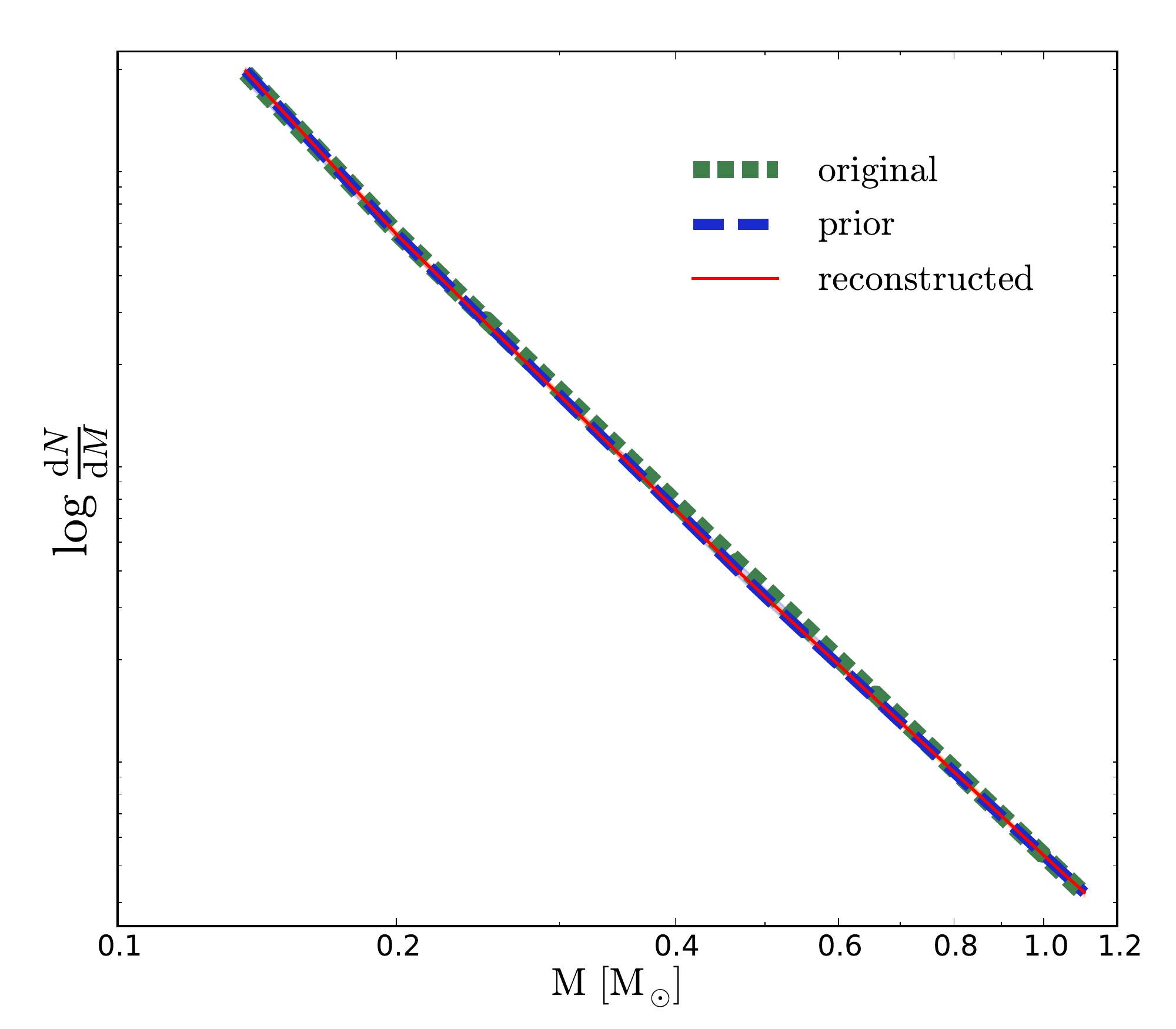}}
	\caption{As in Fig. \ref{summary1} for \texttt{mock11} with $t = 8.5$ Gyr, $\mathrm{[Fe/H]=0.0}$ and a bottom-heavy IMF.}
	\label{summary10}
\end{figure}

\begin{figure}
	\centering
		{\includegraphics[width=0.99\columnwidth]{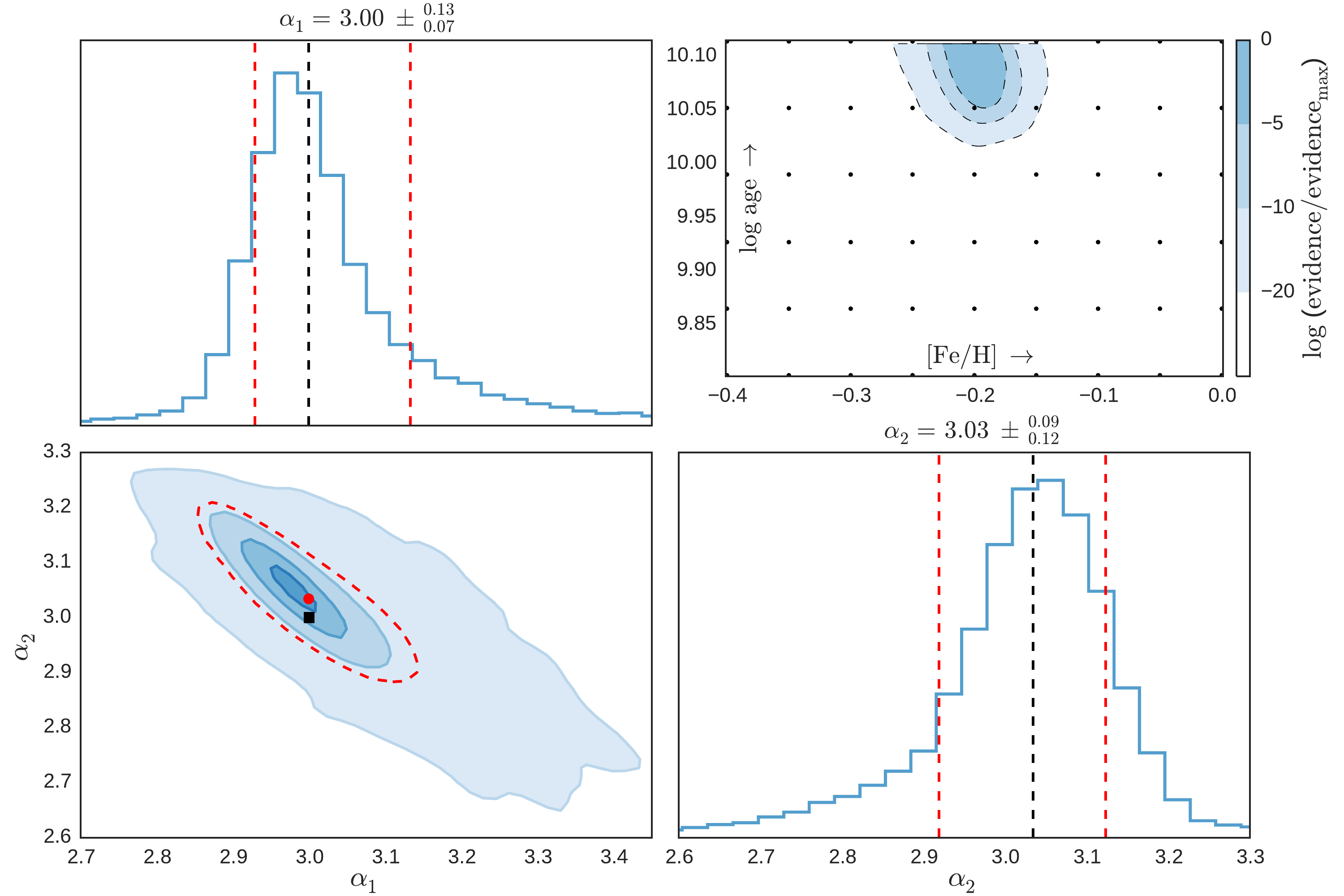}}
		{\includegraphics[width=0.9\columnwidth]{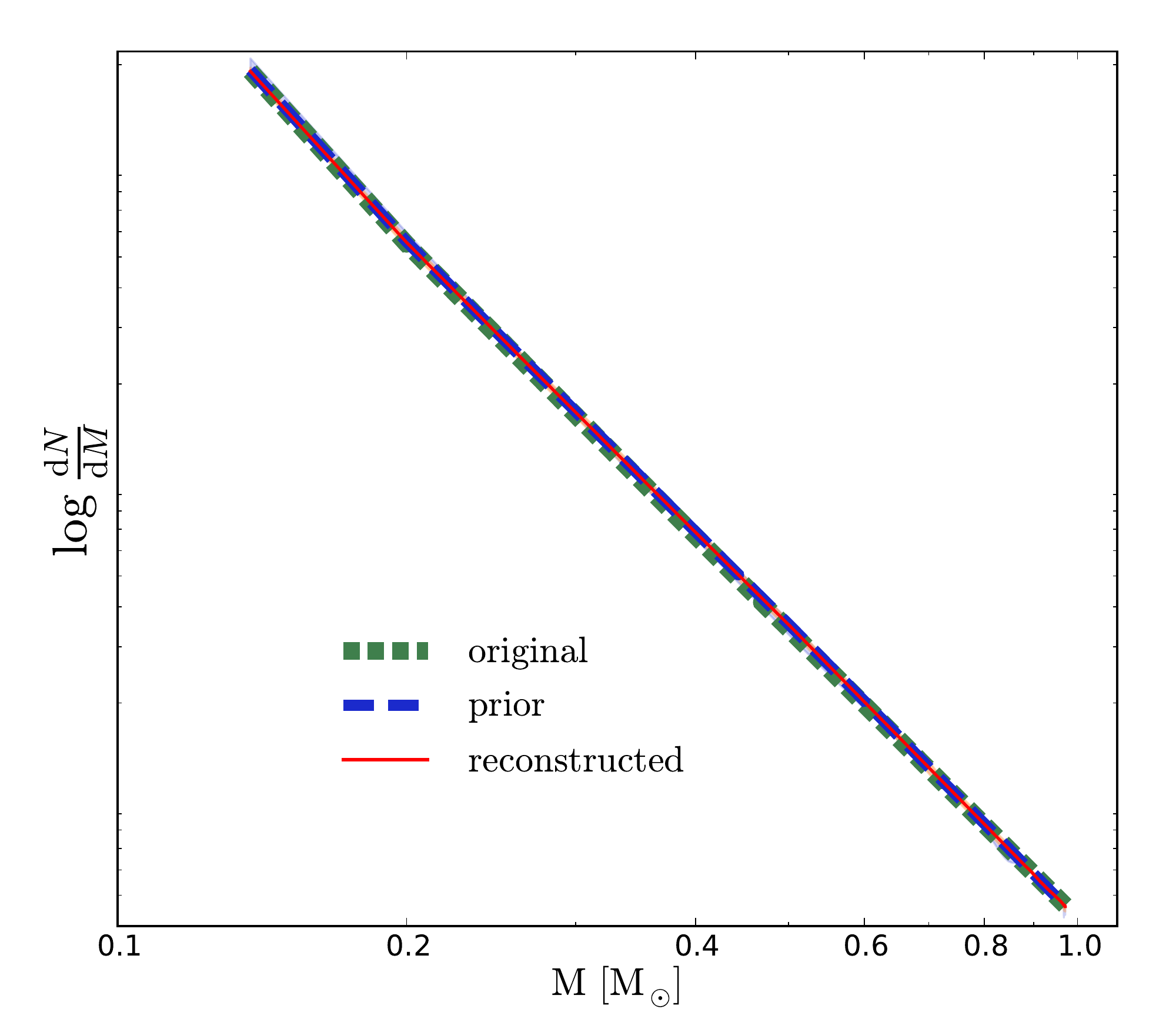}}
	\caption{As in Fig. \ref{summary1} for \texttt{mock12} with $t = 13.0$ Gyr, $\mathrm{[Fe/H]=-0.2}$ and a bottom-heavy IMF.}
	\label{summary11}
\end{figure}

\clearpage

\section{Results for MILES SSPs}
\label{sec:MILES_results}
In Section \ref{sec:modelvsmodel} we have discussed the application of our model to three MILES SSPs. The results for \texttt{MILES2} have been shown in Section \ref{sec:modelvsmodel}. Here we show the results for \texttt{MILES1} and \texttt{MILES3}

\begin{figure}
	\centering
		{\includegraphics[width=0.99\columnwidth]{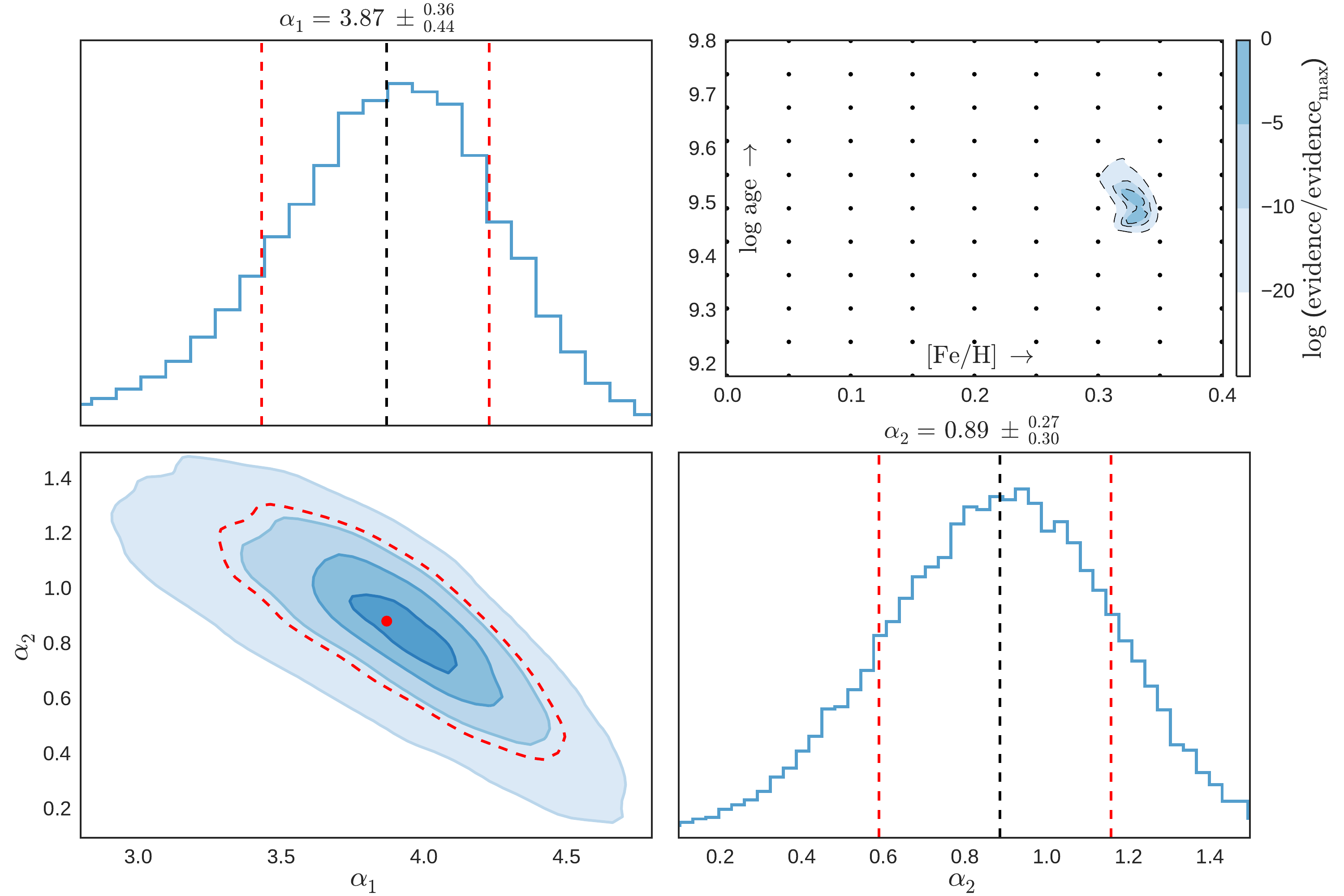}}
		{\includegraphics[width=0.9\columnwidth]{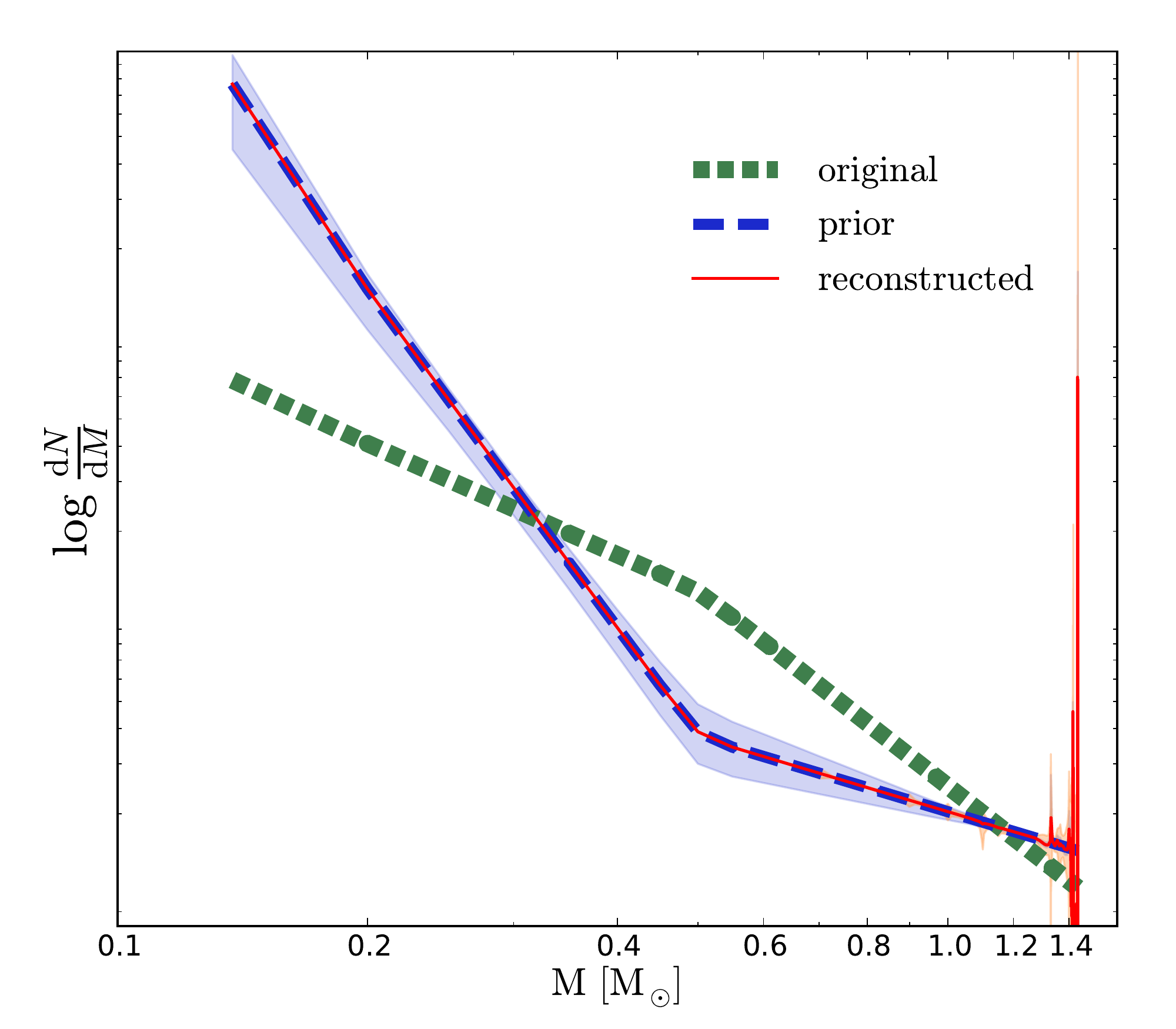}}
	\caption{As in Fig. \ref{summary1} for \texttt{MILES1} with $t = 3.2$ Gyr and $\mathrm{[Fe/H]=0.22}$.}
	\label{summary_MILES1}
\end{figure}

The reconstructed IMF for \texttt{MILES1} (Fig. \ref{summary_MILES1}) suggests a very steep low-mass IMF combined with a high-mass slope that is much flatter than it is for a Kroupa IMF. This result is inconsistent with the input IMF. The over-abundance of the lowest-mass templates ($M \lesssim 0.3$ $\mathrm{M}_{\odot}$) with respect to the input Kroupa IMF produces a relatively small amount of light. Therefore, the under-abundance of the intermediate-mass templates ($0.3$ $\mathrm{M}_{\odot} \lesssim M \lesssim 1.2$ $\mathrm{M}_{\odot}$) is almost completely compensated by a slight over-abundance of the highest mass templates ($M \gtrsim 1.2$ $\mathrm{M}_{\odot}$). 

The reconstructed IMFs for the MILES SSPs show a spiky distribution for the high-mass end. There are three reasons that might explain this behaviour. First, at the high-mass end the difference between two subsequent mass bins may be very small. To convert the reconstructed weights into an IMF, we divide the weights by the width of the mass bin. As a consequence, a relatively small deviation of the weights may result in a large deviation of the IMF if the mass bin is small. Second, the regularization scheme that we use for the MILES SSPs penalizes the absolute deviation of the weights. This implies that a deviation of one star from the prior for a low-mass template is penalized as much as a deviation of one star for a high-mass template. However, since the number of stars for higher-mass templates is in general much lower (because of the shape of the IMF and the width of the mass bins) the effect on the IMF will be much larger. Third, the use of NNLS may result in some of the templates being equal to zero. If this is the case, the model has to compensate for this by increasing the weights of the templates that surround the one that is set to zero in the fit.

\begin{figure}
	\centering
		{\includegraphics[width=0.99\columnwidth]{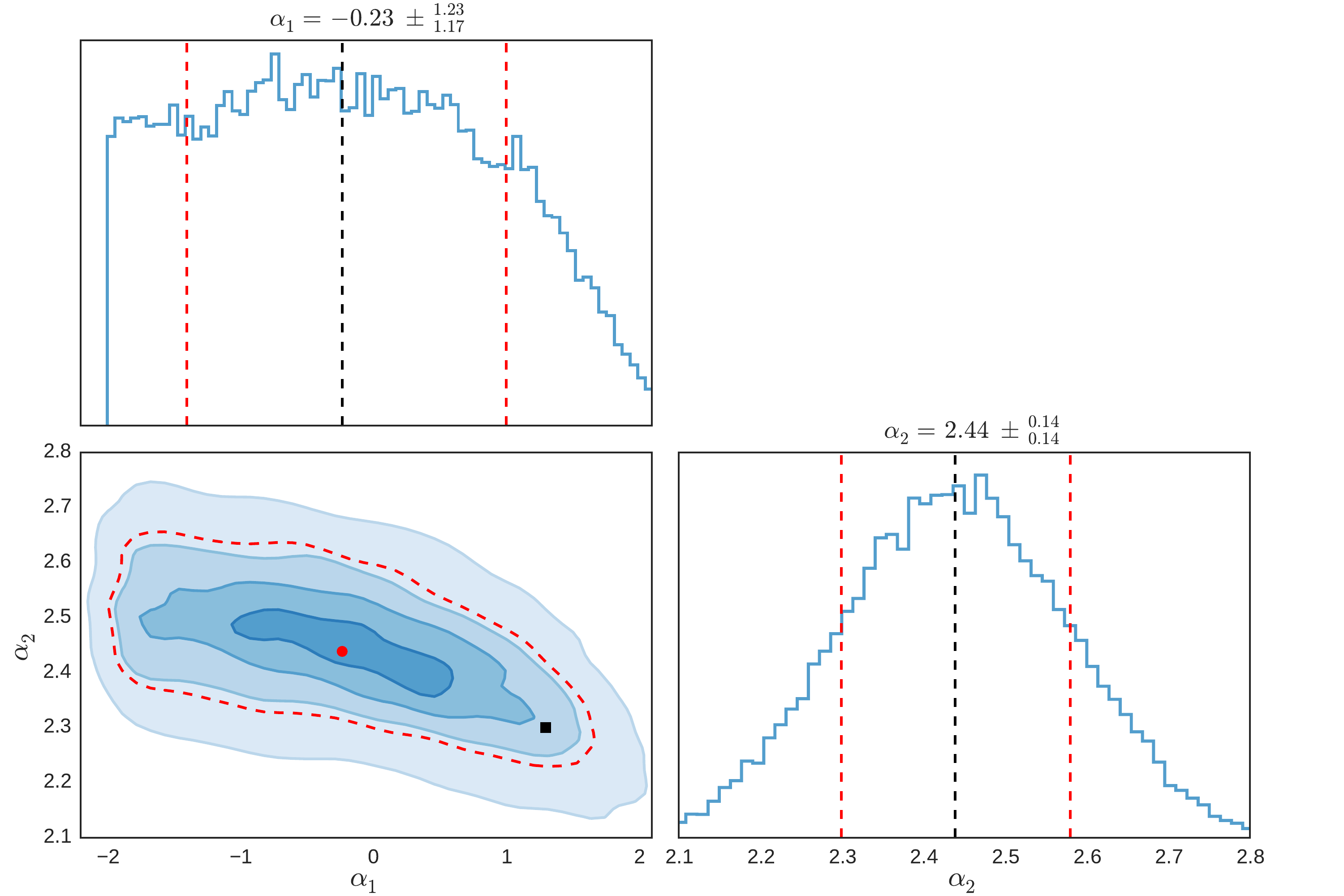}}
		{\includegraphics[width=0.9\columnwidth]{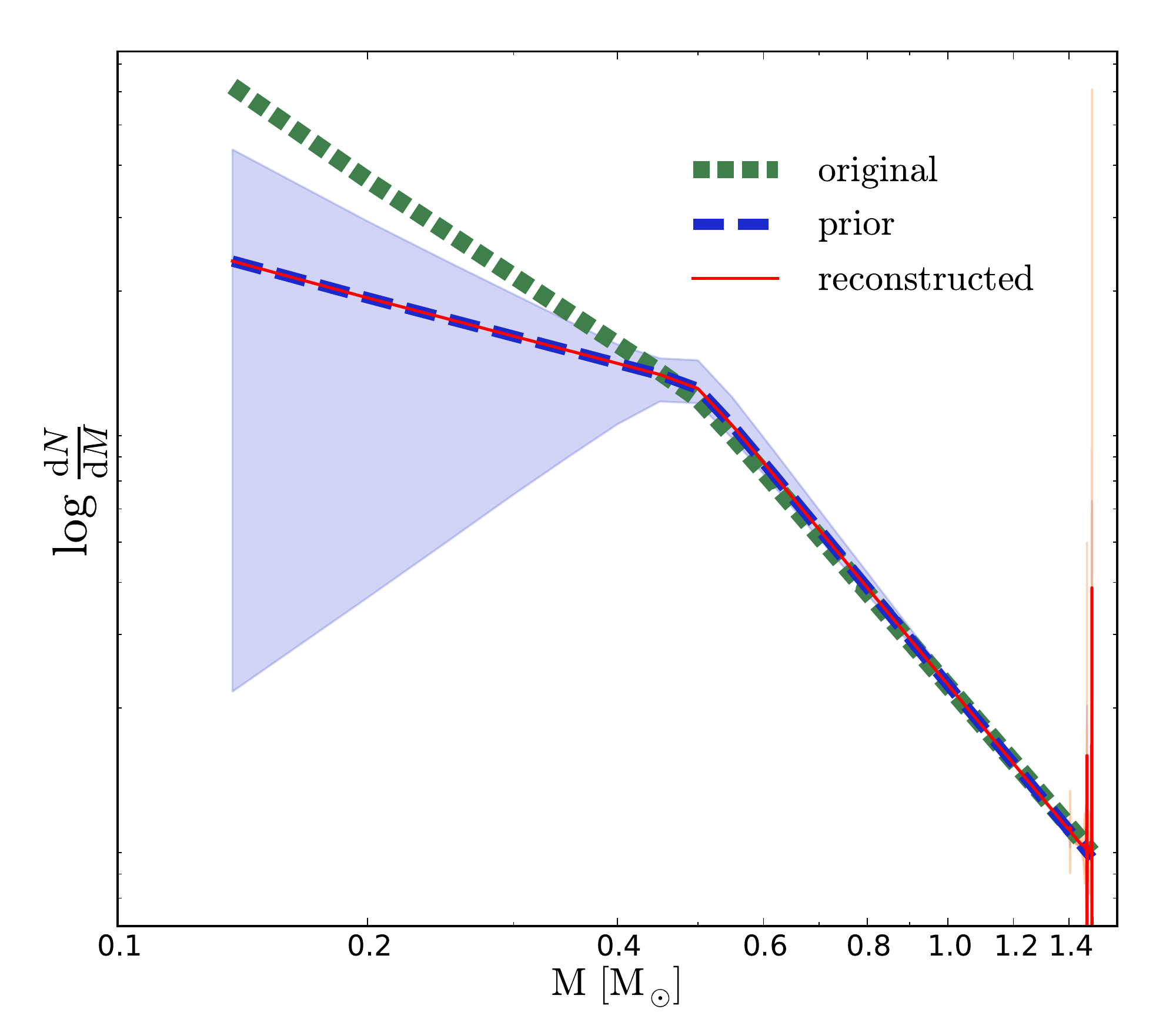}}
	\caption{As in Fig. \ref{summary_MILES1} for \texttt{MILES1} with $t = 3.2$ Gyr and $\mathrm{[Fe/H]=0.22}$ but now using the stellar templates created using the interpolator on the MILES website.}
	\label{summary_MILES1_MILES-interpolator}
\end{figure}

\begin{figure}
	\centering
		{\includegraphics[width=0.99\columnwidth]{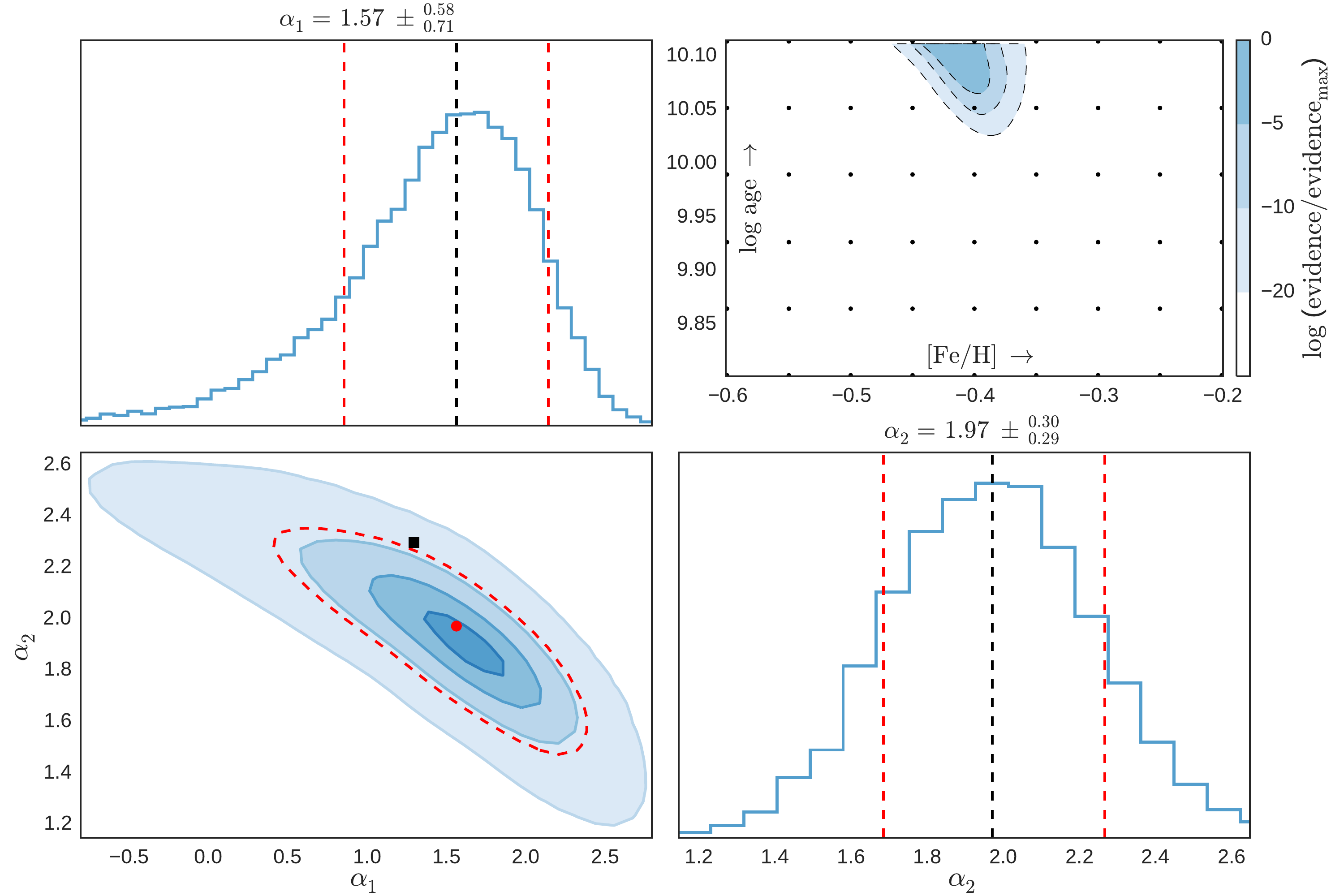}}
		{\includegraphics[width=0.9\columnwidth]{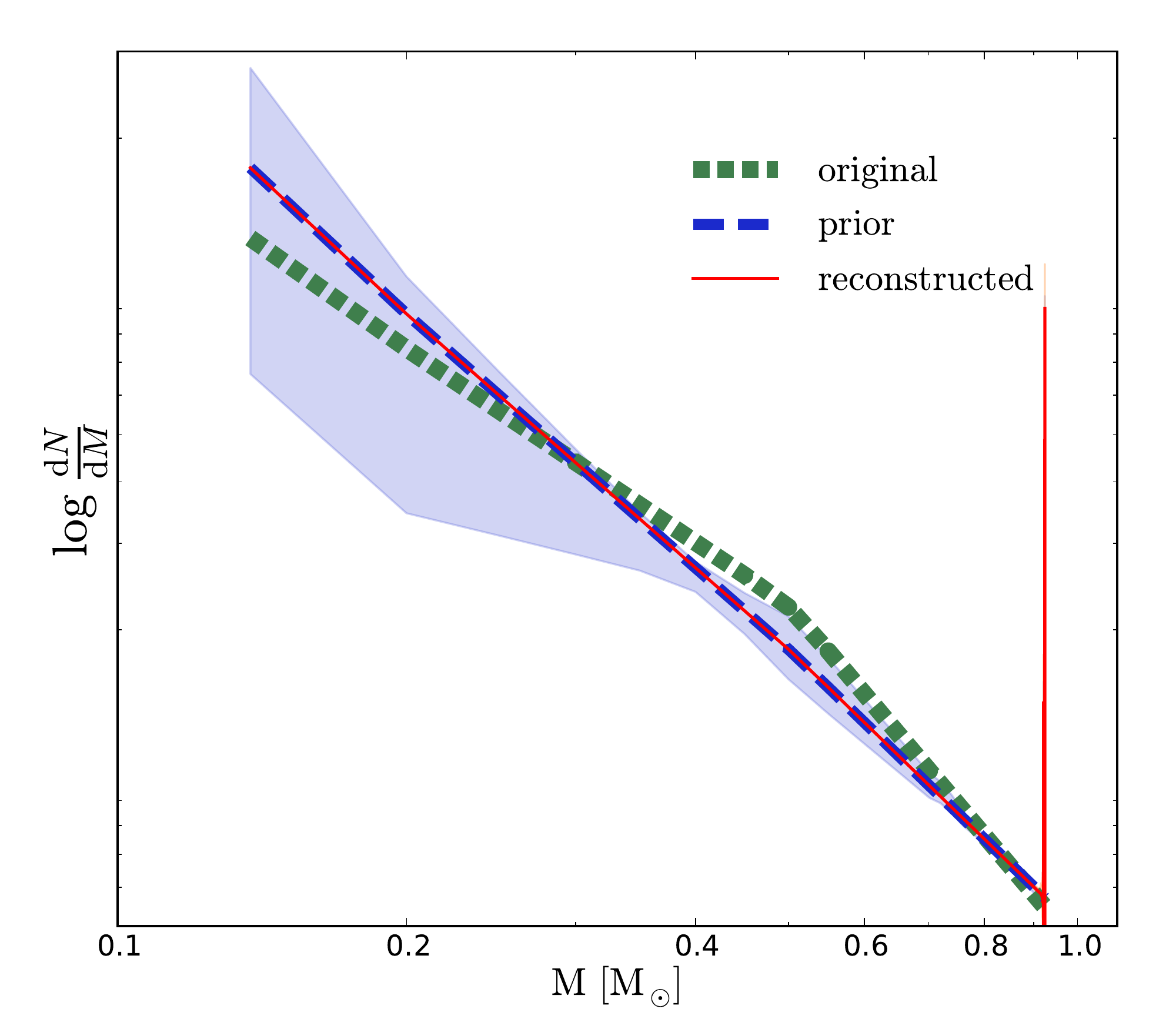}}
	\caption{As in Fig. \ref{summary1} for \texttt{MILES3} with $t = 12.6$ Gyr and $\mathrm{[Fe/H]=-0.40}$.}
	\label{summary_MILES3}
\end{figure}

\begin{figure}
	\centering
		{\includegraphics[width=0.99\columnwidth]{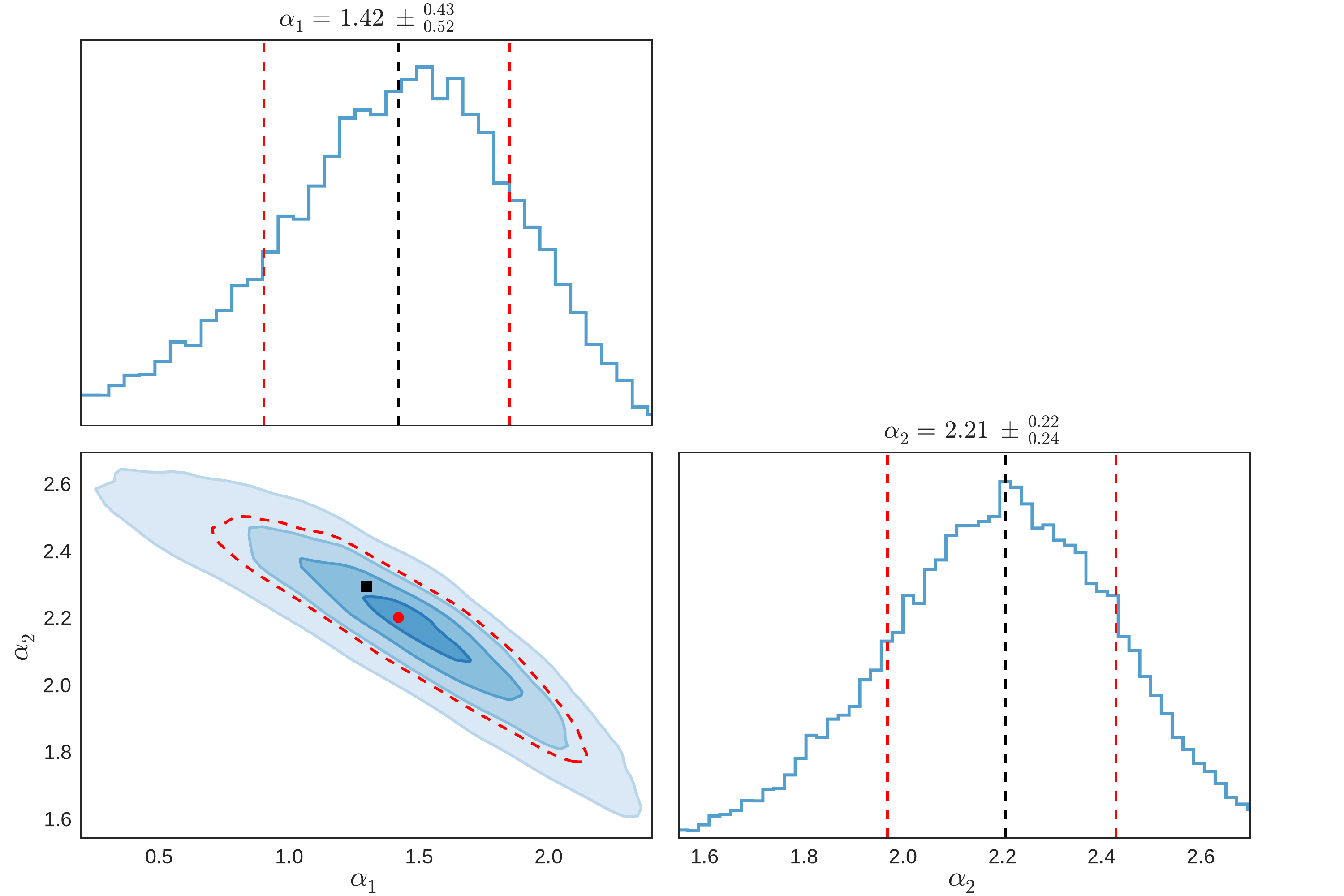}}
		{\includegraphics[width=0.9\columnwidth]{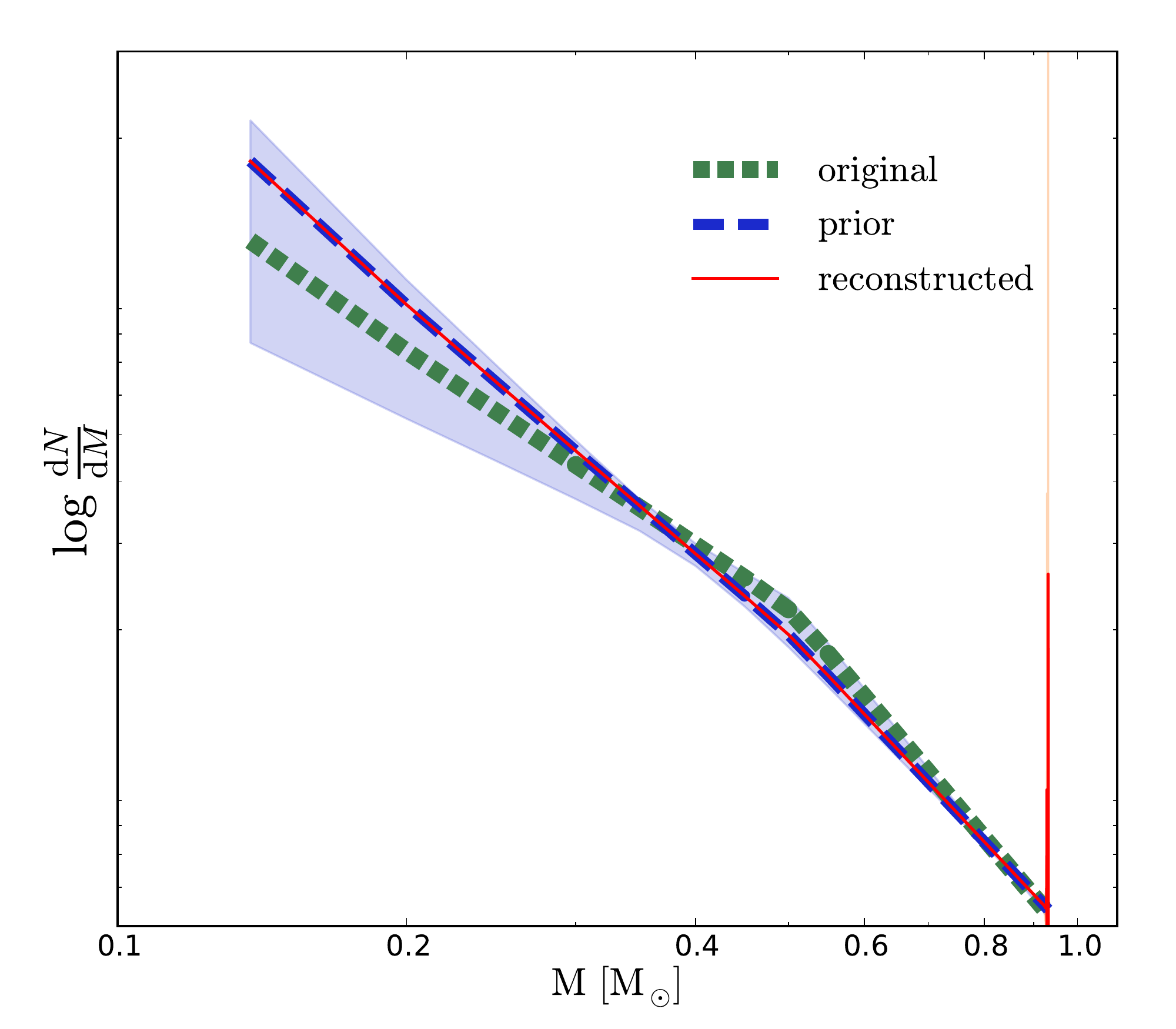}}
	\caption{As in Fig. \ref{summary_MILES3} for \texttt{MILES3} with $t = 12.6$ Gyr and $\mathrm{[Fe/H]=-0.40}$ but now using the stellar templates created using the interpolator on the MILES website.}
	\label{summary_MILES3_MILES-interpolator}
\end{figure}


\bsp	

\label{lastpage}
\end{document}